\documentclass[twocolumn]{autart}    

\usepackage{graphicx}          

\usepackage[utf8]{inputenc}
\usepackage{textcomp}

\usepackage{graphicx}
\usepackage{amsmath}

\usepackage{siunitx}
\usepackage{longtable,tabularx}
\usepackage{subfigure}
\usepackage{algorithm}
\usepackage{algpseudocode}
\usepackage{xcolor}
\usepackage{comment}
\usepackage{upgreek}
\usepackage{kotex}
\usepackage{graphicx}
\usepackage{booktabs}
\usepackage{hyperref}
\usepackage{setspace}
\usepackage{mathtools}
\usepackage{amsfonts}
\usepackage{amssymb}

\newcommand{\bs}[1]{\boldsymbol{#1}}

\graphicspath{{images/images_revised/}, {images/results_vF}}

\begin{document}

\begin{frontmatter}

\title{Constraint--Aware Mesh Refinement Method by Reachability Set Envelope of Curvature Bounded Paths\thanksref{footnoteinfo}} 
\thanks[footnoteinfo]{This paper was not presented at any conference.}
\author[KAIST]{Juho Bae}\ead{johnbae1901@kaist.ac.kr},    
\author[Nearthlab]{Ji Hoon Bai\thanksref{corrautinfo}}\ead{jihoon.bai@nearthlab.com},               
\author[Nearthlab]{Byung--Yoon Lee}\ead{byungyoon.lee@nearthlab.com},  
\author[Nearthlab]{Jun--Yong Lee}\ead{junyong.lee@nearthlab.com}
\thanks[corrautinfo]{Corresponding author at: Nearthlab, Seoul, 06246, Republic of Korea. Tel +821063088045.}
\address[KAIST]{Korea Advanced Institute of Science and Technology, Daejeon, 34141, Republic of Korea}                                               
\address[Nearthlab]{Nearthlab, Seoul, 06246, Republic of Korea}


\begin{abstract}                          
This paper presents an enhanced direct-method-based approach for the real-time solution of optimal control problems to handle path constraints, such as obstacles. The principal contributions of this work are twofold: first, the existing methods for constructing reachability sets in the literature are extended to derive the envelope of these sets, which determines the region swept by all feasible trajectories between adjacent sample points. Second, we propose a novel method to guarantee constraint violation--free between discrete states in two dimensions through mesh refinement approach. To illustrate the effectiveness of the proposed methodology, numerical simulations are conducted on real-time path planning for fixed--wing unmanned aerial vehicles. 
\end{abstract}

\end{frontmatter}

\section{Introduction}
Over the past decade, the need for rapid and efficient real--time algorithms to solve optimal control problems has intensified, particularly for onboard guidance problems such as the powered descent landing problem in~\cite{accikmecse2011lossless}, \cite{sagliano2016onboard}, and \cite{sagliano2018pseudospectral}, the re--entry guidance problem in~\cite{liu2016entry}, \cite{wang2017constrained}, \cite{wang2020improved}, and \cite{bae2022}, and planning of unmanned--aerial vehicle(UAV) trajectories in~\cite{oh2019} and \cite{tordesillas2019}. 

Recent advancements in real--time optimal control involve direct methods combined with sequential convex programming (SCP)~\cite{accikmecse2011lossless}, \cite{liu2016entry}, \cite{szmuk2017successive}, \cite{mao2018}, \cite{szmuk2020}, and \cite{bae2022}. The direct method partitions the domain into multiple intervals separated by sample points, or \textit{nodes}. Subsequently, polynomial interpolation is employed to determine the values of state and control variables between these nodes, with the specific interpolation method chosen based on the circumstances. In a typical optimal control problem(OCP) that involves path constraints, the \textit{forbidden regions}, where the path constraint restrains the state variables to enter, is a key focus of our study. Obstacles from the UAV trajectory planning problems, dynamics pressure bound from the re--entry guidance problem are specific examples of such forbidden regions. 

Direct method formulation applies path constraints at each mesh points, preventing state variables from entering forbidden regions. However, the generated trajectory in between the mesh points may trespass the forbidden regions. This is commonly known as the \textit{inter--sample collision problem}~\cite{dueri2017}. Not only confined in the SCP framework based approaches, it is seriously considered as one of the most critical difficulties in practical application of the general direct method based algorithms~\cite{tordesillas2019}. Despite the term \textit{collision}, path constraints extend beyond collision avoidance, as seen in re-entry guidance problem in~\cite{liu2016entry} and \cite{bae2022}. In this context, we will refer such an issue as the problem of obtaining a \textit{path constraint violation--free} trajectory. 

To address this challenge, a method for implementing path constraints involving cylindrical obstacles was proposed in~\cite{dueri2017}. The core idea involved estimating the shortest distance between an obstacle's center and the path connecting two adjacent nodes. By ensuring that this minimum distance exceeds the obstacle's radius, a collision-free constraint was established. However, this method relies on linearization of the dynamics, which introduces truncation errors. Moreover, the most significant drawback of this method is the loss of optimality. When there are only a limited number of nodes in close proximity to an obstacle, the resulting collision-free constraints tend to enforce a trajectory that maintains a significant distance from the obstacle in an overly cautious manner. This can result in discrete trajectories that are far from continuous time optimal trajectory, the actual correct solution. 

Motivated by these trends, our research offers improved solutions for the general inter--sample collision problems, guaranteeing path constraint violation--free. Considering the aforementioned issues, we recognized the importance of incorporating additional nodes, or refining mesh, to effectively apply collision-free constraints. 

Previous studies have adopted additional nodes in direct methods, primarily to reduce dynamics interpolation errors during discretization. Several collocation methods have been explored, primarily centered around the \textit{Pseudo-Spectral method}~\cite{garg2010unified}. Various frameworks have been developed in~\cite{sagliano2019generalized}, \cite{liu2015adaptive}, \cite{liu2017adaptive}, and \cite{patterson2015ph} to adaptively refine the mesh intervals between the nodes by strategically inserting additional nodes in suitable locations. It is worth noting that the conventional mesh refinement methods primarily focus on addressing interpolation errors arising from specific attributes of the dynamics constraints, such as smoothness, but fail to consider the arbitrary placement of forbidden regions during the mesh refinement process~\cite{liu2015adaptive}. Consequently, potential risk of violation into the forbidden region cannot be detected using these methods. Additionally, arbitrarily increasing the number of nodes is not a viable solution as it leads to a significant increase in computational costs. 

Our proposed method targets path constraint violation--free trajectories, while minimizing mesh insertion for computational efficiency. The desired property is attained by analytically determining the bounds within which the continuous-time trajectory can exist between two fixed adjacent nodes. Such concept is an expansion the idea of reachability set, or attainability set, of curvature bounded paths in~\cite{patsko2003}, \cite{cockayne1975}, and \cite{gusev2017} into the envelope of multiple reachability sets. Whenever a solution is obtained from the direct method, the violation of the aforementioned trajectory bound into the forbidden region indicates the potential risk of path constraint violation at such mesh interval. Our approach starts with formulating and proving a method to construct such trajectory bounds, followed by a technique to assess any bound violations into forbidden regions. 
Furthermore, a proof of the convergence of the overall proposed algorithm is provided, along with a numerical demonstration of its computational cost, particularly in light of its intended use in onboard applications. 

Let us first introduce some notations in the related literatures regarding curvature bounded paths in plane. From hereon, whenever a planar curve has an upper bound of $\kappa_m$ on its curvature, let us denote the part of the trajectory with constant signed curvature of $\pm\kappa_m$ by \textit{C}, as it is part of a circular arc. The part of the trajectory with zero curvature is denoted by \textit{S}, as it is a straight segment. For instance, if a trajectory initially follows a circular arc and then subsequently follows a segment, then such trajectory is denoted by CS. It was further proved in~\cite{patsko2003} that the corresponding control input that brings the endpoint of the trajectory to the boundary of the reachability set in a sense of terminal location and direction, contains at most two switchings in the control. Moreover, it was also proved that if the trajectory contains two switchings, then it is in the class of CSC or CCC. For simpler case when dealing only terminal location, it was proved that the boundary of the reachability set can be reached by trajectories of class CC and CS~\cite{cockayne1975}. 

Unlike the ingenious geometric arguments in~\cite{dubins1957} and \cite{cockayne1975}, the optimal control theory based approaches in~\cite{Lee1967} and \cite{patsko2003} further motivates us to formulate the considered problem of obtaining the trajectory bound in a form of optimal control problem. By solving the necessary conditions of optimality given by the Pontryagin Maximum Principle(PMP), it is proved that the boundary of the trajectory bound can be constructed using trajectories consist of at most five C and S segments. In addition to the mathematical proof, further intuitive insight is also provided. Consequently, we adopt the idea of covering the obtained trajectory bound by multiple \textit{rectangular patches}, thereby developing a method to distinguish whether the constructed bounds encroach upon the forbidden region. This analytical approach enables efficient and precise assessment of whether additional mesh insertion is required at a certain mesh interval, thereby achieving path constraint violation--free trajectories while minimizing the number of additional mesh insertion. The proposed method can be easily integrated into general direct formulation of optimal control problems, and the termination of the algorithm in a finite number of iteration is proved. The performance and computational efficiency of the algorithm is validated through numerical simulations with 2D fixed--wing dynamics. 
%

The remaining sections of this paper are organized as follows. Section~\ref{sec:02} presents the problem statement and direct formulation of the Bolza optimal control problem. In Section~\ref{sec:03}, the construction method of the reachability set envelope of curvature bounded paths is derived. Section~\ref{sec:04} describes the methodology to detect intersection between the forbidden region and the obtained envelope. Section~\ref{sec:05} provides numerical demonstration results to further validate the effectiveness of the proposed approach. Finally, in Section~\ref{sec:06}, we conclude this work by highlighting the key contributions of this paper. 




\section{Direct Formulation of the Optimal Control Problem}\label{sec:02}
In this paper, the below form of Bolza optimal control problem is considered: 
\begin{equation} \label{eq:1}
\begin{array}{l}
{\rm{minimize}}\quad J = \Phi \left( {\boldsymbol{z}\left( {-1} \right), t_0, \boldsymbol{z}\left( {+1} \right), t_f} \right) \\
+  \frac{t_f - t_0}{2} \int_{-1}^{+1} {F\left( {\boldsymbol{z}\left( \tau \right), \boldsymbol{u}(\tau), \tau} \right)d\tau}  \\ 
 \text{subject to} \quad \frac{d\boldsymbol{z}}{d\tau} = \frac{t_f - t_0}{2} f\left( {\boldsymbol{z}\left( \tau \right), \boldsymbol{u}\left( \tau \right), \tau} \right), \\
 h\left( {\boldsymbol{z}\left( \tau \right)} \right) \ge 0,\quad \Psi \left( {\boldsymbol{z}\left( {-1} \right), t_0, \boldsymbol{z}\left( {+1} \right), t_f} \right) \ge 0 \\ 
 \end{array}
\end{equation}
where the state variables are denoted by $\boldsymbol{z}(\tau) \in \mathbb{R}^{n_{\boldsymbol{z}}}$ and the control by $\boldsymbol{u}(\tau) \in \mathbb{R}^{n_{\boldsymbol{u}}}$. The functions $f: \mathbb{R}^{n_{\boldsymbol{z}} + n_{\boldsymbol{u}} + 1} \mapsto \mathbb{R}^{n_{\boldsymbol{z}}}$, $F: \mathbb{R}^{n_{\boldsymbol{z}} + n_{\boldsymbol{u}} + 1} \mapsto \mathbb{R}$, $\Phi: \mathbb{R}^{2n_{\boldsymbol{z}} + 2} \mapsto \mathbb{R}$, $h: \mathbb{R}^{n_{\boldsymbol{z}}} \mapsto \mathbb{R}^{n_h}$, $\Psi: \mathbb{R}^{2n_{\boldsymbol{z}} + 2} \mapsto \mathbb{R}^{n_{\Psi}}$ denotes the dynamics, cost integrand, boundary cost, path inequality constraint, and boundary constraint functions where $n_{\Psi}$ and $n_h$ denotes the number of each constraints. If the final time $t_f$ is considered as a free variable, it is also considered as one of the control variable. Throughout the paper, the $i$--th component function of any function $g(\cdot)$ is denoted by $g_i(\cdot)$. An assumption on the path constraints is that each component functions are of bivariate form. (i.e. For each $k \in \{1, \dots, n_h\}$, the function $h_k$ depends on two state variables and remains constant with respect to the other remaining state variables.) Such formulation includes numerous practical examples such as the re-entry guidance problem in~\cite{liu2016entry}, problems with 2--dimensional dynamics under holonomic constraints~\cite{dueri2017}, and missile guidance problems with no--fly--zones. Extension of the proposed method to higher dimensions is left for the future work. 
Suppose now the interval $[-1, +1]$ is discretized by \textit{mesh points(nodes)} : $-1 = \tau_1, \tau_2, \dots, \tau_{N-1}, \tau_N = +1$ into total of $N-1$ number of mesh intervals(\textit{meshes}) $[\tau_{j}, \tau_{j+1}], \;\; j = 1,\dots, N-1$. The lengths of the mesh intervals are denoted as $\Delta \tau_j = \tau_{j+1} - \tau_j, \;\; j = 1,\dots, N-1$. Let $\boldsymbol{z}_j$ and $\boldsymbol{u}_j$ denote the values at $\tau_j$, $\boldsymbol{z}(\tau_j)$ and $\boldsymbol{u}(\tau_j)$, respectively. Then the discretization step based on numerical integration with trapezoidal rule yields the below discretized problem Eq.~\eqref{eq:2}. 
%
%
\begin{equation} \label{eq:2}
\begin{array}{l}
{\rm{minimize}}\quad J = \Phi \left( {\boldsymbol{z}_1, t_0, \boldsymbol{z}_N, t_f } \right) \\ 
+ \Bigg[\frac{{\Delta \tau_1 }}{2}F\left( {\boldsymbol{z}_1, \boldsymbol{u}_1, \tau_1} \right) + \left( {\sum\limits_{j = 2}^{N - 1} {\frac{{\Delta \tau_{j - 1}  + \Delta \tau_j }}{2}}} F\left( {\boldsymbol{z}_j, \boldsymbol{u}_j, \tau_j} \right) \right) \\ 
 + \frac{{\Delta \tau_{N - 1} }}{2}F\left( {\boldsymbol{z}_N, \boldsymbol{u}_N, \tau_N} \right) \Bigg] \times \frac{t_f - t_0}{2} \\ 
 \text{subject to} \quad \boldsymbol{z}_{j + 1}  - \boldsymbol{z}_j  = \frac{t_f - t_0}{2} \frac{{\Delta \tau_j }}{2}\left[ f\left( {\boldsymbol{z}_j, \boldsymbol{u}_j, \tau_j} \right) \right. \\
 \left. + f\left( {\boldsymbol{z}_{j+1}, \boldsymbol{u}_{j+1}, \tau_{j+1}} \right) \right], \\ 
 h\left( {\boldsymbol{z}_j} \right) \ge 0, \quad j = 1, \ldots ,N - 1, \\
 \Psi \left( {\boldsymbol{z}_1, t_0, \boldsymbol{z}_N, t_f } \right) \ge 0 \\ 
\end{array}
\end{equation}

Among the various interpolation methods available, we opted for the standard collocation method instead of the well-known Pseudo--Spectral method in this paper. Such decision is based on two main reasons. Firstly, the Pseudo--Spectral method tends to significantly improve dynamics interpolation error at the cost of increase in computational burden~\cite{sagliano2018pseudospectral}. However, in the context of our method, minimizing dynamics interpolation error is not the primary objective. As a result, incorporating the Pseudo-Spectral method would only add computational overhead without the contribution of enhancing the efficiency of mesh refinement method by minimizing the number of additional meshes. Secondly, the scope of application of the proposed method in this paper need not to be limited by the collocation methods. Therefore, we aimed to keep the basic problem formulation as simple as possible. By adopting the standard collocation method within the SCP framework, we strike a balance between computational efficiency and achieving path constraint violation--free trajectories, while maintaining flexibility to extend the method to Pseudo-Spectral formulations if needed.

\section{Reachability Set Envelope of Curvature Bounded Paths}\label{sec:03}
In this section, we develop a construction methodology for the region in which a trajectory connecting two adjacent mesh points can exist while the state variables at the two mesh points are assumed to be given. This construction method is developed by introducing the concept of the \textit{Reachability Set} and its envelope. Let us commence by presenting the problem settings and associated notations. 
\subsection{The Reachability Set of Curvature Bounded Paths} \label{sec3subsec:00}
\begin{figure}[ht] 
	\begin{center}
	\resizebox{64mm}{!}{\includegraphics{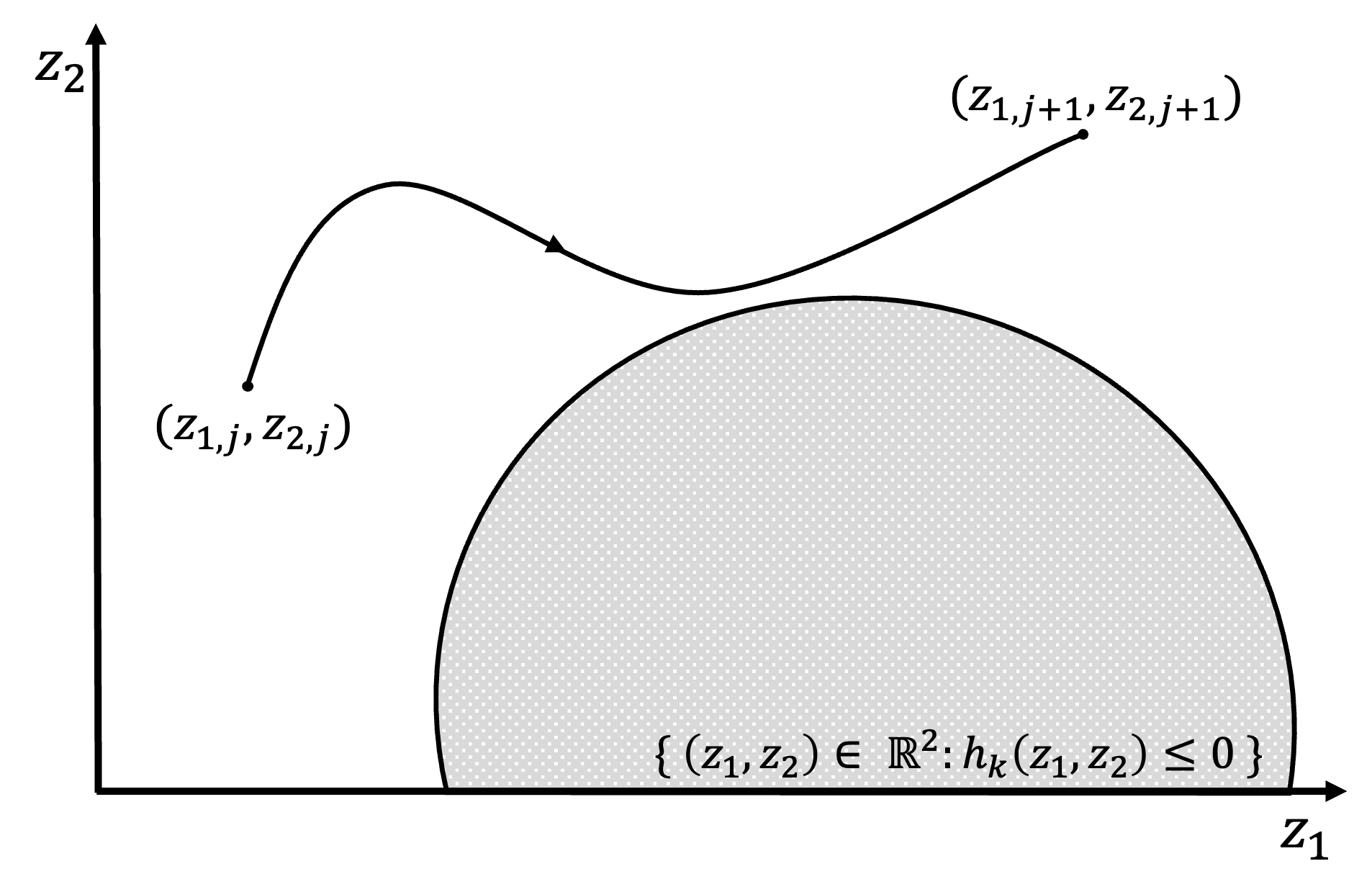}}
	\caption{Example of a curve $(z_{1}(\tau), z_{2}(\tau))$ and a forbidden region} \label{fig_sec3_1}
	\end{center}
\end{figure}

Let us denote $z_i (\tau_j)$, the $i$--th state variable at $\tau_j$, by $z_{i,j}$. Since the path inequality constraints are of bivariate form, we may further assume without loss of generality that the path constraints are applied on the first and second state variables, $h_k(z_{1,j}, z_{2,j}) \ge 0$ and $h_k(z_{1,j+1}, z_{2,j+1}) \ge 0$, at two adjacent mesh points $\tau_j$ and $\tau_{j+1}$. Define a projection map $\pi: \mathbb{R}^{n_{\boldsymbol{z}}} \mapsto \mathbb{R}^2$, $\pi(z_1, z_2, \dots , z_{n_{\boldsymbol{z}}}) = (z_1, z_2)$. 
Whenever the states, $\boldsymbol{z}_j$ and $\boldsymbol{z}_{j+1}$, and the controls, $u_j$ and $u_{j+1}$, at the mesh points, $\tau_j$ and $\tau_{j+1}$, are determined respectively, we can consider a two--dimensional curve $(\pi \circ \boldsymbol{z})(\tau) = (z_1(\tau), z_2(\tau)) : [\tau_j, \tau_{j+1}] \mapsto \mathbb{R}^2$ that connects the two points $(\pi \circ \boldsymbol{z})(\tau_j) = (z_{1,j}, z_{2,j})$ and $(\pi \circ \boldsymbol{z})(\tau_{j+1}) = (z_{1,j+1}, z_{2,j+1}) \in \mathbb{R}^2$, which is depicted in Fig.~\ref{fig_sec3_1}. 
Let us denote the curvature of such curve by $\kappa(\tau)$. 
We additionally assume that the curve is regular, the function $f$ is $C^1$, and the control input is piecewise $C^1$, which guarantees that the curve is piecewise $C^2$. Furthermore, it is assumed that an upper bound on the curvature exists, denoted by $\kappa_m \geq \sup\limits_{\tau \in [\tau_j, \tau_{j+1}]} \kappa(\tau)$. The ultimate goal of this section is to find the reachable region by the curve $(\pi \circ \boldsymbol{z})$ under such curvature constraint. For further procedure, we normalize the curve $\pi \circ \boldsymbol{z}$ by its arc length $L = \int_{\tau_j}^{\tau_{j+1}} \| (\pi \circ \boldsymbol{z})'(\tau) \| d\tau$ for sake of simplicity. Then the trajectory, or the image of the normalized curve, $\frac{1}{L}(\pi \circ \boldsymbol{z})([\tau_j, \tau_{j+1}])$, of length one can be parametrized by a regular, curvature bounded curve parametrized by arc length. We further represent such curve as $\alpha_u = (x, y) : [0, 1] \mapsto \mathbb{R}^2$, where $\alpha_u([0, 1]) = \frac{1}{L}(\pi \circ \boldsymbol{z})([\tau_j, \tau_{j+1}])$. 
Then $\alpha_u$ can be characterized by the below dynamics: 
\begin{equation} \label{eq:2-1}
	\dot{\boldsymbol{\chi}} = \begin{pmatrix} \dot{x} \\ \dot{y} \\ \dot{\gamma} \end{pmatrix} = \begin{pmatrix} \cos\gamma \\ \sin\gamma \\ u \end{pmatrix} \equiv f_{\chi}(\boldsymbol{\chi}, u)
\end{equation}
where $\gamma$ is the direction of the curve $\alpha_u$, and the state variables $[x \; y \; \gamma]^T$ are denoted by $\boldsymbol{\chi}$. $u$ is regarded as the signed curvature of the planar curve $\alpha_u$, and the upper bound on curvature $|u| \leq \kappa_m' = L\kappa_m$ is considered. Let $\Omega$ denote the admissible control set, $\Omega = \{ u : [0, 1] \mapsto [-\kappa_m', \kappa_m'] \enspace | \enspace \text{piecewise } C^1 \}$. 
 Since $\dot{z}_1(\tau) = f_1(\boldsymbol{z}, \boldsymbol{u}, \tau)$ and $\dot{z}_2(\tau) = f_2(\boldsymbol{z}, \boldsymbol{u}, \tau)$, it follows directly that the initial and final directions, $\gamma(0)$ and $\gamma(1)$, of the trajectory depicted in Fig.~\ref{fig_sec3_1} can be determined by appropriately applying inverse tangent to the ratios $f_2(\boldsymbol{z}_j, \boldsymbol{u}_j, \tau_j) / f_1(\boldsymbol{z}_j, \boldsymbol{u}_j, \tau_j)$ and $f_2(\boldsymbol{z}_{j+1}, \boldsymbol{u}_{j+1}, \tau_{j+1}) / f_1(\boldsymbol{z}_{j+1}, \boldsymbol{u}_{j+1}, \tau_{j+1})$. Hence, the values of $\boldsymbol{\chi}(0)$ and $\boldsymbol{\chi}(1)$ can be specified whenever the state and controls at the both mesh points are given. This dynamics is often referred as \textit{Dubin's Car} dynamics~\cite{dubins1957}. For the projection $\pi_1 : \mathbb{R}^3 \mapsto \mathbb{R}^2$, $\pi_1 (x, y, \gamma) = (x, y)$, it is evident that firstly, $\pi_1(\boldsymbol{\chi}) = \alpha_u$ and secondly, the image $\pi_1(\boldsymbol{\chi})([0, 1]) = \alpha_u([0, 1])$ is the image $(\pi \circ \boldsymbol{z})([\tau_j, \tau_{j+1}])$ scaled to $\frac{1}{L}$ size. Hence, we aim to first find the reachable region of the unit length curve $\alpha_u$, and scale it back to find the reachable region of $\pi \circ \boldsymbol{z}$. 
 By applying an appropriate rigid motion, we may presume that the initial state $\boldsymbol{\chi}_0 = \left[ {x_0 ,\;y_0 ,\;\gamma_0} \right]^T = \boldsymbol{0}$. (i.e. $\alpha_{u}(0) = (0, 0)$ and $\alpha_{u}'(0) = (1, 0)$). Consequently, we can introduce the below definition of \textit{Reachability Set}, which was introduced and studied extensively in~\cite{Lee1967} and \cite{patsko2003}.
\begin{defn}
	Reachability Set~\cite{Lee1967} and \cite{patsko2003} \\
	\begin{equation}
		\mathcal{G}(T) \equiv \{ \boldsymbol{z} : \exists u \in \Omega \text{ s.t. } \boldsymbol{\chi}(T) = \boldsymbol{z} \}
		\nonumber
	\end{equation}
\end{defn}
The set $\mathcal{G}(T)$ represents the extent to which the state variables can either deviate from or remain close to the initial point over a given time $T$. It is worth noting that the terminal state $\boldsymbol{\chi}(1)$ is not considered in this definition; instead, the primary emphasis is placed on determining the region of feasible terminal states. It was further stated and proved in~\cite{Lee1967} that any trajectory $\{ \bar{\boldsymbol{\chi}}, \bar{u} \}$ reaching the boundary point of $\mathcal{G}(T)$ at time $T$ satisfies the below PMP: 
\begin{equation} \label{def1PMP}
	\begin{array}{l}
		\dot{\boldsymbol{p}}(s) = -\frac{\partial \mathcal{H}}{\partial \boldsymbol{\chi}} \\
		\mathcal{H}(\boldsymbol{p}, \bar{\boldsymbol{\chi}}, \bar{u}) = \max\limits_{u \in \Omega} \mathcal{H}(\boldsymbol{p}, \bar{\boldsymbol{\chi}}, u) \quad \text{almost everywhere}
	\end{array}
\end{equation}
where the Hamiltonian is defined as $\mathcal{H} = \boldsymbol{p}^T f_{\chi}$. 
The aforementioned statement is not restricted to the dynamics of Dubins' car but is applicable to general dynamics of class $C^1$ that have bounded response for a measurable admissible control set~\cite{Lee1967}. For the Dubins' car dynamics specifically, it was proved in~\cite{patsko2003} that any trajectory reaching the boundary of $\mathcal{G}(T)$ is of class CCC or CSC. 

An additional useful information about $\mathcal{G}(T)$ is that it is compact. Such conclusion is obtained from the development in~\cite{Lee1967} and that $V(\bs{\chi}) = \{ f_{\bs{\chi}}(\bs{\chi}, u) : u \in \Omega \}$ is obviously convex for each fixed $\bs{\chi}$. 

When the focus of reachability is solely on the locations, the following set, which is essentially the projection of $\mathcal{G}(T)$ onto the $xy$-plane, can be defined: 
\begin{defn}
	\begin{equation}
		\mathcal{G}'(T) \equiv \{ \pi_1(\boldsymbol{z}) : \exists u \in \Omega \text{ s.t. } \boldsymbol{\chi}(T) = \boldsymbol{z} \}
		\nonumber
	\end{equation}
\end{defn}
It was proved in~\cite{cockayne1975} that any boundary point of $\mathcal{G}'(T)$ can be reached by a trajectory of class CC or CS. With such considerations, let us define the set below, which is of our primary interest in this paper. Additionally, we denote the tangent bundle of a regular surface $S$ by $\mathbb{T}S$. 
\begin{defn}
	Given $\mathbb{X} \in \mathbb{T}\mathbb{R}^2$ and $\kappa_m \in \mathbb{R}_{>0}$, define the below sets where the curve $\chi$ is defined by $\alpha_u$ as Eq.~\eqref{eq:2-1}. 
	\begin{equation}
	\begin{array}{l}
		\mathcal{B}(\mathbb{X}, \kappa_m) \equiv \cup  \boldsymbol{\chi}([0, 1]) \text{ s.t. } u \in \Omega, \\
		(\alpha_u(0), \dot{\alpha}_u(0)) = ((0, 0), (1, 0)), (\alpha_u(1), \dot{\alpha}_u(1)) = \mathbb{X}
		\nonumber
	\end{array}
	\end{equation}
	\begin{equation}
	\begin{array}{l}
		\mathcal{B}'(\mathbb{X}, \kappa_m) \equiv \cup \alpha_u([0, 1]) \text{ s.t. } u \in \Omega, \\
		(\alpha_u(0), \dot{\alpha}_u(0)) = ((0, 0), (1, 0)), (\alpha_u(1), \dot{\alpha}_u(1)) = \mathbb{X}
		\nonumber
	\end{array}
	\end{equation}
\end{defn}
The set $\mathcal{B}(\mathbb{X}, \kappa_m)$ is the collection of all points such that there exists some curve passing through the point that connects $((0,0), (1,0)) \in \mathbb{T}\mathbb{R}^2$ and $\mathbb{X}$, with a curvature bounded above $\kappa_m$. The boundary of $\mathcal{B}(\mathbb{X}, \kappa_m)$ is the envelope of those possible curves. The set $\mathcal{B}'(\mathbb{X}, \kappa_m)$ is the projection of $\mathcal{B}(\mathbb{X}, \kappa_m)$ onto $xy$--plane, neglecting the bound on the last component $\gamma$. By the aforementioned definitions, the task of identifying the bounded region in which the state variables of interest can exist is transformed into the problem of determining the set $\mathcal{B}'(\mathbb{X}, \kappa_m)$. Therefore, the remaining subsections are focused on developing a methodology to construct the boundary of $\mathcal{B}'(\mathbb{X}, \kappa_m)$. 
%
%
\subsection{The Original Optimal Control Problem}\label{sec3subsec:01}
Consider the OCP [$\textbf{P0}$] below, where $f_{\chi}(\cdot)$ is as defined in Eq.~\eqref{eq:2-1}. 

[$\textbf{P0}$] : maximize
\begin{equation} \label{eq:3}
	\enspace J  =  \Phi \left(\boldsymbol{\chi} \left( T \right) \right) 
\end{equation}
subject to
\begin{equation} \label{eq:4}
\begin{array}{l}
\dot{\boldsymbol{\chi}}\left( \tau \right) = f_{\chi}\left( \boldsymbol{\chi} \left( \tau \right), u \left( \tau \right) \right), \\
\boldsymbol{\chi} \left( 0 \right) = \boldsymbol{\chi}_0, \\
-\kappa_{m} \leq u \leq \kappa_{m}
\end{array}
\end{equation}
In the above formulation, a terminal cost function $\Phi : \mathbb{R}^3 \mapsto \mathbb{R}$ with a nonzero constant gradient field in an arbitrary direction is considered. The OCP [$\textbf{P0}$] is defined per every such $\Phi$. The below proposition provides some intuition for Eq.~\eqref{def1PMP}. 

\begin{prop} \label{prop:1}
The collection of PMP sense solutions of every [$\textbf{P0}$] is the set of solutions of Eq.~\eqref{def1PMP}. 
\end{prop}
\begin{pf*}{Proof.}
	The PMP condition of [$\textbf{P0}$] consists of the conditions of Eq.~\eqref{def1PMP}, the transversality condition $\boldsymbol{p}(T) = p_0 \nabla_{\boldsymbol{\chi}} \Phi$, $p_0 \geq 0$, and the nontriviality condition applied to $\{ \boldsymbol{p}, p_0 \}$. Hence for any solution of Eq.~\eqref{def1PMP}, nontriviality of $\boldsymbol{p}$ directly implies nontriviality of the pair $\{ \boldsymbol{p}, p_0 \}$. The transversality condition $\boldsymbol{p}(T) = p_0 \nabla_{\boldsymbol{\chi}} \Phi$ can be met by choosing appropriate $p_0 \geq 0$ and $\nabla_{\boldsymbol{\chi}} \Phi \in \mathbb{R}^{3} \setminus \{ \boldsymbol{0} \}$. Therefore, if $\{ \boldsymbol{\chi}, u, \boldsymbol{p} \}$ is a solution of Eq.~\eqref{def1PMP}, then it is also a PMP sense solution of some [$\textbf{P0}$]. 
	Conversely, suppose the pair $\{ \boldsymbol{\chi}, u, \boldsymbol{p}, p_0 \}$ is a solution of some [$\textbf{P0}$]. If $\boldsymbol{p} \equiv 0$, then the transversality condition further implies $p_0 = 0$ since $\nabla_{\boldsymbol{\chi}} \Phi \neq \boldsymbol{0}$. This violates the nontriviality condition of [$\textbf{P0}$]. Therefore, the nontriviality condition must hold for $\boldsymbol{p}$ alone, and hence the pair $\{ \boldsymbol{\chi}, u, \boldsymbol{p} \}$ becomes a solution of Eq.~\eqref{def1PMP}. 
\qed
\end{pf*}
The fact that any boundary point of $\mathcal{G}(T)$ is a solution of [$\textbf{P0}$] is intuitive in a sense that [$\textbf{P0}$] measures how far or close the endpoint of the trajectory can be located in a certain direction. To invoke the concept of `direction', it should be noted that the nonvanishing gradient of $\Phi$ must be presumed. Indeed, such condition played an important role related to the nontriviality condition in the above proof. 
It is important to note some specific discriminations between the existing literatures~\cite{patsko2003}, \cite{gusev2017}, and \cite{cockayne1975}, and our situation. 
The bound we aim to obtain is not the reachability set itself at a certain moment, but it is the envelope, or the collection, of the reachability sets per each terminal times in $[0, 1]$. In other words, the problem considered in the existing literatures utilized the necessary conditions that optimize the \textit{terminal} state, while the considered problem in this paper requires optimizing the state at a certain moment, $T$, \textit{during the trajectory}. Then collecting all such optimized state over $T \in [0, 1]$ would form an envelope, which is the desired bound for mesh refinement. The subsequent subsection is devoted to handle such differences.  
\subsection{Augmentation of Two Optimal Control Problems}\label{sec3subsec:02}
Let us denote the desired terminal state by $\boldsymbol{\chi}_f = \left[ {x_f ,\;y_f ,\; \gamma_f} \right]^T$ where $\gamma_f \in (-\pi, \pi]$. Then considering potential cycles, any curve with terminal value $\left[ {x_f ,\;y_f ,\;\gamma_f + 2k\pi} \right]^T$ where $k \in \mathbb{Z}$ satisfies the terminal state constraint. 
Consider the below region where the points on a trajectory can be located at a specific time $T \in [0, 1]$. 
\begin{defn}
	
	\begin{equation}
	\begin{array}{l}
		\mathcal{R}_k(T) \equiv \{ \boldsymbol{z} : \exists u \in \Omega \text{ s.t. } \boldsymbol{\chi}(0) = \boldsymbol{\chi}_0, \\
		\boldsymbol{\chi}(T) = \boldsymbol{z}, \boldsymbol{\chi}(1) = \left[ {x_f ,\;y_f ,\;\gamma_f + 2k\pi} \right]^T \}
		\nonumber
	\end{array}
	\end{equation}
	
\end{defn}
Similar definition as $\mathcal{B}'(\mathbb{X}, \kappa_m)$ is made as follows: $\mathcal{R}_k'(T) \equiv \pi_1(\mathcal{R}_k(T))$. 
In relation to finding such regions, we take into account the following OCP: 

[$\textbf{P0}_{T, k}$] : maximize
\begin{equation} \label{eq:P0T-1}
	\enspace J  =  \Phi \left(\boldsymbol{\chi} \left( T \right) \right) 
\end{equation}
subject to
\begin{equation} \label{eq:P0T-2}
\begin{split}
& \dot{\boldsymbol{\chi}}\left( \tau \right) = f_{\chi}\left( \boldsymbol{\chi} \left( \tau \right), u \left( \tau \right) \right), \\
& \boldsymbol{\chi} \left( 0 \right) = \boldsymbol{\chi}_0, \\
& \boldsymbol{\chi} \left( 1 \right) = \left[ {x_f ,\;y_f ,\; \gamma_f + 2k\pi} \right]^T, \\
& -\kappa_{m} \leq u \leq \kappa_{m}
\end{split}
\end{equation}
It is evident that $\bigcup\limits_{k \in \mathbb{Z}} \bigcup\limits_{T \in [0, 1]} \mathcal{R}_k(T) = \mathcal{B}(\mathbb{X}, \kappa_m)$ and hence we devote the remaining part of this subsection to identify the boundary of $\mathcal{R}_k(T)$. It is noteworthy that it is not yet trivial that any trajectory reaching the boundary point of $\mathcal{R}_k(T)$ is a PMP sense solution of $[\textbf{P0}_{T, k}]$. 

Whenever the time $\tau$ reaches $T$, the remaining time--to--go must be $1-T$, as the final time is set to be 1. Since the conventional PMP only permits the consideration of cost functions at the terminal time, and not at any point along the trajectory, we will address two OCPs. One is defined over the time domain $[0, T]$ and the other is defined over $[0, 1-T]$. For the second part of the trajectory on interval $[0, 1-T]$, we consider a trajectory with orientation--reversing reparametrization, starting from initial state $\left[ {x_f ,\;y_f ,\;\gamma_f + 2k\pi} \right]^T$. By adding a terminal state constraints that the two trajectories have same terminal location and direction at $T$, we can define an augmented OCP. 

Now, let us define the two aforementioned OCPs as follows. After appropriate reparametrization, the both OCPs are parametrized by $s \in [0, 1]$ instead of $\tau$, where $T$ is some constant in $(0, 1)$. $\boldsymbol{\chi}_1(s) = \left[ { x_1(s) ,\;y_1(s) ,\;\gamma_1(s) } \right]^T $ and $\boldsymbol{\chi}_2(s) = \left[ { x_2(s) ,\;y_2(s) ,\;\gamma_2(s) } \right]^T$ represent the state variables of the two OCPs respectively. 

[$\textbf{P1}$] : maximize
\begin{equation} \label{eq:5}
	\enspace J  =  \Phi \left(\boldsymbol{\chi}_1 \left( 1 \right) \right) 
\end{equation}
subject to
\begin{equation} \label{eq:6}
\begin{split}
& \frac{d}{ds} \boldsymbol{\chi}_1\left( s \right) = T f_{\chi}\left( \boldsymbol{\chi}_1 \left( s \right), u_1 \left( s \right) \right), \\
& \boldsymbol{\chi}_1 \left( 0 \right) = \boldsymbol{\chi}_0, \\
& -\kappa_{m} \leq u \leq \kappa_{m}
\end{split}
\end{equation}
[$\textbf{P1}$] is the OCP for the time interval $[0, T]$. We define the OCP [$\textbf{P2}$] corresponding to the time interval $[T, 1]$ as below, with consideration of the terminal state constraints. 

[$\textbf{P2}_k$] : maximize
\begin{equation} \label{eq:7}
	\enspace J  =  \Phi \left(\boldsymbol{\chi}_2 \left( 1 \right) \right) 
\end{equation}
subject to
\begin{equation} \label{eq:8}
\begin{split}
& \frac{d}{ds} \boldsymbol{\chi}_2\left( s \right) = -(1 - T) f_{\chi}\left( \boldsymbol{\chi}_2 \left( s \right), u_2 \left( s \right) \right), \\
& \boldsymbol{\chi}_2 \left( 0 \right) = \left[ {x_f ,\;y_f ,\; \gamma_f + 2k\pi} \right]^T, \\
& -\kappa_{m} \leq u \leq \kappa_{m}
\end{split}
\end{equation}
It is noteworthy that the curvature constraint remains the same in the normalized domain of $s \in [0, 1]$ since curvature is invariant under reparametrization. 
Consequently, the augmented OCP below is considered. 

[$\textbf{P3}_k$] : maximize

\begin{equation} \label{eq:9}
	\enspace J  =  \Phi \left(\boldsymbol{\chi}_1 \left( 1 \right), \boldsymbol{\chi}_2 \left( 1 \right) \right) 
\end{equation}
subject to 
\begin{equation} \label{eq:10}
\frac{d}{ds} \boldsymbol{\chi}\left( s \right) = \begin{bmatrix} \dot{\boldsymbol{\chi}}_1\left( s \right) \\ \dot{\boldsymbol{\chi}}_2\left( s \right) \end{bmatrix} = \begin{bmatrix} T f_{\chi}\left( \boldsymbol{\chi}_1 \left( s \right), u_1 \left( s \right) \right) \\ -(1 - T) f_{\chi}\left( \boldsymbol{\chi}_2 \left( s \right), u_2 \left( s \right) \right) \end{bmatrix}
\end{equation}
\begin{equation} \label{eq:11}
	\boldsymbol{\chi} \left( 0 \right) = \begin{bmatrix} \boldsymbol{\chi}_1\left( 0 \right) \\ \boldsymbol{\chi}_2\left( 0 \right) \end{bmatrix} = \left[ {x_0 ,\;y_0 ,\;\gamma_0 ,\;x_f ,\;y_f ,\; \gamma_f + 2k\pi} \right]^T 
\end{equation}
\begin{equation} \label{eq:12}
	-\kappa_{m} \leq u_1,\; u_2 \leq \kappa_{m}
\end{equation}
\begin{equation} \label{eq:13}
	m(\boldsymbol{\chi}(1)) \equiv \boldsymbol{\chi}_1\left( 1 \right) - \boldsymbol{\chi}_2\left( 1 \right) = \boldsymbol{0}
\end{equation}
where $\Phi(\boldsymbol{\chi}_1, \boldsymbol{\chi}_2)$ in Eq.~\eqref{eq:9} is the extension of the cost function in Eq.~\eqref{eq:5} to the domain $\mathbb{R}^6$ and $m : \mathbb{R}^6 \mapsto \mathbb{R}^3$ is the terminal state constraint function. Similar to the definition of [$\textbf{P0}$], $\Phi$ is assumed to have constant gradient in $\mathbb{R}^6$, nonzero on the manifold defined by Eq.~\eqref{eq:13}. The rationale behind such augmentation is to enable the consideration of cost functions applied at an arbitrary time $T$, not only at the terminal time $1$. Among the concatenated components, the problem [$\textbf{P2}_k$] part is regarded as an orientation--reversed one of the original curve. It is important to note that the invariance property of curvature under reparametrization implies that if any curve on $[0, 1]$ is in the search space of [$\textbf{P3}_k$] in sense of concatenating $\boldsymbol{\chi}_1 ([0, 1])$ and $\boldsymbol{\chi}_2 ([0, 1])$, then it is contained in the search space of [$\textbf{P0}_{T, k}$] and vice versa. Therefore, the search space of [$\textbf{P3}_k$] is identical with the original OCP, [$\textbf{P0}_{T, k}$]. Such preservation of search space may not hold if additional constraints besides curvature bound that discriminates the orientation of the trajectory are applied. 

Now, consider the dynamical system defined by Eqs.~\eqref{eq:10}$\sim$\eqref{eq:12}. Let us denote the reachability set of such dynamical system at final time $1$ as $\mathcal{D}_k(1)$. The boundary points of $\mathcal{D}_k(1)$ can be obtained by Eq.~\eqref{def1PMP} applied to the dynamics Eq.~\eqref{eq:10}. Among the points in $\mathcal{D}_k(1)$, the set of points that are on the plane defined by Eq.~\eqref{eq:13} is $\mathcal{R}_k(T)$. Then the below theorem and corollary can be proved. 

\begin{thm}\label{thm0}
	The below sets are identical for each $k \in \mathbb{Z}$. 
\begin{itemize}
	\item The set of solutions of Eq.~\eqref{def1PMP} applied to obtain the boundary points of $\mathcal{D}_k(1)$, such that the endpoint of the trajectory is on the plane defined by Eq.~\eqref{eq:13}
	\item The set of PMP sense solutions of [$\textbf{P3}_k$] 
\end{itemize}
\end{thm}
\begin{pf*}{Proof.}
Let us begin by introducing the necessary conditions of optimality(PMP) of [$\textbf{P3}_k$]. The costate variables are denoted by $\boldsymbol{p} = [ p_{x, 1} \; p_{y, 1} \; p_{\gamma, 1} \; p_{x, 2} \; p_{y, 2} \; p_{\gamma, 2} ]^T$. 
Then the necessary conditions are as follows~\cite{hartl1995survey}: 
\begin{enumerate}
	\item Costate Differential \\
		\begin{equation} \label{eq:14}
			\begin{split}
				& \dot{p}_{x, 1}(s) 			= 	0 	\\
				& \dot{p}_{y, 1}(s) 			= 	0 	\\
				& \dot{p}_{\gamma, 1}(s) 	= 	T [ p_{x, 1}(s) \sin\gamma_1(s) - p_{y, 1}(s) \cos\gamma_1(s) ] 	\\
				& \dot{p}_{x, 2}(s) 			= 	0 	\\
				& \dot{p}_{y, 2}(s) 			= 	0 	\\
				& \dot{p}_{\gamma, 2}(s) 	= 	-(1 - T) [p_{x, 2}(s) \sin\gamma_2(s) - p_{y, 2}(s) \cos\gamma_2(s) ]
			\end{split}
		\end{equation}
	\item Pointwise Maximum Condition \\
		\begin{equation} \label{eq:15}
		\begin{split}
			& T\left[ p_{x, 1}(s) \cos\gamma_1(s) + p_{y, 1}(s) \sin\gamma_1(s) + p_{\gamma, 1}(s) u_1^*(s) \right] \\
			& - (1-T) \left[ p_{x, 2}(s) \cos\gamma_2(s) + p_{y, 2}(s) \sin\gamma_2(s) \right. \\
			& \left. + p_{\gamma, 2}(s) u_2^*(s) \right] \\
			& = \max\limits_{|u_1|, |u_2| \leq \kappa_m} T\left[ p_{x, 1}(s) \cos\gamma_1(s) + p_{y, 1}(s) \sin\gamma_1(s) \right. \\
			& \left. + p_{\gamma, 1}(s) u_1(s) \right] - (1-T)\left[ p_{x, 2}(s) \cos\gamma_2(s) \right. \\
			& \left. + p_{y, 2}(s) \sin\gamma_2(s) + p_{\gamma, 2}(s) u_2(s) \right]
		\end{split}
		\end{equation}
		almost everywhere on the interval $[0, 1]$. \\
	\item Transversality Condition \\
		\begin{equation} \label{eq:16}
			\boldsymbol{p}(1^-) = \lambda_0 \nabla_{\boldsymbol{\chi}} \Phi(1) + \left[ I_3 ;\; -I_3 \right]^T \boldsymbol{\beta} 
		\end{equation}
		where $\lambda_0 \geq 0$, $\boldsymbol{\beta} \equiv [ \beta_x \; \beta_y \; \beta_{\gamma} ]^T \in \mathbb{R}^3$, and $I_n$ denotes the $n \times n$ identity matrix. 
\end{enumerate}
The substitution $\nabla_{\boldsymbol{\chi}} m^T = \left[ I_3 ;\; -I_3 \right]^T$ was made for the transversality condition. The solution pair of the above necessary conditions is denoted by $\{ \boldsymbol{\chi}, \boldsymbol{u}^*, \boldsymbol{p}, \lambda_0, \boldsymbol{\beta} \}$. 

It is evident that when the conditions of Eq.~\eqref{def1PMP} are applied to obtain $\mathcal{D}_k(1)$, it reduces to the analogous costate differential and maximum condition as above. The only difference is the transversality condition, expressed as: 
\begin{equation} \label{eq:transversality}
	\tilde{\boldsymbol{p}}(1^-) = \tilde{\lambda}_0 \nabla_{\boldsymbol{\chi}} \tilde{\Phi}(1)
\end{equation}
where $\{ \tilde{\boldsymbol{\chi}}, \tilde{\boldsymbol{u}}, \tilde{\boldsymbol{p}}, \tilde{\lambda}_0 \}$ denotes the solution pair of the conditions of Eq.~\eqref{def1PMP}. $\tilde{\Phi}$ denotes the terminal cost function as a counterpart of $\Phi$, utilizing the idea of Proposition~\ref{prop:1} interpreting Eq.~\eqref{def1PMP} as [$\textbf{P0}$].

Now suppose a fixed vector $\tilde{\lambda}_0 \nabla_{\boldsymbol{\chi}} \tilde{\Phi}(1)$ is given, where $\nabla_{\boldsymbol{\chi}} \tilde{\Phi}(1)$ is nonzero. 
First suppose that $\nabla_{\boldsymbol{\chi}} \tilde{\Phi}(1)$ is in the column space of $\left[ I_3 ;\; -I_3 \right]^T$. 
If $\tilde{\lambda}_0 \neq 0$, then $\exists \tilde{\boldsymbol{\beta}} \neq \boldsymbol{0}$ such that $ \tilde{\lambda}_0 \nabla_{\boldsymbol{\chi}} \tilde{\Phi}(1) = \left[ I_3 ;\; -I_3 \right]^T\tilde{\boldsymbol{\beta}}$. Then choice of $\lambda_0 = 0$, $\boldsymbol{\beta} = \tilde{\boldsymbol{\beta}}$, $\boldsymbol{p} = \tilde{\boldsymbol{p}}$, $\boldsymbol{\chi} = \tilde{\boldsymbol{\chi}}$, $\boldsymbol{u}^* = \tilde{\boldsymbol{u}}$, and arbitrary $\Phi$ having nonvanishing gradient on the manifold Eq.~\eqref{eq:13} satisfies the condition $\tilde{\lambda}_0 \nabla_{\boldsymbol{\chi}} \tilde{\Phi}(1) = \lambda_0 \nabla_{\boldsymbol{\chi}} \Phi(1) + \left[ I_3 ;\; -I_3 \right]^T \boldsymbol{\beta}$. The nontriviality condition of the pair $\{ \boldsymbol{p}, \lambda_0, \boldsymbol{\beta} \}$ is satisfied by $\boldsymbol{\beta}$. 
If $\tilde{\lambda}_0 = 0$, then $\tilde{\boldsymbol{p}}$ satisfies the nontriviality condition alone. Consequently, we can set $\lambda_0 = 0$, $\boldsymbol{\beta} = \boldsymbol{0}$, $\boldsymbol{p} = \tilde{\boldsymbol{p}}$, $\boldsymbol{\chi} = \tilde{\boldsymbol{\chi}}$, and $\boldsymbol{u}^* = \tilde{\boldsymbol{u}}$ to meet the transversality and the nontriviality conditions. Again, $\Phi$ is considered arbitrary with nonzero constant gradient on the manifold Eq.~\eqref{eq:13}. 
If $\nabla_{\boldsymbol{\chi}} \tilde{\Phi}(1)$ is not in the column space of $\left[ I_3 ;\; -I_3 \right]^T$, then $\tilde{\Phi}(1)$ has nonvanishing gradient on the manifold Eq.~\eqref{eq:13}. Then the choice of $\lambda_0 = \tilde{\lambda}_0$, $\boldsymbol{\beta} = \boldsymbol{0}$, $\boldsymbol{p} = \tilde{\boldsymbol{p}}$, $\boldsymbol{\chi} = \tilde{\boldsymbol{\chi}}$, $\boldsymbol{u}^* = \tilde{\boldsymbol{u}}$, and $\Phi = \tilde{\Phi}$ satisfies the transversality condition. For this case, the nontriviality condition of $\{ \tilde{\boldsymbol{p}}, \tilde{\lambda}_0 \}$ implies the nontriviality of such choice of $\{ \boldsymbol{p}, \lambda_0, \boldsymbol{\beta} \}$. 
Hence, any PMP sense solution of Eq.~\eqref{def1PMP} applied to obtain $\mathcal{D}_k(1)$, with its endpoint of the trajectory satisfying Eq.~\eqref{eq:13}, is also a solution of some [$\textbf{P3}_k$]. 

Conversely, for a fixed vector $\lambda_0 \nabla_{\boldsymbol{\chi}} \Phi(1) + \left[ I_3 ;\; -I_3 \right]^T \boldsymbol{\beta} \neq \boldsymbol{0}$, $\exists \{ \tilde{\boldsymbol{\chi}}, \tilde{\boldsymbol{u}}, \tilde{\boldsymbol{p}}, \tilde{\lambda}_0 \}$ and $\nabla_{\boldsymbol{\chi}} \tilde{\Phi}$ such that $\lambda_0 \nabla_{\boldsymbol{\chi}} \Phi(1) + \left[ I_3 ;\; -I_3 \right]^T \boldsymbol{\beta} = \tilde{\lambda}_0 \nabla_{\boldsymbol{\chi}} \tilde{\Phi}(1)$. This is directforward from the fact that the choice of nonzero $\nabla_{\boldsymbol{\chi}} \tilde{\Phi}$ is arbitrary. The nontriviality condition is also directly satisfied by choosing nonzero $\tilde{\lambda}_0$ arbitrarily. Hence, any PMP sense solution of [$\textbf{P3}_k$] is also a solution of Eq.~\eqref{def1PMP} applied to obtain $\mathcal{D}_k(1)$. 
\qed
\end{pf*}
\begin{cor}\label{cor0}
	Any trajectory that reaches the boundary point of $\mathcal{R}_k(T)$ is a PMP sense solution of [$\textbf{P3}_k$]. 
\end{cor}
In the subsection followed, we solve the necessary conditions of optimality(PMP) of [$\textbf{P3}_k$] to derive the subsequent conclusions. 
\subsection{Necessary Conditions of the Augmented Problem}\label{sec3subsec:03}
By dropping the constant terms from Eq.~\eqref{eq:15}, the maximum condition reduces to 
\begin{equation}
\begin{split}
	& p_{\gamma, 1}(s) T u_1^*(s) - p_{\gamma, 2}(s) (1 - T) u_2^*(s) \\
	& = \max\limits_{|u_1|, |u_2| \leq \kappa_m} p_{\gamma, 1}(s) T u_1(s) - p_{\gamma, 2}(s) (1 - T) u_2(s)
\nonumber
\end{split}
\end{equation}
%
%
As the control variables are independent, it is evident that 
\begin{equation} \label{eq:18}
\begin{split}
	& u_1^*(s) = \begin{cases}
		\kappa_m, \quad  \left( p_{\gamma, 1}(s) > 0 \right) \\
		\text{(Indetermediate)}, \quad \left( p_{\gamma, 1}(s) = 0 \right) \\
		-\kappa_m, \quad \left( p_{\gamma, 1}(s) < 0 \right)
	\end{cases} \\
	& u_2^*(s) = \begin{cases}
		-\kappa_m, \quad  \left( p_{\gamma, 2}(s) > 0 \right) \\
		\text{(Indetermediate)}, \quad \left( p_{\gamma, 2}(s) = 0 \right) \\
		\kappa_m, \quad \left( p_{\gamma, 2}(s) < 0 \right)
	\end{cases}
\end{split}
\end{equation}
Then by substituting Eq.~\eqref{eq:10} and integrating the costate differential Eq.~\eqref{eq:14}, it follows 
\begin{equation} 
\begin{split}
	& p_{\gamma, 1}(s) = T p_{x, 1} y_1(s) - T p_{y, 1} x_1(s) + C_1 \\
	& p_{\gamma, 2}(s) = (1 - T) p_{x, 2} y_2(s) - (1 - T) p_{y, 2} x_2(s) + C_2
\nonumber
\end{split}
\end{equation}
Therefore, whenever $p_{\gamma, 1}(s)$ or $p_{\gamma, 2}(s) = 0$, the below equations hold respectively. 
\begin{equation} \label{eq:19}
\begin{split}
	& T p_{x, 1} y - T p_{y, 1} x + C_1 = 0 \\
	& (1 - T) p_{x, 2} y - (1 - T) p_{y, 2} x + C_2 = 0
\end{split}
\end{equation}
In other words, $p_{\gamma, 1}(s) = 0$(or $p_{\gamma, 2}(s) = 0$) if and only if the location $(x_1(s), y_1(s))$(or $(x_2(s), y_2(s))$) is on the line Eq.~\eqref{eq:19}. Similar observation for the problem [$\textbf{P0}$] was made in~\cite{patsko2003}. 
\begin{rem} \label{rem:1}
	The costate differential condition of Eq.~\eqref{eq:14} and the maximum condition of Eq.~\eqref{eq:18} are the same necessary conditions (3.1) and (3.2) in~\cite{patsko2003}. Consequently, the nontriviality condition of [$\textbf{P3}_k$] indicates that either one of the decomposed trajectories $(x_1, y_1, \gamma_1)$ or $(x_2, y_2, \gamma_2)$ meet the switching condition outlined in the theorem from~\cite{patsko2003}. Hence, any boundary point of the set $\mathcal{R}_k(T)$ can be reached by a trajectory which at least one of the decomposed trajectory is of class CSC or CCC or their subsegments. In the subsequent lemma, we utilize such observation to construct the boundary of $\mathcal{B}'(\mathbb{X}, \kappa_m)$. 
\end{rem}
As mentioned in Sec.~\ref{sec3subsec:00}, we are interested in the bounds applied for $x$ and $y$, or the set $\mathcal{B}'(\mathbb{X}, \kappa_m)$. Hence, we can set $\nabla \Phi = [ (\cdot) \; (\cdot) \; 0 \; (\cdot) \; (\cdot) \; 0 ]^T$ where $(\cdot)$ are unspecified values. 
The third and last components are zero as we are not considering the cost dependency on $\gamma_1$ and $\gamma_2$. It is important to note that when there are such restrictions on $\nabla \Phi$, the `converse' part of the proof of Theorem~\ref{thm0} does not hold trivially. However, the nontrivial part of the transversality condition of Eq.~\eqref{eq:16} reduces to $p_{\gamma, 1}(1^-) + p_{\gamma, 2}(1^-) = \beta_\gamma - \beta_\gamma = 0$. Then, for $\nabla_{\boldsymbol{\chi}} \tilde{\Phi}$ such that $\lambda_0 \nabla_{\boldsymbol{\chi}} \Phi(1) + \left[ I_3 ;\; -I_3 \right]^T \boldsymbol{\beta} = \tilde{\lambda}_0 \nabla_{\boldsymbol{\chi}} \tilde{\Phi}(1)$, it is evident that $\tilde{\lambda}_0 \nabla_{\boldsymbol{\chi}} \tilde{\Phi} = [ (\cdot) \; (\cdot) \; \beta_{\gamma} \; (\cdot) \; (\cdot) \; -\beta_{\gamma} ]^T$. Consequently, the constraint Eq.~\eqref{eq:13} ensures that $\tilde{\lambda}_0 \tilde{\Phi} = (\cdot) + \beta_{\gamma}\gamma_1 - \beta_{\gamma}\gamma_2 = (\cdot) + \beta_{\gamma}\gamma_1 - \beta_{\gamma}\gamma_1 = (\cdot)$. Hence, the cost dependency on $\gamma$ is not present in $\tilde{\Phi}$ as well and Theorem~\ref{thm0} still holds. In light of such observations, we first state and prove the below Lemma~\ref{lemma:2}, followed by derivation of the construction method of $\mathcal{B}'(\mathbb{X}, \kappa_m)$ in Theorem~\ref{thm:1}. 
\begin{lem} \label{lemma:2}
	Suppose both of the trajectories $\boldsymbol{\chi}_1$ and $\boldsymbol{\chi}_2$ satisfy the nontriviality condition with $p_{\gamma, i}(1) = 0$, $i = 1, 2$. If the point $\pi_1(\boldsymbol{\chi}_1(1)) = \pi_1(\boldsymbol{\chi}_2(1))$ is on the boundary of $\mathcal{R}_k'(T)$, it can be reached by curves belonging to the following classes: CCCCC, CSCCC, CCCSC, CSCSC, or their subsegments. 
\end{lem}
\begin{pf*}{Proof.}
As stated in Remark~\ref{rem:1}, both of the trajectories $\boldsymbol{\chi}_1$ and $\boldsymbol{\chi}_2$ are of classes CSC or CCC or their subsegments. Among the possible classes, it suffices to show only the cases where concatenation is done among CSC or CCC, with the two C's that are being concatenated having opposite winding directions. Otherwise, the two C's can be reduced to a single arc, or a single C, and the overall concatenation belongs to the classes stated. 
The condition $p_{\gamma, i}(1) = 0$ further implies that the points of switching and the endpoints are on the line Eq.~\eqref{eq:19} for both trajectories. 

First, consider the case when the concatenation is done with two CCC curves. If a CCC curve contains no cycles, then assertion (a) of lemma 1 from~\cite{patsko2003} implies that the second and last C have same length. Then lemma 2 from ~\cite{patsko2003} implies that there exists an auxiliary trajectory that has the same endpoint but cannot satisfy the nontriviality condition. Consequently, if the concatenation consists of two CCC curves without a cycle, then both trajectories will have such auxiliary trajectories. As a result, the endpoint must lie within the interior of $\mathcal{D}_k(T)$, and therefore also within the interiors of both $\mathcal{R}_k(T)$ and $\mathcal{R}_k'(T)$. 
If a CCC curve has a cycle, assertion (b) of lemma 1 from~\cite{patsko2003} implies that the geometric coordinates of the endpoint and the switching points coincide. This implies that the second and last C are full cycles. Then reversing the winding direction of the last C component does not alter the endpoints, $\pi_1(\boldsymbol{\chi}_1(1))$ or $\pi_1(\boldsymbol{\chi}_2(1))$. Hence, the CCCCCC curve reaching the endpoint can be reduced to a CCCCC curve. 

For CSC curves, the endpoint must lie on the line containing the S segment, because the line Eq.~\eqref{eq:19} subsumes the S segment.  Hence, assertion (d) of lemma 1 from~\cite{patsko2003} implies that the last C is a cycle. Therefore, analogous steps by reversing the winding direction of the last C component reduces the concatenation with any CCC or CSC curve into the stated classes. 
\qed
\end{pf*}
\begin{thm} \label{thm:1}
	Any boundary point of $\mathcal{B}'(\mathbb{X}, \kappa_m)$ can be reached by a curve belonging to one of the following classes: CCCCC, CSCCC, CCCSC, CSCSC, or their subsegments. 
\end{thm}
\begin{pf*}{Proof.}
First, suppose that the nontriviality condition is applied to both of the concatenated trajectories. Then it suffices to show only the cases where concatenation is done among CSC or CCC. 
By the argument in Lemma~\ref{lemma:2}, we may presume that $p_{\gamma, 1}(1) > 0$ and $p_{\gamma, 2}(1) < 0$ without loss of generality. Then continuity of the adjoint variables implies that $\exists \varepsilon > 0$ such that $p_{\gamma, 1}(s) > 0$ and $p_{\gamma, 2}(s) < 0$ on $(1 - \varepsilon, 1]$. Therefore, $p_{\gamma, 1}(s)$ and $p_{\gamma, 2}(s)$ have opposite signs on $(1 - \varepsilon, 1]$. This implies that the substrings on the interval $(1 - \varepsilon, 1]$ are both of class C, with $u_1$ and $u_2$ same signs. Hence, the last C part of the two plane curves are of a single arc. Therefore, the concatenated curves CCCCCC, CSCCCC, CCCCSC, and CSCCSC can be reduced to CCCCC, CSCCC, CCCSC, and CSCSC respectively. 

Before proving the remaining part when nontriviality condition is applied to only one of the concatenated trajectories, let us introduce some additional notations. For a set $S$, let us denote the interior points of $S$ by $int(S)$ and the boundary points by $bd(S)$. 
For some fixed $T$, let us denote the reachability set for the $\boldsymbol{\chi}_1$ component of Eq.~\eqref{eq:10} by $\mathcal{G}_1(T)$, the forward reachability set. Similarly, the reachability set for the $\boldsymbol{\chi}_2$ component of Eq.~\eqref{eq:10}, with its initial condition at $\left[ {x_f ,\;y_f ,\; \gamma_f + 2k\pi} \right]^T$, the backward reachability set, is denoted as $\mathcal{G}_{2, k}(1-T)$. 
Similar definitions are made for $\mathcal{G}_1'(T)$ and $\mathcal{G}_{2, k}'(1-T)$ as well, by applying the projection map $\pi_1$ to $\mathcal{G}_1(T)$ and $\mathcal{G}_{2, k}(1-T)$. It follows trivially that $\mathcal{R}_k(T) = \mathcal{G}_1(T) \cap \mathcal{G}_{2, k}(1-T)$ and $\mathcal{R}_k'(T) \subset \mathcal{G}_1'(T) \cap \mathcal{G}_{2, k}'(1-T)$. 

Without loss of generality, suppose some $\boldsymbol{x} \in bd(\mathcal{R}_k(T))$ is in the interior of $\mathcal{G}_{2, k}(1-T)$ and boundary of $\mathcal{G}_1(T)$. In other words, $\boldsymbol{x}$ is reached by a trajectory such that the nontriviality condition is met for the $\boldsymbol{\chi}_1$ part, but not necessarily on the $\boldsymbol{\chi}_2$ part. 

The problem of identifying the boundary of $\mathcal{G}_1'(T)$ was first solved in~\cite{cockayne1975} by geometric means, that any boundary point can be reached by curves of class CS or CC, or their subsegments. 
Analogous statement can be derived by solving the PMP conditions of OCP [$\textbf{P0}$] by neglecting the cost dependency of $\Phi$ on $\gamma$, or with an additional condition, $p_{\gamma}(T) = 0$. Hence, if $\boldsymbol{x}$ cannot be reached by curves of class CS or CC, or their subsegments, there exists a neighborhood $V$ of $\boldsymbol{x}$ in $\mathcal{G}_1(T)$ such that the point $\pi_1(\boldsymbol{x})$ is in the interior of $\pi_1(V)$. 
Then if $V$ is small enough that $V \subset int(\mathcal{G}_{2, k}(1-T))$, it follows $V \subset \mathcal{R}_k(T)$ and hence $\pi_1(V) \subset \mathcal{R}_k'(T)$. Therefore, $\pi_1(\boldsymbol{x}) \in int(\pi_1(V)) \subset int(\mathcal{R}_k'(T))$ and it follows $\pi_1(\boldsymbol{x}) \in int(\mathcal{B}'(\mathbb{X}, \kappa_m))$. Hence, it suffices to consider the case when $\boldsymbol{x}$ can be reached from the initial point by curves of class CS or CC, or their subsegments. 
It remains to show that if $\pi_1(\boldsymbol{x}) \in bd(\mathcal{B}'(\mathbb{X}, \kappa_m))$, then $\boldsymbol{x}$ can be reached by trajectories belonging to one of the classes of CCCCC, CSCCC, CCCSC, CSCSC, or their subsegments. 

Whenever $\boldsymbol{x}$ is reached by a curve of CS, CC or their subsegments, we define an auxiliary trajectory, $\boldsymbol{x} + \hat{\bs{x}}(\tau)$, where $\tau \in [0, 1-T]$. $\hat{\bs{x}}(\tau)$ is assumed to be C or S, so that the overall auxiliary trajectory is an elongation of the original one that reaches $\bs{x}$, without altering the class. For instance, if $\bs{x}$ is reached by a CS curve, then $\hat{\bs{x}}(\tau)$ is assumed to be S, and the overall trajectory is also a CS curve but with a longer length of S. We will prove that $\exists \tau \in [0, 1-T]$ such that the point $\boldsymbol{x} + \hat{\bs{x}}(\tau)$ can be reached by curves of CSC, CCC, or their subsegments from the endpoint, $\mathbb{X}$.

Assume the converse, that for $\forall \tau \in [0, 1-T]$, $\boldsymbol{x} + \hat{\bs{x}}(\tau)$ cannot be reached by curves of CSC, CCC or their subsegments with lengths $1-T-\tau$ from $\mathbb{X}$. Such assumption implies that $\boldsymbol{x} + \hat{\bs{x}}(\tau)$ does not touch the boundary of the compact set $\mathcal{G}_{2, k}(1-T-\tau)$. Otherwise, as the boundary point can be reached by CSC, CCC or their subsegments, the overall concatenated trajectory reduces to the desired four classes. 

Now, let us define the two sets, $T_1 \equiv \{ \tau \in [0, 1-T] : \boldsymbol{x}+ \hat{\bs{x}}(\tau) \in int(\mathcal{G}_{2, k}(1-T-\tau)) \}$ and $T_2 \equiv \{ \tau \in [0, 1-T] : \boldsymbol{x}+ \hat{\bs{x}}(\tau) \in \mathcal{G}_{2, k}(1-T-\tau)^C \}$. It follows trivially that $T_1 \cap T_2 = \emptyset$ and $T_1 \cup T_2 = [0, 1-T]$. 

For each $\tau$, the set of points reached from $\mathbb{X}$ by curves of class CSC, CCC, or their subsegments, with lengths equal to $1-T-\tau$, are denoted as $D(1-T-\tau)$. Now we briefly show that $D(1-T-\tau)$ is compact. 
Once the class of the curve and the lengths of the two components in the front are determined, the endpoint is determined uniquely. For instance, if it is given that the curve is of CSC and the winding directions are given as well as the lengths of the first C and S, then the endpoint of the CSC curve is determined uniquely. Such relationship between the two lengths and the endpoint is obviously continuous. Moreover, the region of feasible lengths of the two components in the front is compact, as it consists of nonnegative pairs with sum $\leq 1-T-\tau$. Hence, $D(1-T-\tau)$ is an image of a compact set under a continuous map, and it is therefore compact. 
In addition, it is trivial that $D(1-T-\tau)$ varies in $\tau$ continuously with respect to Hausdorff metric. 

Now, we will show that $T_1$ is closed. 
To begin with, it follows from definition that $D(1-T-\tau) \subset \mathcal{G}_{2, k}(1-T-\tau)$. Since all boundary points of $\mathcal{G}_{2, k}(1-T-\tau)$ can be reached by CSC, CCC, or their subsegments, we have $bd(\mathcal{G}_{2, k}(1-T-\tau)) \subset D(1-T-\tau)$. Compactness of $\mathcal{G}_{2, k}(1-T-\tau)$ implies that if $\bs{x} + \hat{\bs{x}}(\tau) \in \mathcal{G}_{2, k}(1-T-\tau)^C$, then $dist(\bs{x} + \hat{\bs{x}}(\tau), \mathcal{G}_{2, k}(1-T-\tau)) = dist(\bs{x} + \hat{\bs{x}}(\tau), D(1-T-\tau)) > 0$. Let us further denote $dist(\bs{x} + \hat{\bs{x}}(\tau), \mathcal{G}_{2, k}(1-T-\tau))$ by $d(\tau)$. Since $\mathcal{G}_{2, k}(1-T-\tau)$ varies continuously with respect to Hausdorff metric~\cite{Lee1967}, $d(\tau)$ is continuous as well. 
Then if $\tau \in T_2$ so that $d(\tau) > 0$, continuity of $d$ implies that $\exists \varepsilon > 0$ such that $d(t') > 0$ for $\forall t' \in (\tau-\varepsilon, \tau + \varepsilon) \cap [0, 1-T]$. Since $d(t') > 0$, $\bs{x} + \hat{\bs{x}}(t') \in \mathcal{G}_{2, k}(1-T-t')^C$ and hence $t' \in T_2$. Therefore, $(\tau-\varepsilon, \tau + \varepsilon) \cap [0, 1-T] \subset T_2$ and $T_2$ is open. Then it follows that its complement, $T_1$ is closed. 

Suppose $t''$ be a boundary point of $T_1$. Since $T_1$ is closed, $t'' \in T_1$ and $d(t'') = 0$. Because $t''$ is a boundary point, for $\forall \varepsilon' > 0$, $\exists t''' \in (t'' - \varepsilon', t'' + \varepsilon') \cap [0, 1-T]$ such that $t''' \in T_2$. Then it follows from definition that $\bs{x} + \hat{\bs{x}}(t''') \in \mathcal{G}_{2, k}(1-T-t''')^C$. Therefore, there exists a sequence $t_n \rightarrow t''$ such that $\bs{x} + \hat{\bs{x}}(t_n) \in \mathcal{G}_{2, k}(1-T-t_n)^C$. Then continuity of $D(1-T-\tau)$ and $d(\tau)$ implies $d(t'') = \lim\limits_{n\rightarrow \infty} dist(\bs{x} + \hat{\bs{x}}(t_n), \mathcal{G}_{2, k}(1-T-t_n)) = \lim\limits_{n\rightarrow \infty} dist(\bs{x} + \hat{\bs{x}}(t_n), D(1-T-t_n)) =  dist(\bs{x} + \hat{\bs{x}}(t''), D(1-T-t''))= 0$. Since $D(1-T-t'')$ is compact, it follows $\bs{x} + \hat{\bs{x}}(t'') \in D(1-T-t'')$. Therefore, $\bs{x} + \hat{\bs{x}}(t'')$ can be reached by curves of CSC, CCC, or their subsegments. This contradicts the assumption. 

If the closed set $T_1$ has no boundary point, then it is either empty or $[0, 1-T]$. However, $T_1$ is nonempty because $\bs{x} + \hat{\bs{x}}(0) \in int(\mathcal{G}_{2, k}(1-T))$. Moreover, $1-T \notin T_1$ as $\mathcal{G}_{2, k}(0)$ is a singleton set, which has no interior point. Hence, $T_1$ must have a boundary point. This completes the proof. 
\qed
\end{pf*}
It is noteworthy that any other combinations can be obtained with some substrings dropped from one of the classes $\{ CCCCC, CSCCC, CCCSC, CSCSC\}$. Hence, we treat these four classes as the \textit{fundamental classes of curves} that construct the boundary of $\mathcal{B}'(\mathbb{X}, \kappa_m)$. Similar to the characteristics of reachability sets and their projections in~\cite{patsko2003} and \cite{cockayne1975}, $\mathcal{B}'(\mathbb{X}, \kappa_m)$ may contain `holes' inside. To clarify, some of the boundaries derived from the envelopes of the fundamental classes might refer to the inner boundary rather than the outer boundary, as $\mathcal{B}'(\mathbb{X}, \kappa_m)$ is not necessarily simply connected. However, it is trivial from the definition that $\mathcal{B}(\mathbb{X}, \kappa_m)$ and $\mathcal{B}'(\mathbb{X}, \kappa_m)$ are both path connected. 
\subsection{Schematic Examples}
\begin{figure}[htbp]
\centering     
\subfigure[Class CSCSC]{\centering \includegraphics[scale=0.17]{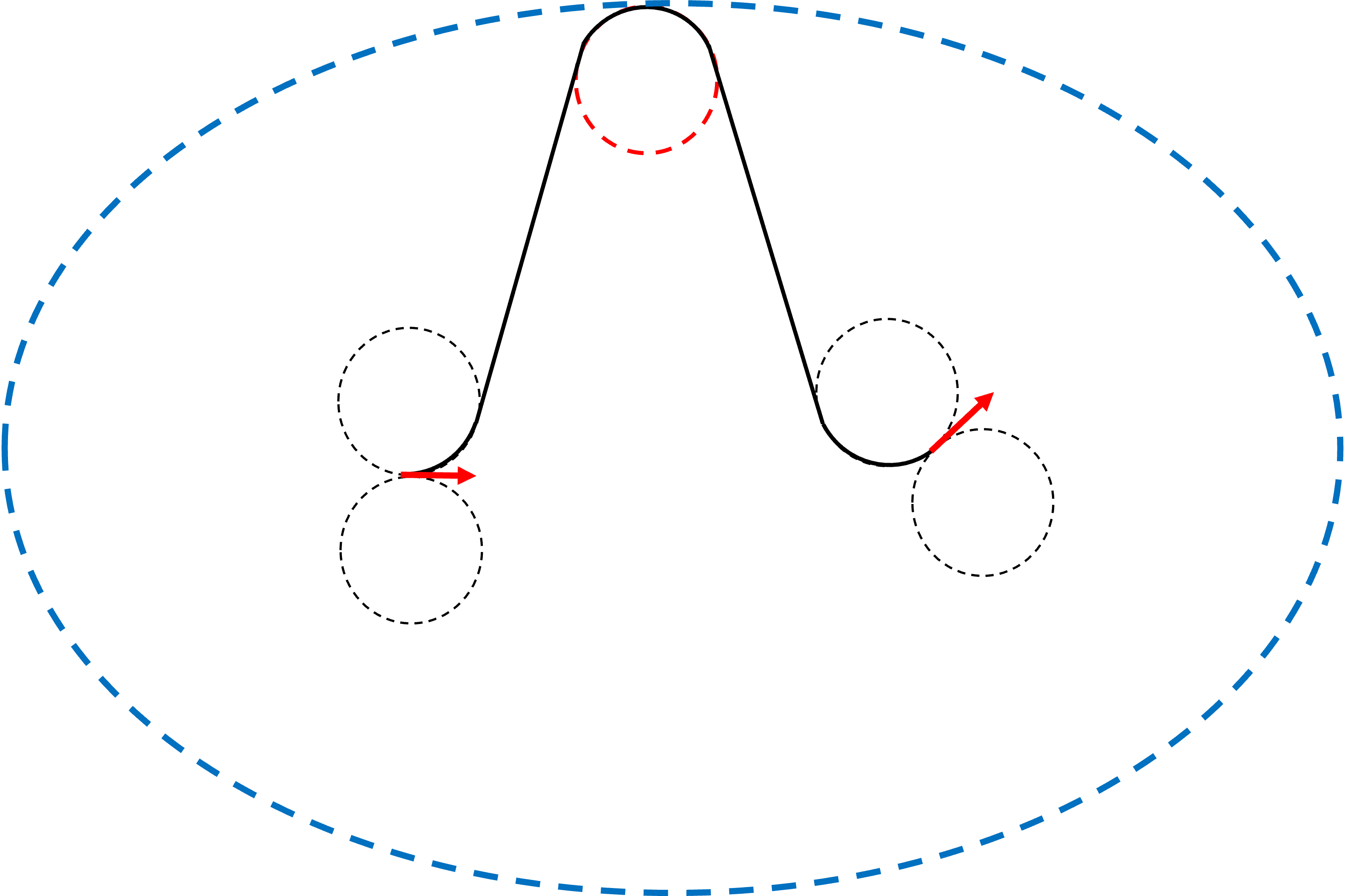}}
\subfigure[Class CCCCC]{\centering \includegraphics[scale=0.17]{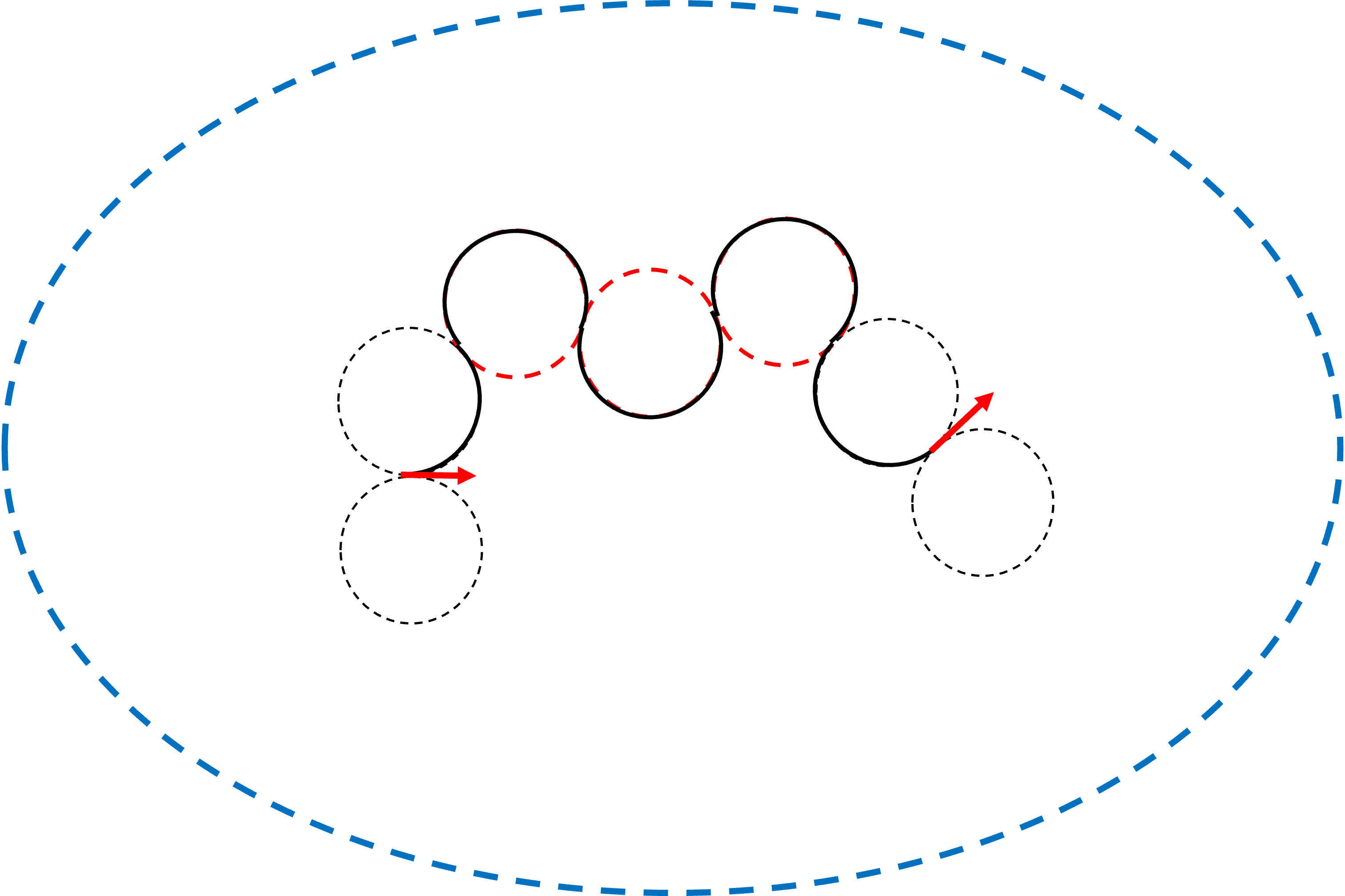}}
\caption{Schematic examples of the curves of fundamental classes}
\label{fig_sec3_2}
\end{figure}

In this subsection, schematic examples of the curves of fundamental classes are outlined. Examples of the curves of class CSCSC and CCCCC are depicted in Fig.~\ref{fig_sec3_2}. The red arrows indicate the initial and terminal locations and directions while the black dotted circles indicate the tangentially bounded region due to curvature bound. The black solid line in Fig.~\ref{fig_sec3_2} (a) represents a curve of class CSCSC. The curve is generated by stretching the red dotted circle to its maximum extent to reach the boundary of $\mathcal{B}'(\mathbb{X}, \kappa_m)$. This curve bears a striking resemblance to the construction of an \textit{ellipse}. As the curvature bound approaches infinity, it becomes clear that there are no curvature constraints. Consequently, the set $\mathcal{B}'(\mathbb{X}, \kappa_m)$ would gradually converge into an ellipse with its foci located at the initial and terminal points, and with a major axis length equal to the prescribed length of the curve. Consequently, the trajectory that reaches a certain boundary point of the ellipse becomes unique, consist of two segments connecting each endpoints and the boundary respectively, as the unique minimum--length trajectory connecting any two points is a straight line linking the two points. This trajectory exhibits infinite curvature at both the initial and terminal points, as well as at the boundary of the ellipse. Thus, we can incorporate circles of curvature $\kappa_m$ at these specific points to ensure the desired curvature constraints. These circles are represented by the black and red dotted circles in Fig.~\ref{fig_sec3_2} (a). Following this, by maneuvering the red dotted circle while maintaining the trajectory's tension (i.e., stretching it to its maximum extent), an envelope can be formed. This envelope is schematically illustrated in Fig.~\ref{fig_sec3_2} using blue dotted lines. 
On the contrary, the black solid line shown in Fig.~\ref{fig_sec3_2} (b) represents a curve of class CCCCC. This curve is formed by compressing the curve to its maximum extent. In a similar manner, by maneuvering the red dotted circles, an envelope will be generated. Depending on the relative length of the curve, it may establish the inner boundary of $\mathcal{B}'(\mathbb{X}, \kappa_m)$. 
%
%
\section{Rectangular Patch Cover and the Mesh Refinement Algorithm} \label{sec:04}
Although we have revealed the four fundamental classes of curves necessary to construct the boundary of $\mathcal{B}'(\mathbb{X}, \kappa_m)$, the challenge remains in establishing whether this boundary actually intersects the forbidden region. In this section, we present a technique for identifying the existence of such intersections by introducing the concept of \textit{rectangular patches}. The fundamental concept is outlined as follows. 

\begin{figure}[ht] 
	\begin{center}
	\vskip -0mm
	\resizebox{64mm}{!}{\includegraphics{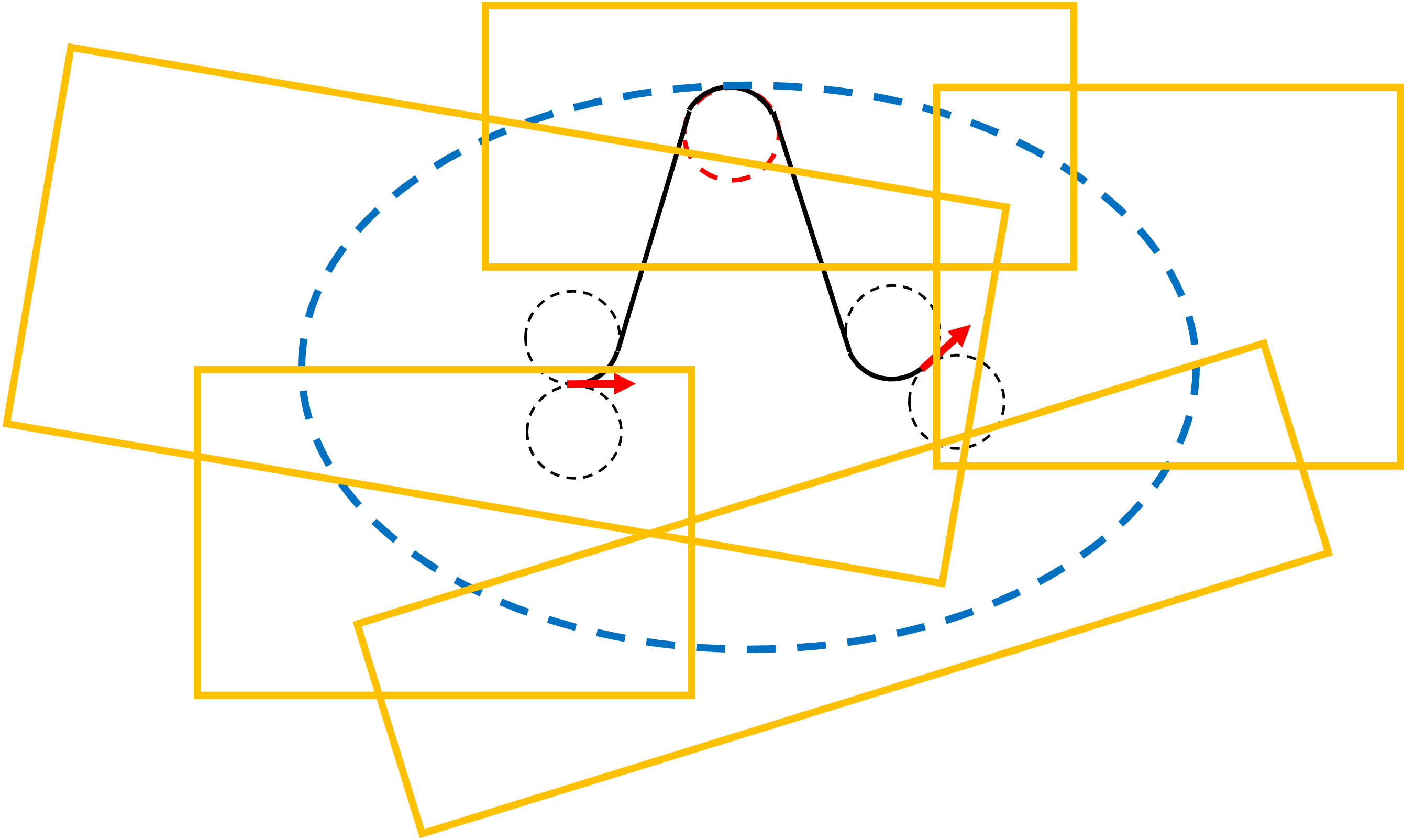}}
	\caption{Example of rectangular patches covering $\mathcal{B}'(\mathbb{X}, \kappa_m)$} \label{fig_sec4_1}
	\end{center}
\end{figure}

Given the challenge of analytically determining the boundary of $\mathcal{B}'(\mathbb{X}, \kappa_m)$, our approach involves identifying a set of \textit{rectangles} that collectively encompass the entire region $\mathcal{B}'(\mathbb{X}, \kappa_m)$. An illustrative representation of this covering concept is depicted in Fig.~\ref{fig_sec4_1}. 
The blue dotted line represents the boundary of the set $\mathcal{B}'(\mathbb{X}, \kappa_m)$, or the envelopes of the curves of fundamental classes. Consequently, the set $\mathcal{B}'(\mathbb{X}, \kappa_m)$ is entirely covered by the yellow rectangular patches. Individually, each patch does not encompass the entire $\mathcal{B}'(\mathbb{X}, \kappa_m)$; however, when considered collectively, the entire set of patches covers the entirety of $\mathcal{B}'(\mathbb{X}, \kappa_m)$. 
The rationale behind the choice of the geometric shapes of such patches by rectangle is as follows. Firstly, whenever the path inequality constraint function $h(\boldsymbol{z})$ is concave, we recognize that the feasible region is convex. Leveraging such prior understanding, we can readily ascertain whether a rectangular patch intersects the forbidden region by evaluating the constraint $h(\boldsymbol{x}) \geq 0$ at its vertices. Secondly, even in scenarios where the concavity of $h(\boldsymbol{z})$ is not confirmed, it is generally more manageable to ascertain whether a segment or a line intersects the boundary of the forbidden region, compared to other geometric shapes. Nevertheless, making a blanket statement about tractability is challenging due to the versatility of the function $h(\boldsymbol{z})$, which varies depending on the specific problem. Therefore, it is highly advisable to explore other shapes in addition to rectangles when applying the methodology presented in this paper to specific situations. 

\begin{figure}[htbp]
\centering     
\subfigure[Left--Left]{\centering \includegraphics[scale=0.27]{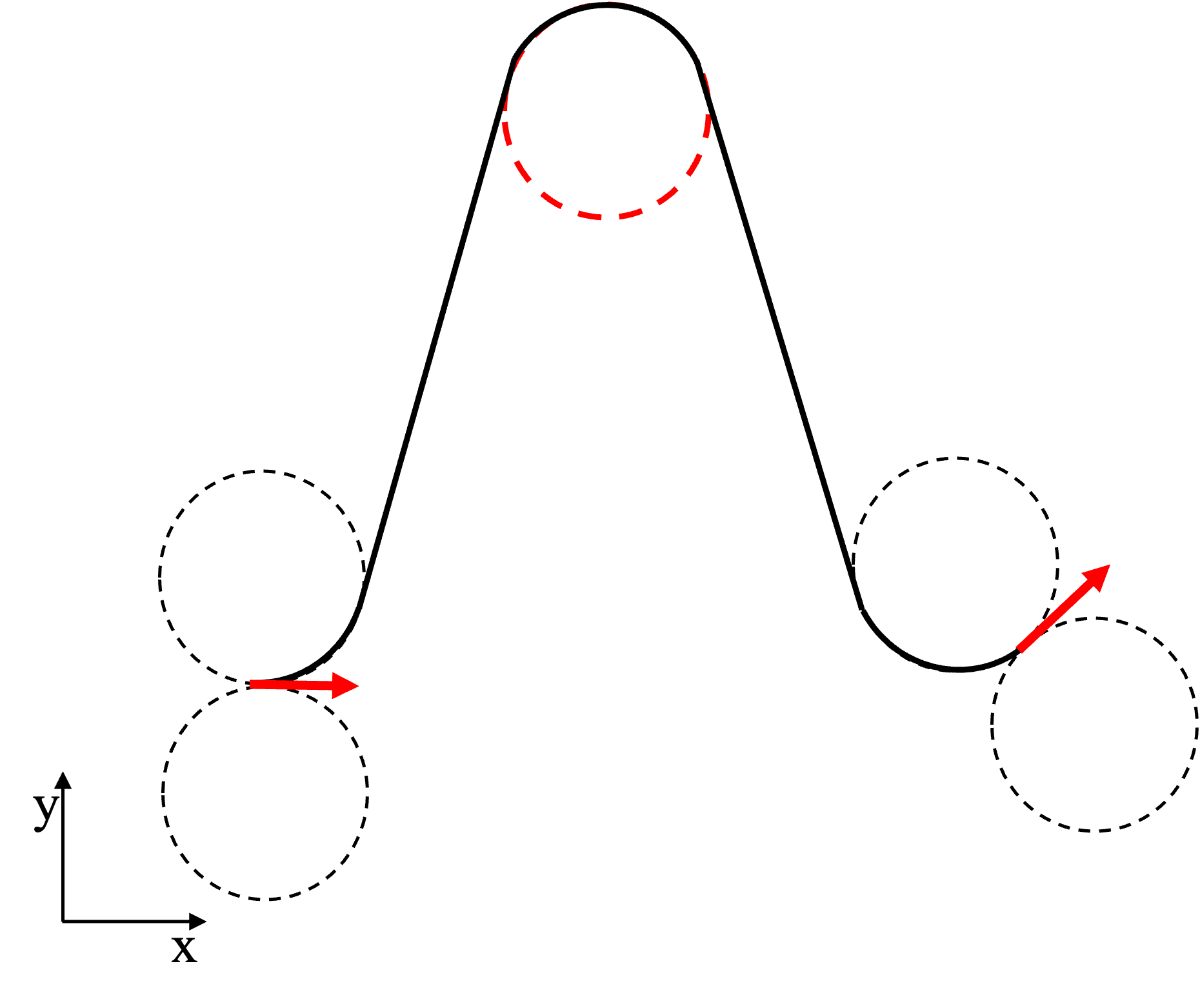}}
\subfigure[Left--Right]{\centering \includegraphics[scale=0.27]{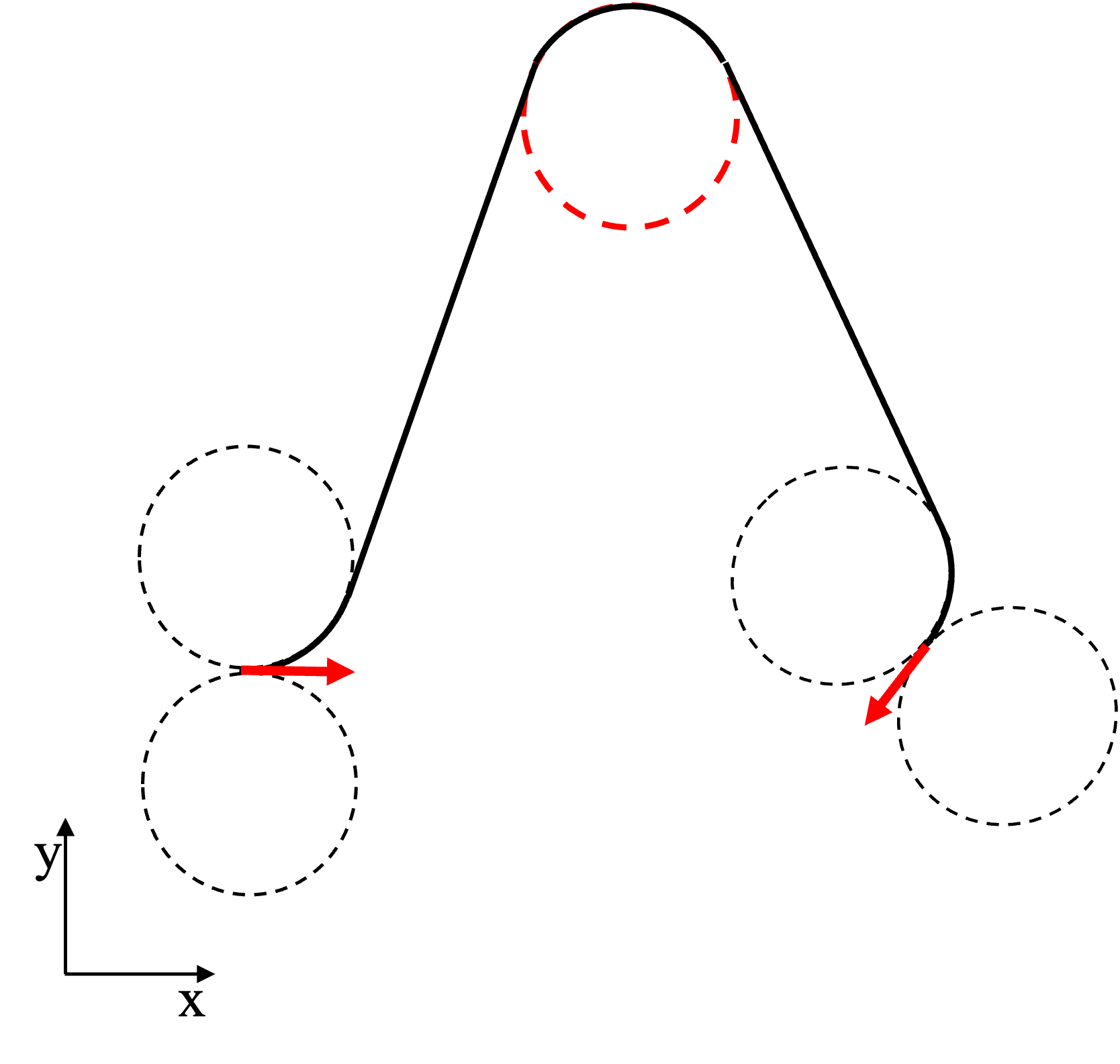}}
\subfigure[Right--Left]{\centering \includegraphics[scale=0.27]{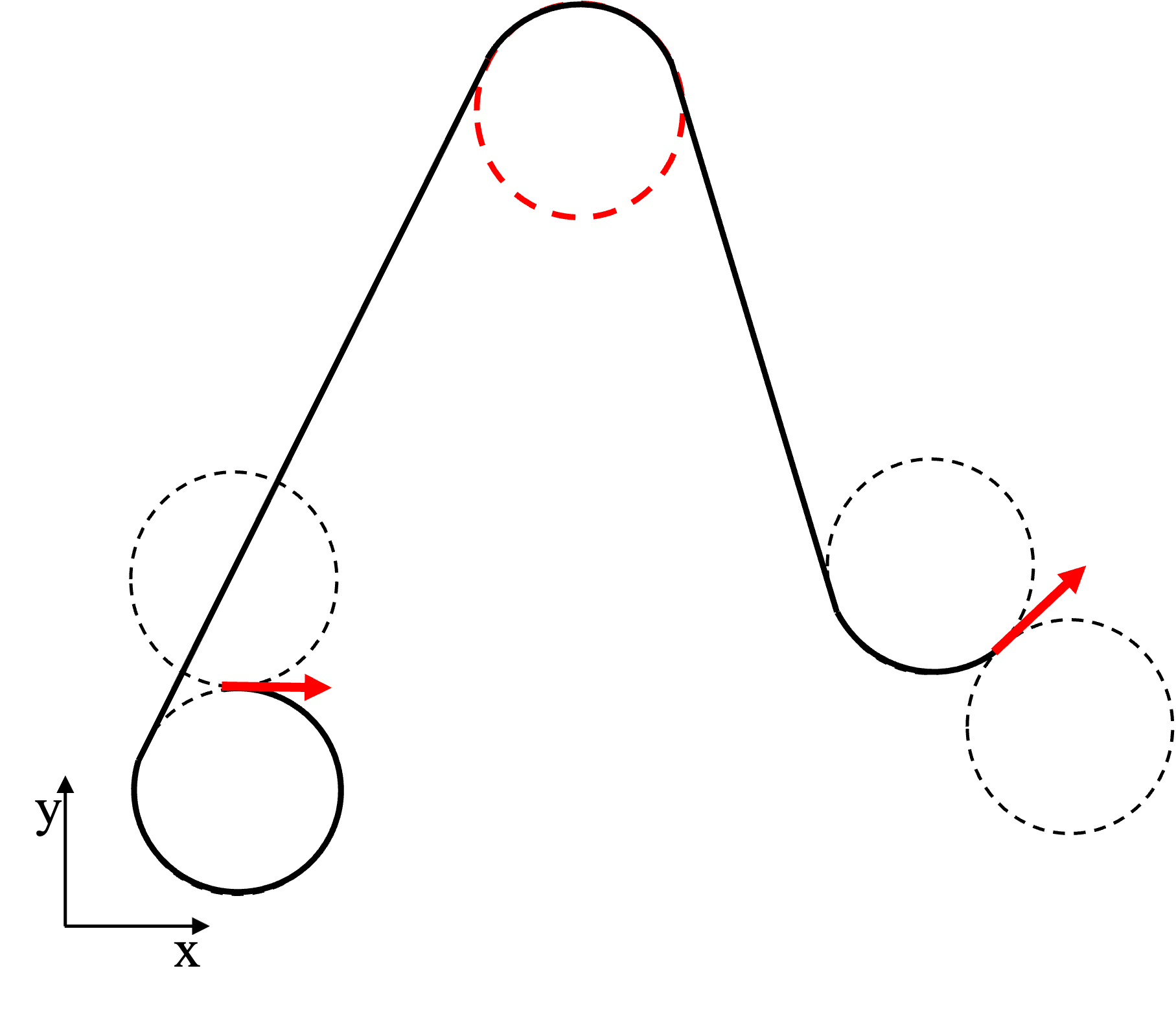}}
\subfigure[Right--Right]{\centering \includegraphics[scale=0.27]{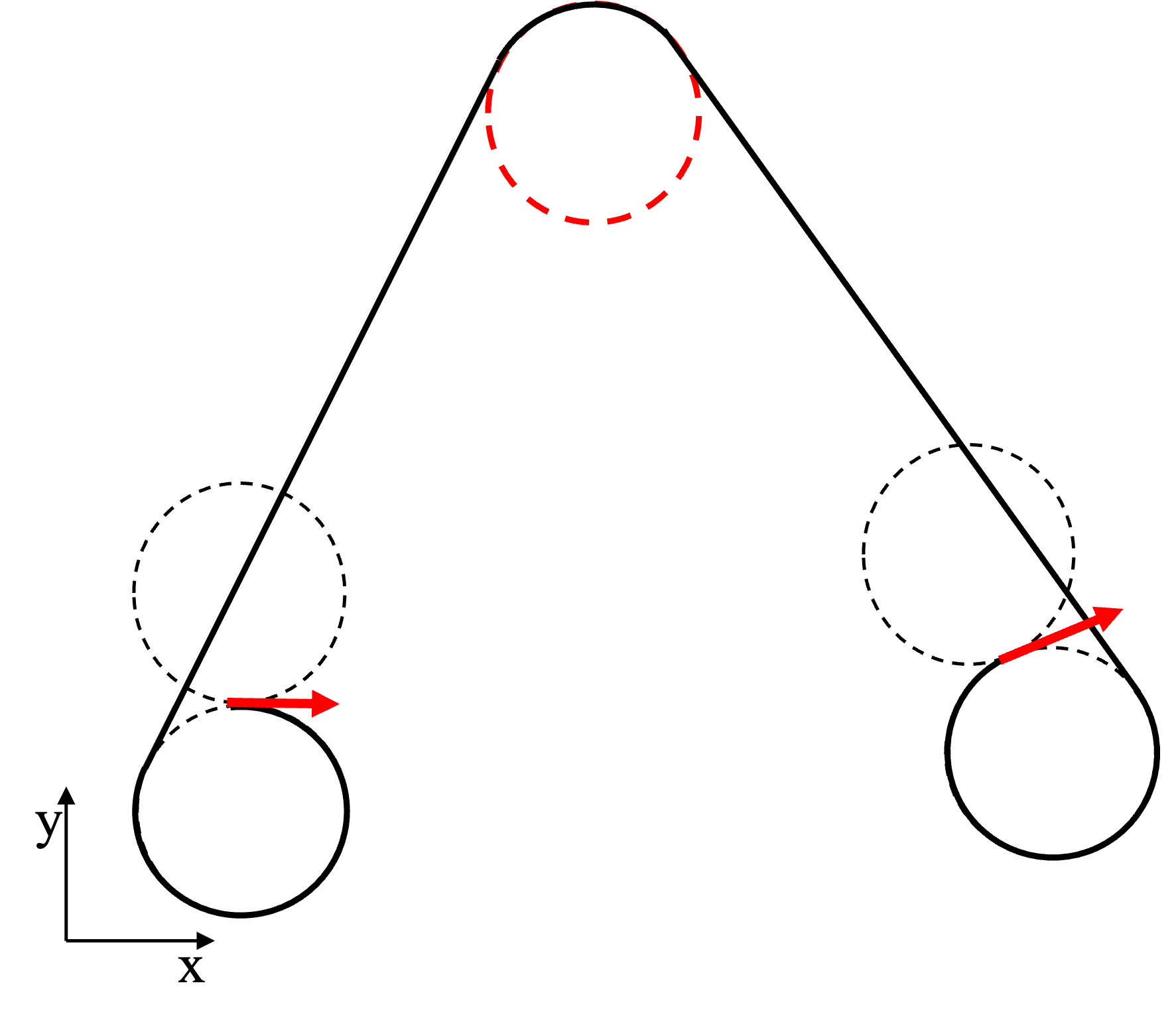}}
\caption{Four scenarios of reaching the upmost boundary of $\mathcal{B}'(\mathbb{X}, \kappa_m)$}
\label{fig_sec4_2}
\end{figure}

At the boundary points of the curve, among the black dotted circles depicted in Fig.~\ref{fig_sec3_2}, we label the circle with a signed curvature of $\kappa_s = \kappa_m$ as the \textit{left circle}, and the circle with $\kappa_s = -\kappa_m$ as the \textit{right circle}. Throughout this paper, the $x$ and $y$ coordinates in the figures are arranged such that the $x$--axis is parallel to the common tangent of the two left circles. The origin is positioned so that both left circles have $y$--coordinates equal to $r = \frac{1}{\kappa_m}$. The below Fig.~\ref{fig_sec4_2} depicts the four possible scenarios that may reach the upmost boundary of $\mathcal{B}'(\mathbb{X}, \kappa_m)$. Scenarios that reach the bottommost boundary can be easily understood by flipping Fig.~\ref{fig_sec4_2}. 
If a curve initiates from its starting point by following the left (right) circle and proceeds to its terminal point via the left (right) circle, it is denoted as `left (right) - left (right)'. 
The fundamental strategy for obtaining rectangular patches is to identify bounding rectangles for each left(right)--left(right) scenarios. In this section, for sake of brevity, we outline a technique for identifying patches in the left--left scenario for illustrative purposes. Nevertheless, the same approach is applicable to the other three cases as well. Additionally, it should be noted that the feasibility of each scenarios varies with the length of the curve, potentially resulting in the absence of corresponding rectangular patches. 
%
\subsection{Upmost Bound of CSCSC Curves}
We begin by obtaining the patch that includes the highest point(largest $y$--coordinate) in $\mathcal{B}'(\mathbb{X}, \kappa_m)$. There are two possible scenarios in which a curve of class CSCSC may attain the maximum $y$--coordinate. The first scenario is when the winding direction of the middle C is opposite to that of the remaining two C's. The second scenario is when the winding directions of all three C's are the same. 
The winding directions of all C's are predetermined in the remaining CCCCC, CSCCC, and CCCSC curves and hence there is no need for such case study. 
For sake of brevity, we defer the discussion of such non--CSCSC curves until Subsection~\ref{subsec:bottommost}. 
\subsubsection{Winding direction of the middle C is opposite} 

The Fig.~\ref{fig_sec4_4} below illustrates an instance of the CSCSC curve that attains the maximum $y$--coordinate where the winding direction of the middle C is opposite to that of the remaining two C's. 
Where $r = \frac{1}{\kappa_m}$, the center of the left circle at departure is positioned at $(0, r)$. Here, $\phi_1$ and $\phi_2$ denote the departure and arrival angles of the curve measured counterclockwise from the $x$--axis, respectively. 
\begin{figure}[htbp]
\centering
\subfigure[Local maximum case]{\centering \includegraphics[scale=0.27]{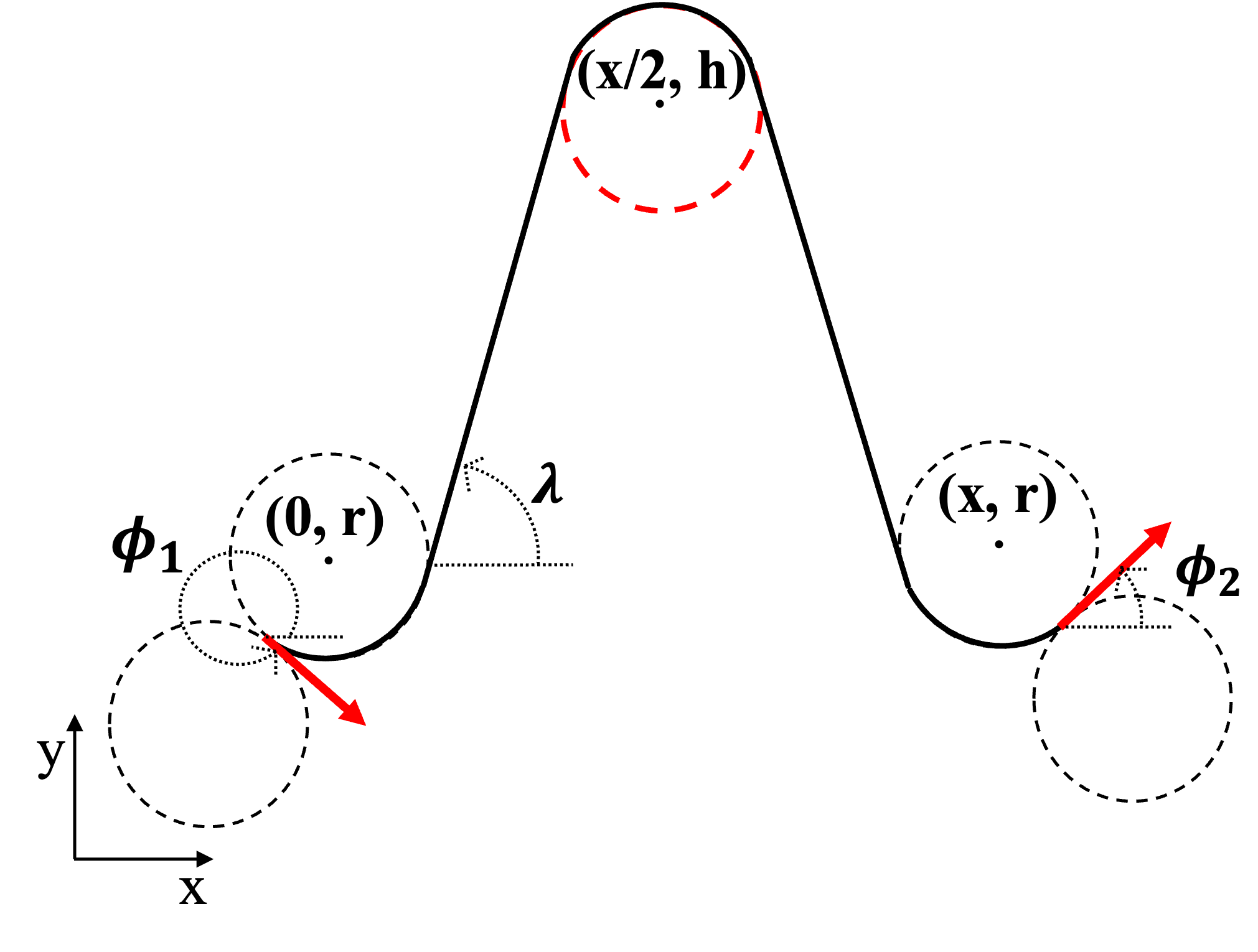}}
\subfigure[Boundary case]{\centering \includegraphics[scale=0.27]{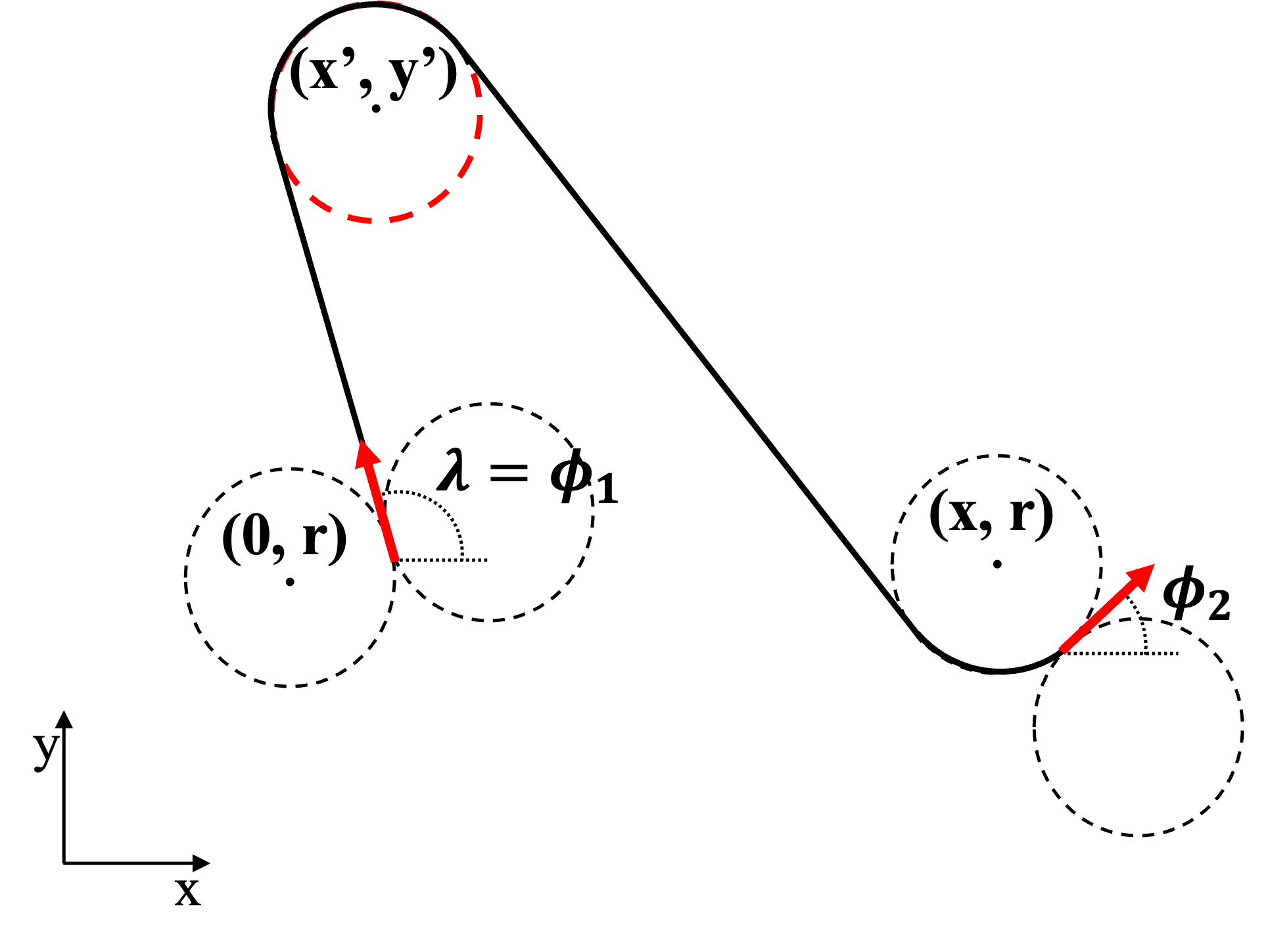}}
\caption{Illustration of the CSCSC curves that attain the maximum $y$ coordinate} 
\label{fig_sec4_4}
\end{figure}
%
%
At the local maximum case in Fig.~\ref{fig_sec4_4} (a), symmetry indicates that the maximum height is achieved at $\frac{x}{2}$. As a result, we can derive the following \textit{length equation} as follows: 
\begin{equation} \label{eq:21}
\begin{split}
	& f(h) = 2 \sqrt{\frac{x^2}{4} + (h-r)^2 - 4r^2} \\
	& + r \left[ 2\lambda + (\lambda - \phi_1 (\mathrm{mod}\ {2\pi})) + (\lambda + \phi_2 (\mathrm{mod}\ {2\pi})) \right] \\
	& - \ell = 0
\end{split}
\end{equation}
where $\ell$ denotes the length of the curve and $\lambda = \sin^{-1} \left( \frac{2r}{\sqrt{\frac{x^2}{4} + (h-r)^2}} \right) + \tan^{-1} \left( \frac{2(h-r)}{x} \right)$. The terms preceding $\ell$ represent the total length of the curve. While the previous sections focused on curves of unit length, this section considers curves of arbitrary length, $\ell$, for convenience. The only distinction lies in the normalization with respect to the scale of $\ell$. The modular operation is used to exclude the possibility of cycles that wrap around the circles. The square root term measures the length of the S part, while the latter term measures the C part. The length equation can be efficiently solved using numerical methods(e.g., Newton--Raphson), typically requiring only a few iterations. 
For the boundary case scenario depicted in Fig.~\ref{fig_sec4_4} (b), the length equation is derived by assuming that the red dotted circle is tangent to the line containing the departure angle and the initial point. 
Then the solution $h$ obtained by solving the Eq.~\eqref{eq:21} must be compared to the value at the boundary cases, such as the one illustrated in Fig.~\ref{fig_sec4_4} (b). This comparison leads to the determination of the maximum $y$--coordinate, given by $h_{max} = \max(h+r, r(1 - \cos\phi_1), r(1 - \cos\phi_2))$. The \textit{max} operation is to consider the cases when the concatenation point is on the first or last C, as outlined in the proof of Theorem~\ref{thm:1}. Similar comparisons need to be conducted when establishing the horizontal and bottommost bounds as well, following the determination of the location of the middle C. 
\subsubsection{Winding directions of all three C's are the same}

\begin{figure}[htbp]
\centering
\subfigure{\centering \includegraphics[scale=0.5]{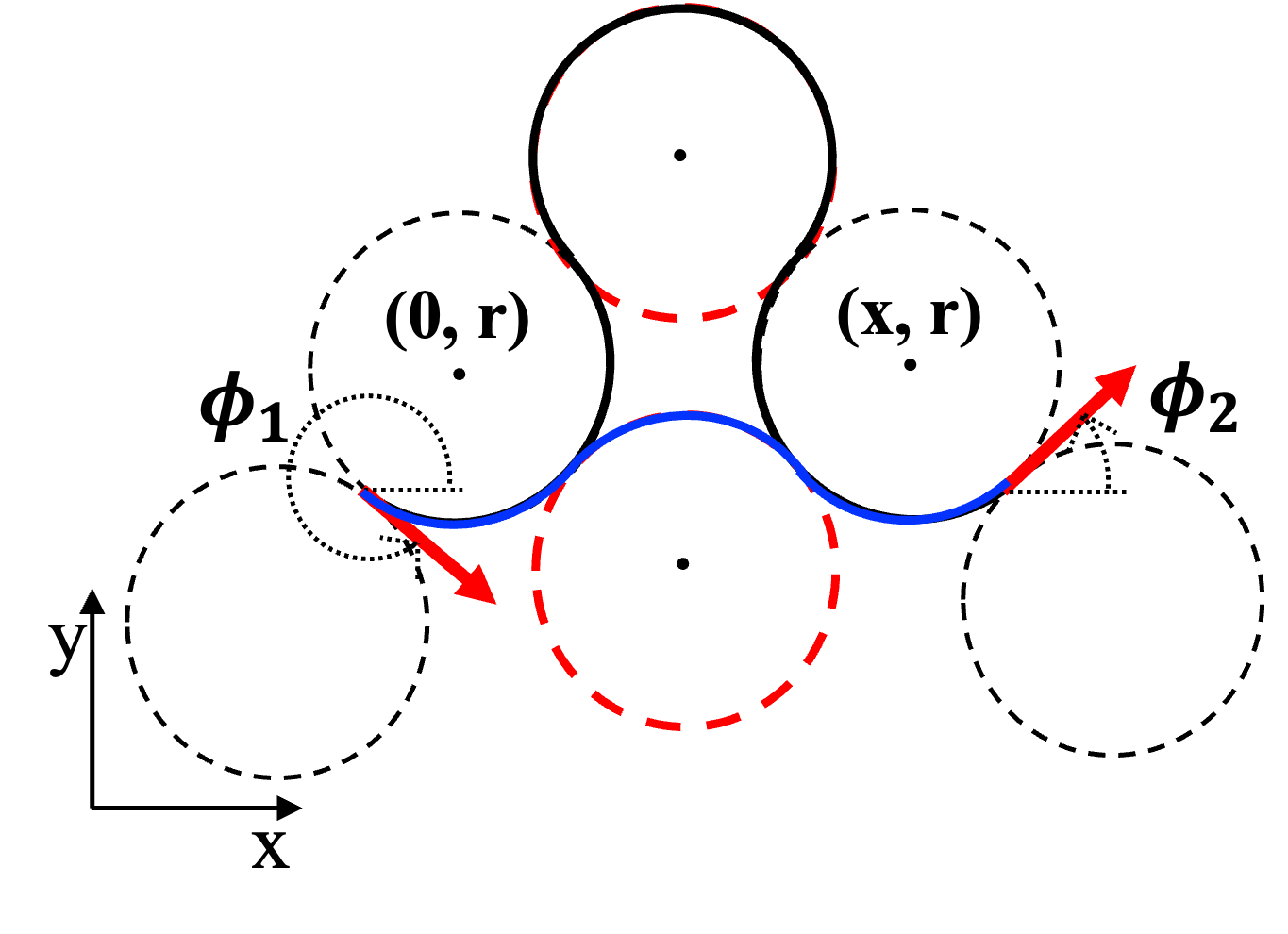}}
\caption{Illustration of a scenario where there exists infeasible lengths.} 
\label{fig_sec4_5-1}
\end{figure}

\begin{figure}[htbp]
\centering
\subfigure[Local maximum case]{\centering \includegraphics[scale=0.27]{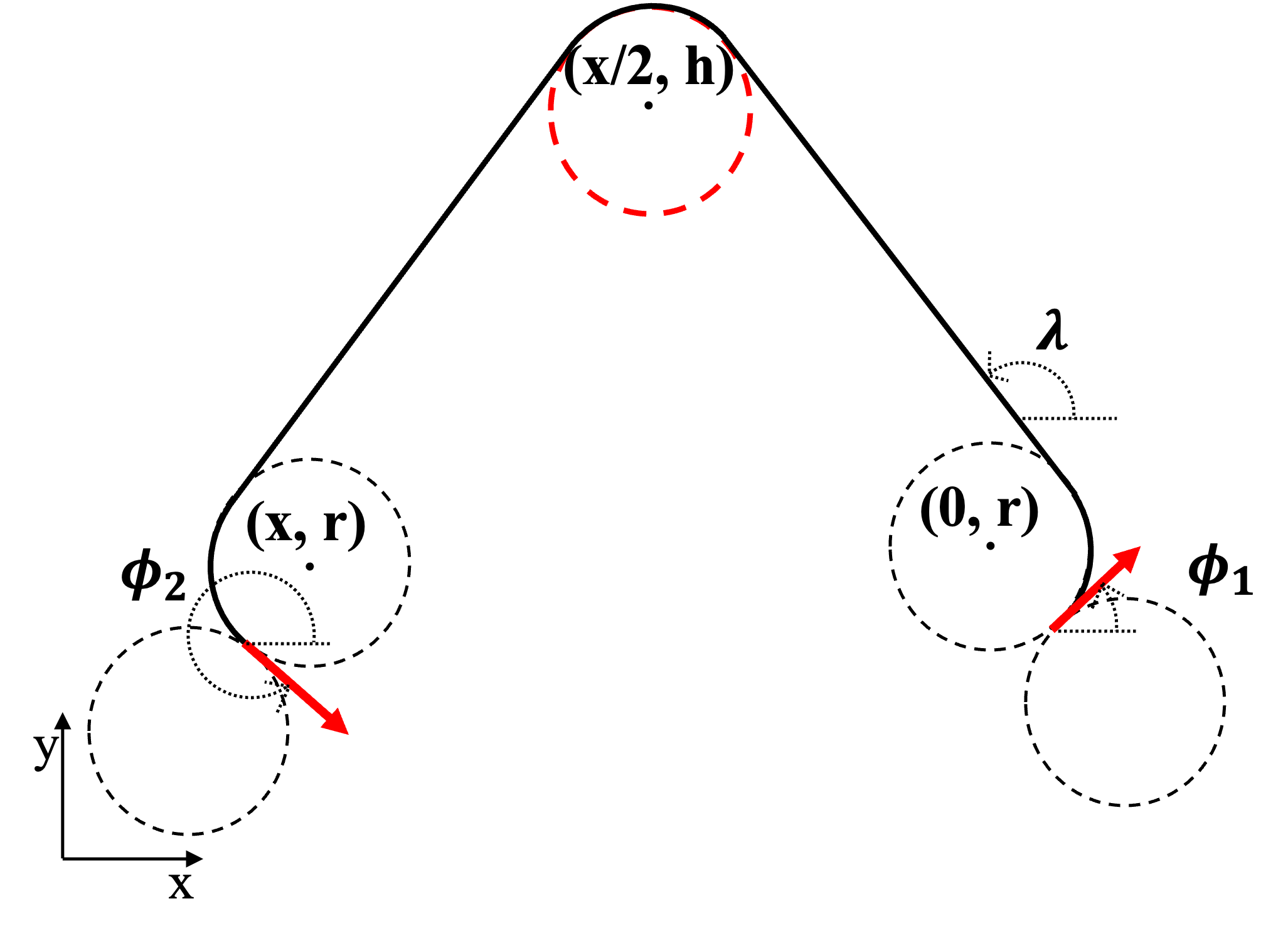}}
\subfigure[Boundary case]{\centering \includegraphics[scale=0.27]{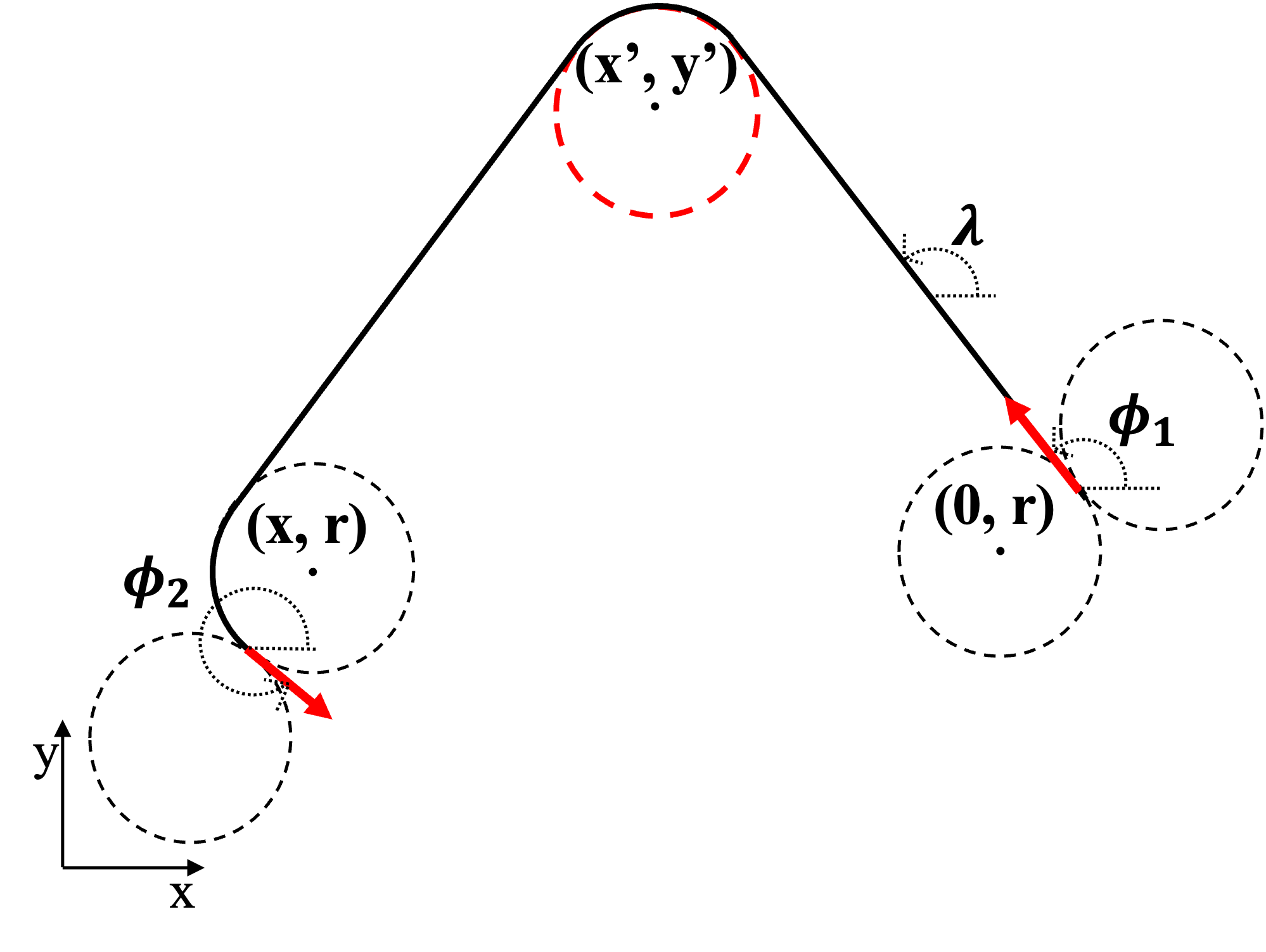}}
\caption{Illustration of a scenario where the initial point is on the right side of the endpoint.} 
\label{fig_sec4_5-2}
\end{figure}

The Fig.~\ref{fig_sec4_5} below illustrates an instance of the CSCSC curve which attains the maximum $y$--coordinate where the winding directions of all three C's are the same. 
The length equation in this case becomes as below: 
\begin{equation} \label{eq:22}
\begin{split}
	& f(h) = 2 \sqrt{\frac{x^2}{4} + (h-r)^2} \\
	& + r \left[ (2\pi - 2\lambda) + (\lambda - \phi_1 (\mathrm{mod}\ {2\pi})) \right. \\
	& + \left. (\lambda + \phi_2 (\mathrm{mod}\ {2\pi})) \right] - \ell = 0
\end{split}
\end{equation}
where $\ell$ denotes the length of the curve and $\lambda = \tan^{-1} \left( \frac{2(h-r)}{x} \right)$. The length equation can be solved by similar means with Eq.~\eqref{eq:21}, followed by routine comparison with the boundary cases as those depicted in Fig.~\ref{fig_sec4_5} (b) and Fig.~\ref{fig_sec4_5-2} (b). 
However, this scenario usually does not contribute to the bound, as the height of the red dotted circles depicted in Fig.~\ref{fig_sec4_4} are higher compared to the ones depicted in Fig.~\ref{fig_sec4_5}. The scenarios where such a curve can achieve the maximum height are limited to two cases. First is where $x < 4r$, and the length of the curve falls between the black and blue lines as depicted in Fig.~\ref{fig_sec4_5-1}. Such situation is known as the \textit{closed region}~\cite{chen2023}. Within this range of $x$ value and length, the CSCSC curve in the previous scenario becomes infeasible. Detailed conditions regarding such length and terminal position is described in~\cite{chen2023}. As a result, the CSCSC curve shown in Fig.~\ref{fig_sec4_5} (a) may attain the highest $y$--coordinate instead. In such scenarios, the CSCSC curves depicted in Fig.~\ref{fig_sec4_4}, with possible cycles, should also be taken into consideration and compared. Second is when the $x$--coordinate of the endpoint is small enough; so that it is close enough to the initial point, or even behind the initial point. Such scenario is depicted in Fig.~\ref{fig_sec4_5-2}. 
\begin{figure}[htbp]
\centering
\subfigure[Local maximum case]{\centering \includegraphics[scale=0.27]{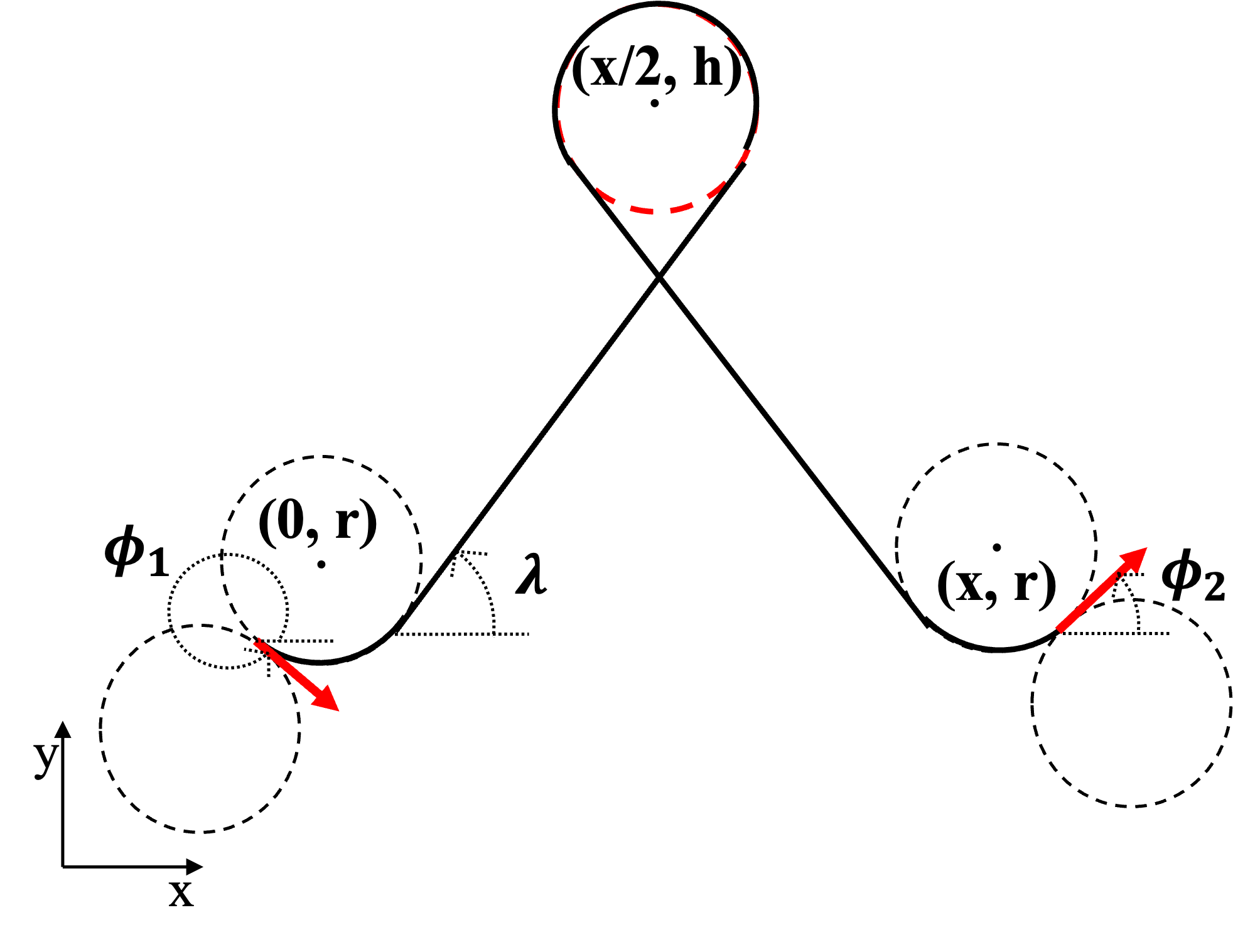}}
\subfigure[Boundary case]{\centering \includegraphics[scale=0.27]{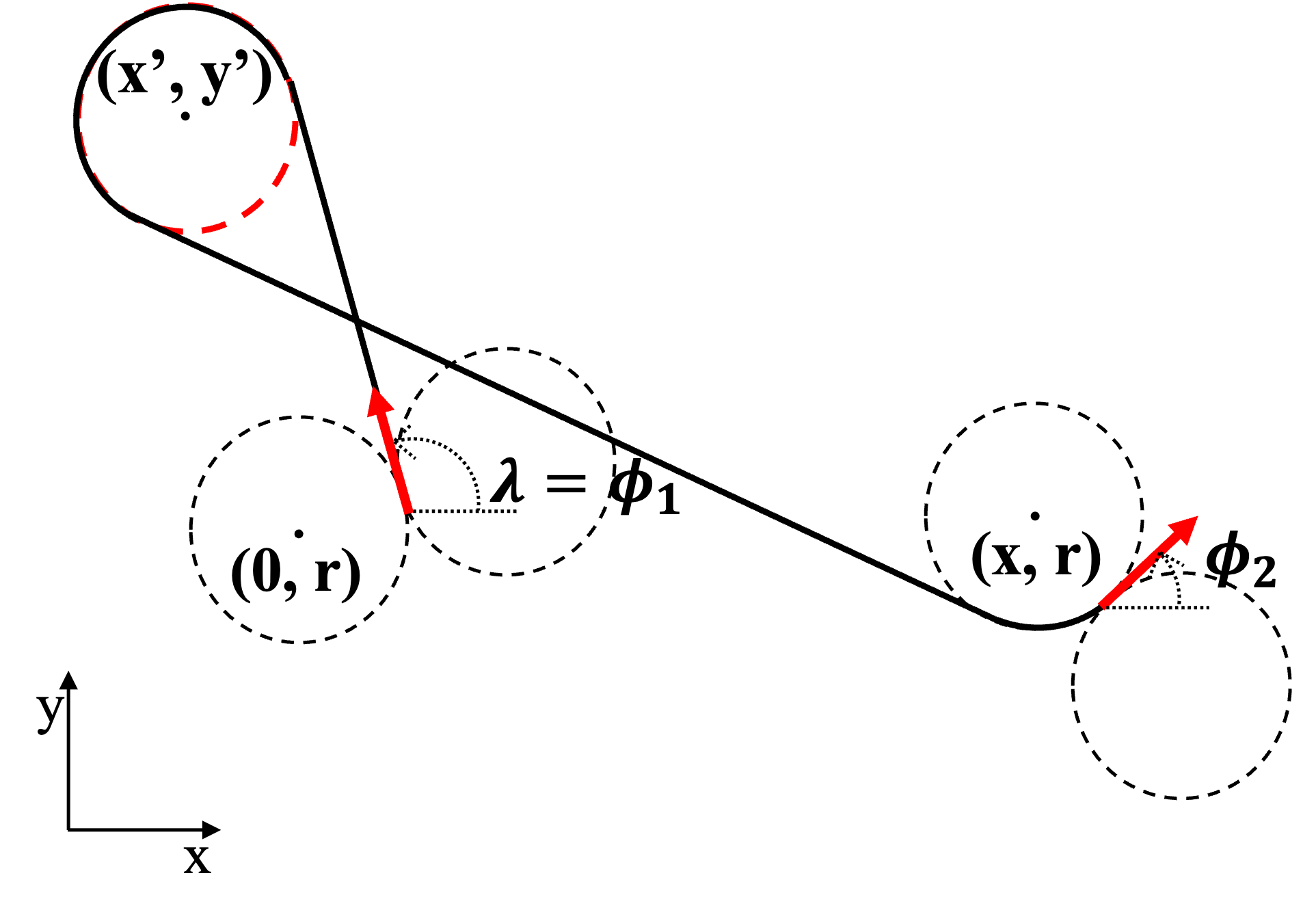}}
\caption{Illustration of the CSCSC curves that attain the maximum $y$--coordinate.} 
\label{fig_sec4_5}
\end{figure}
In the subsequent subsection, we introduce a method for determining horizontal bounds, the maximum and minimum $x$--coordinates. 
\subsection{Horizontal Bound of CSCSC Curves}
\begin{figure}[htbp]
\centering
\subfigure[Local minimum case, large $\ell$, $\beta = 0$]{\centering \includegraphics[scale=0.26]{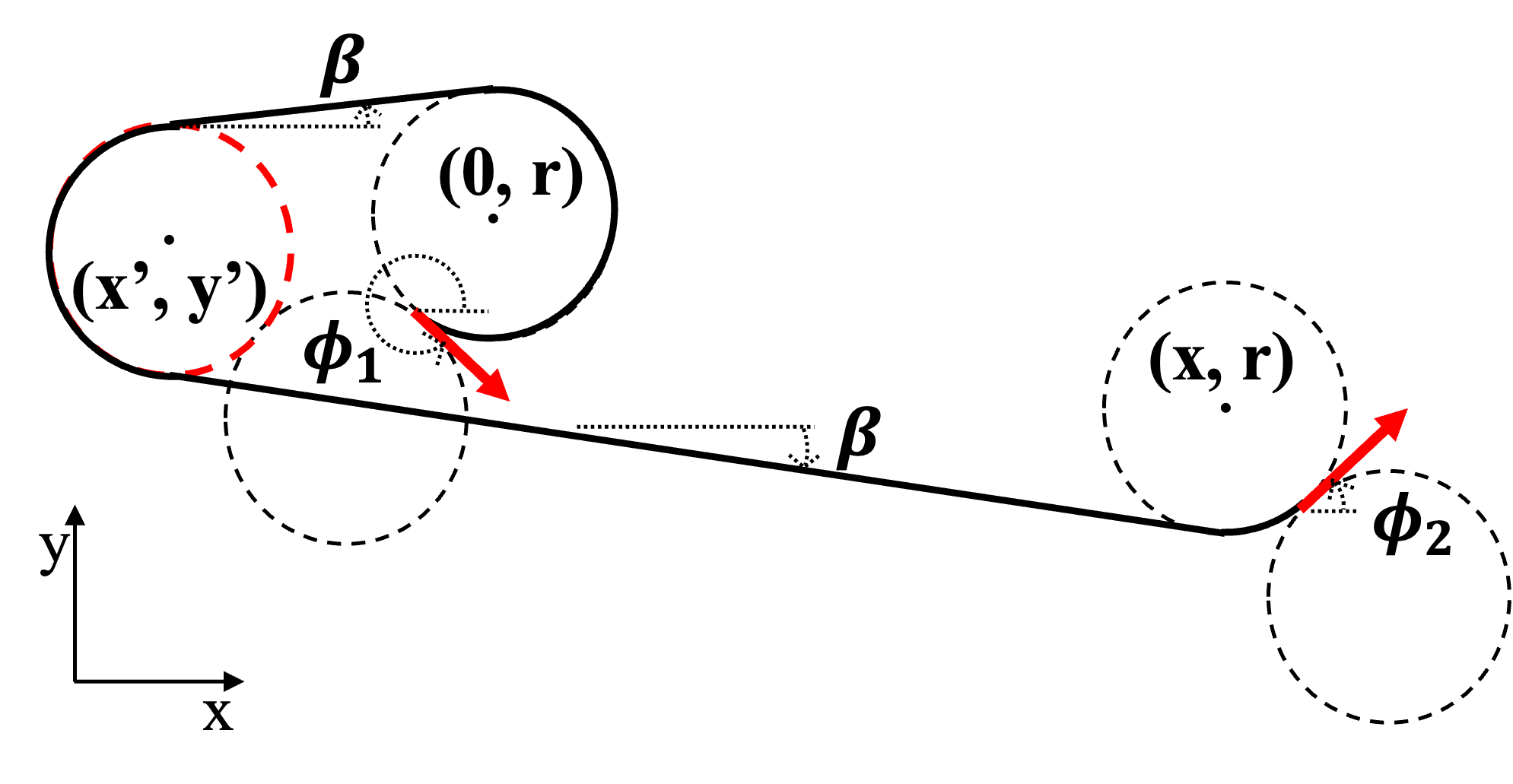}}
\subfigure[Small $\ell$]{\centering \includegraphics[scale=0.25]{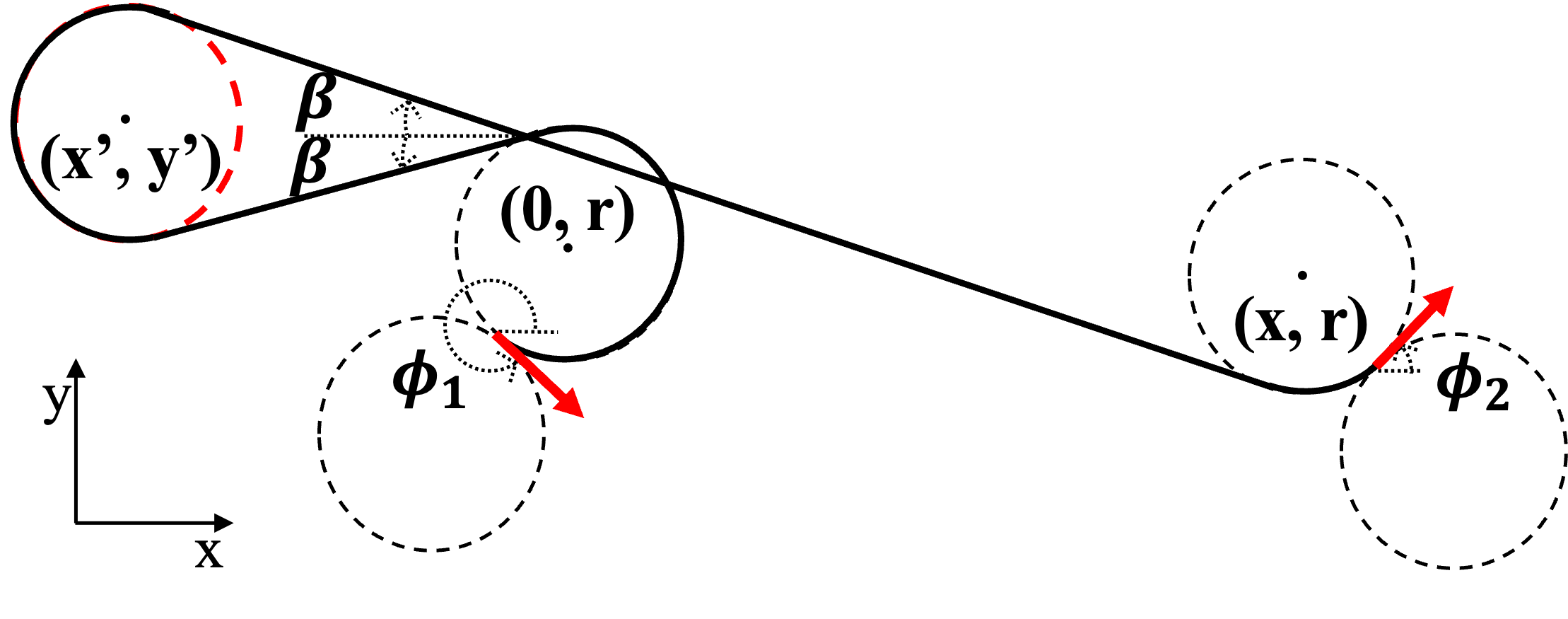}}
\subfigure[Small $\ell$, CCSC]{\centering \includegraphics[scale=0.27]{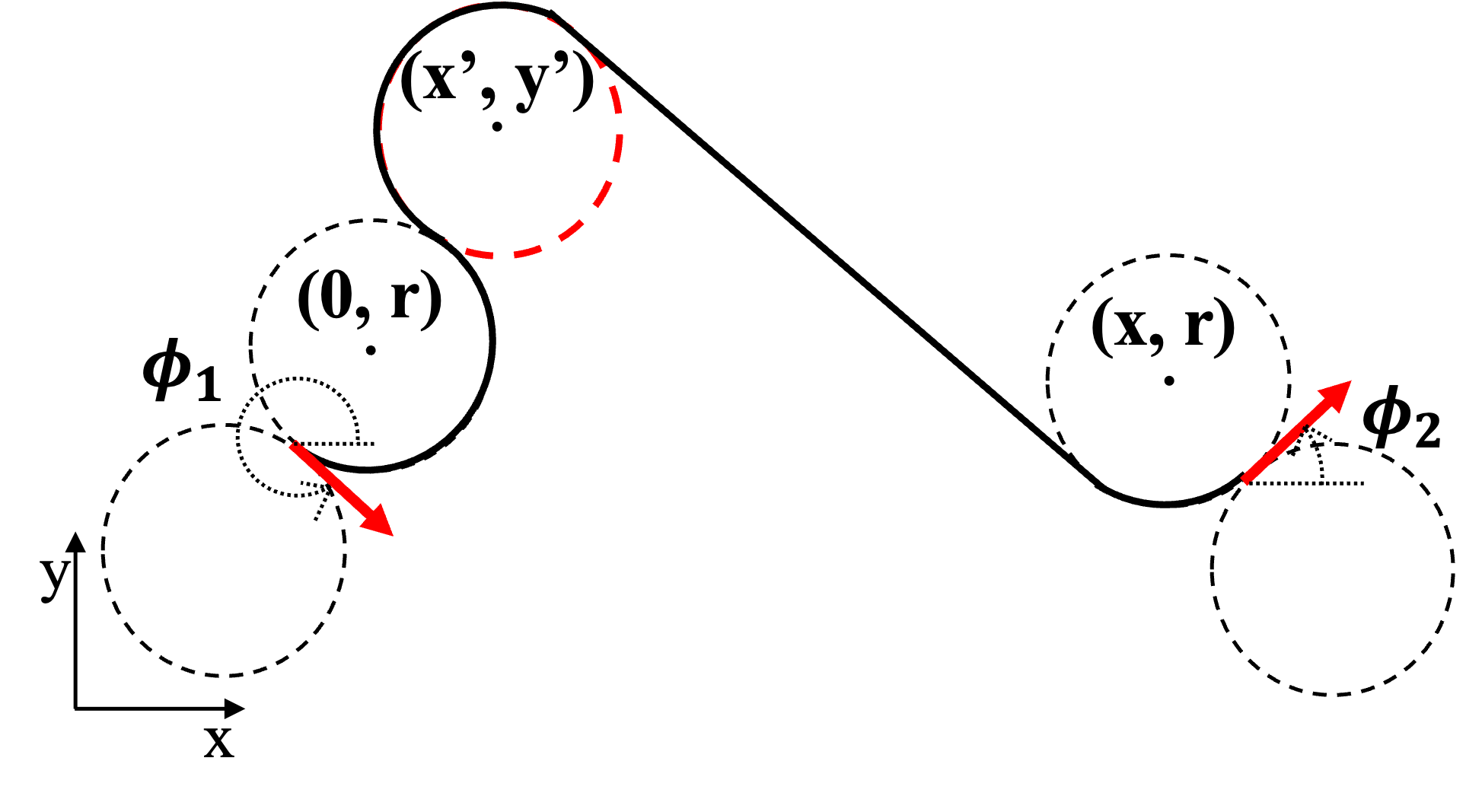}}
\subfigure[Boundary case]{\centering \includegraphics[scale=0.27]{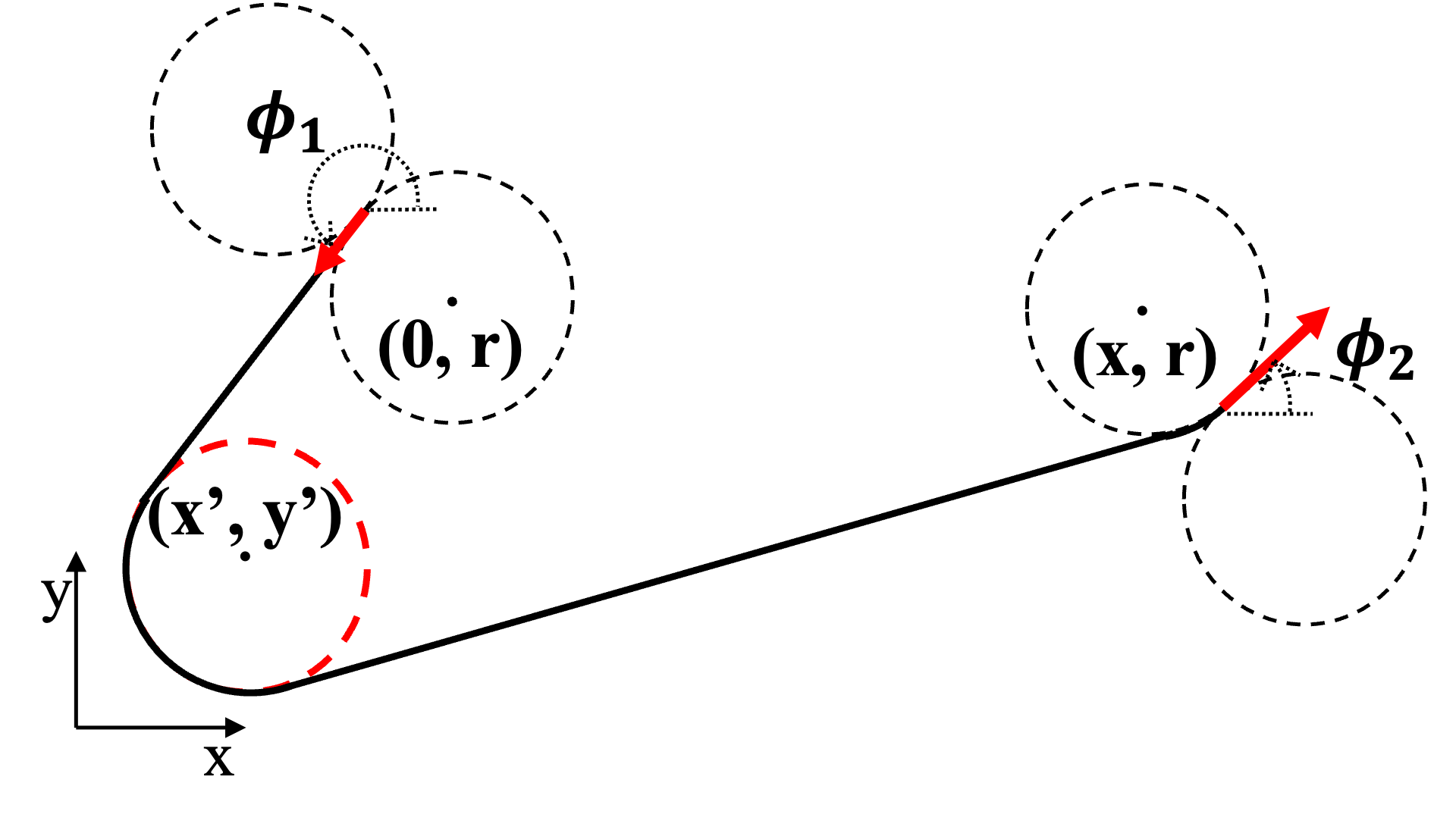}}
\caption{Illustration of the CSCSC curves that attain the optimal $x$--coordinates} 
\label{fig_sec4_6}
\end{figure}

%
Illustrative examples CSCSC curves that achieve the minimum $x$--coordinate in the left--left scenario are depicted in Fig.~\ref{fig_sec4_6}. Similar to the previous subsection, such scenario is when the three C's have same winding directions. For sufficient lengths that such case is feasible, symmetry implies that the curve attains its minimum $x$--coordinate where the two S's have same slopes of different signs, denoted by $\beta$, as depicted in Fig.~\ref{fig_sec4_6} (a). Such condition simplifies to $\beta = 0$. It is important to note that the location of the terminal left circle in Fig.~\ref{fig_sec4_6} (a) is adjusted for illustrative purposes; in reality, the two left circles and the red dotted circle must be aligned with the $x$--axis. 
The remaining case is when the middle C has opposite winding direction. For shorter lengths where the previous shape is not feasible, the minimum $x$ coordinate is achieved by the curves shown in Fig.~\ref{fig_sec4_6} (b) and (c). Therefore, we can formulate a similar length equation as Eq.~\eqref{eq:21} and compare the local and boundary cases to determine the minimum $x$--coordinate on the middle C. 
It should be noted that for the curves consist of four or lesser components as in Fig.~\ref{fig_sec4_6} (c), the location of the middle C is uniquely determined only by the length equation, irrelevant to the optimal condition. Hence, only a finite number of scenarios need to be obtained and compared. 
In practical applications, the scenario depicted in Fig.~\ref{fig_sec4_6} (c) is highly desirable than others. This is because a large distance between adjacent mesh points implies a significant time (or any other parametrization variable depending on the problem) difference between these points. Consequently, this can lead to potential errors in dynamics interpolation, irrespective of any violation of path constraints. As a result, the obtained mesh points themselves might be inaccurate in terms of fulfilling the dynamics constraint. 
%
%
%
\subsection{Bottommost Bound and the Non--CSCSC Curves}\label{subsec:bottommost}
\begin{figure}[htbp]
\centering
\subfigure[Local minimum case]{\centering \includegraphics[scale=0.27]{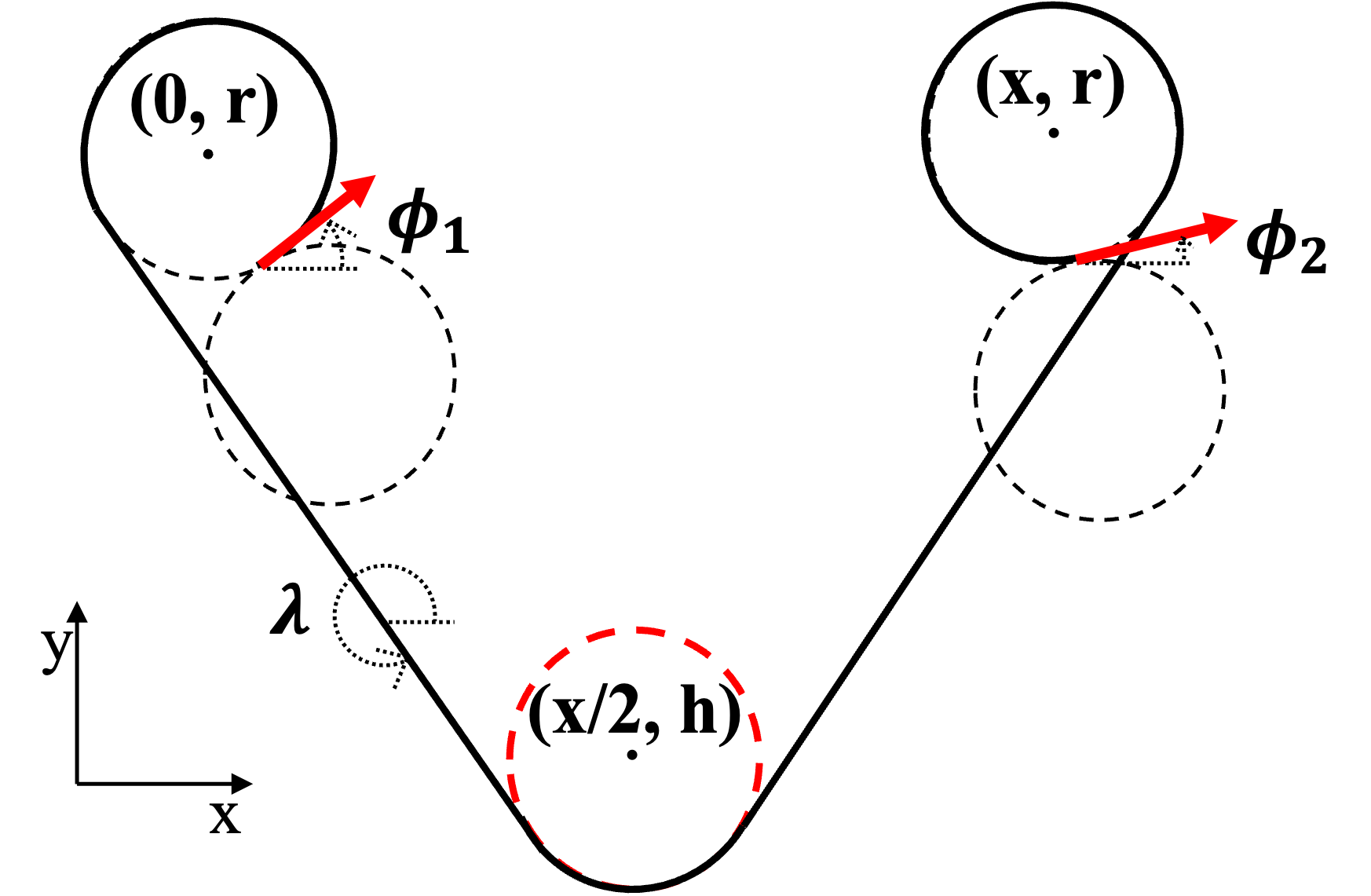}}
\hskip 3mm
\subfigure[Boundary case]{\centering \includegraphics[scale=0.27]{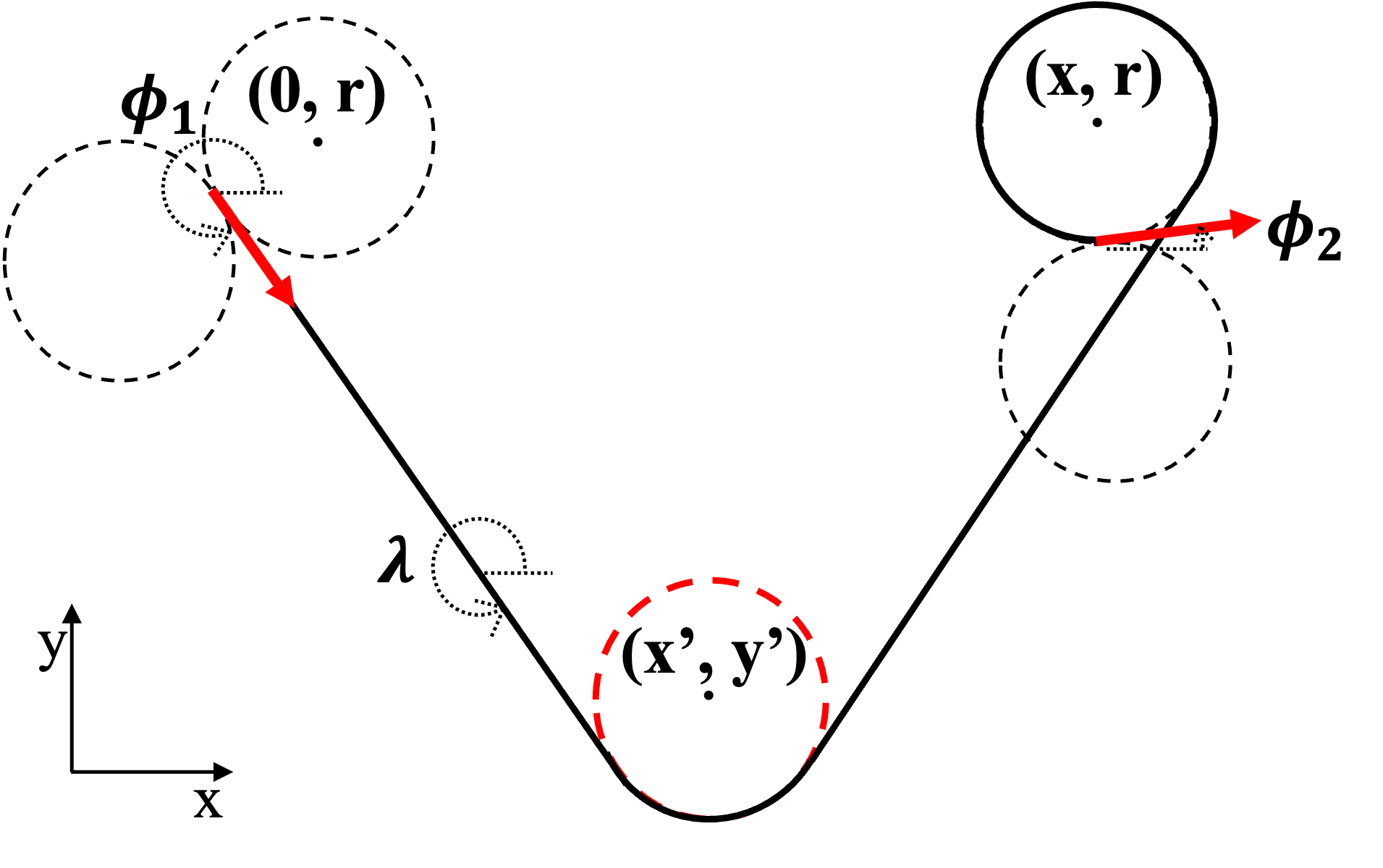}}
\subfigure[Local minimum case]{\centering \includegraphics[scale=0.27]{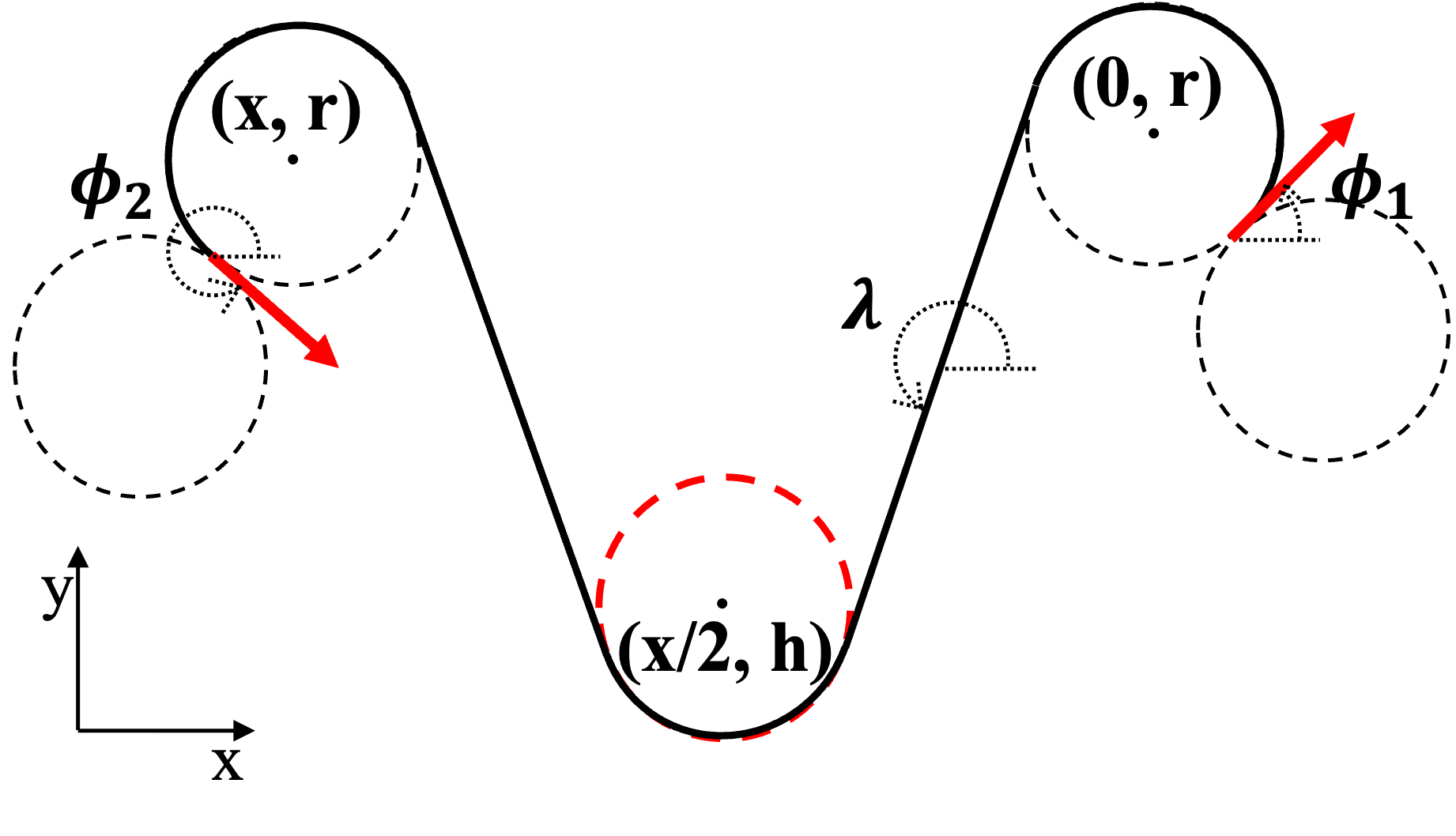}}
\hskip 0mm
\subfigure[Boundary case]{\centering \includegraphics[scale=0.27]{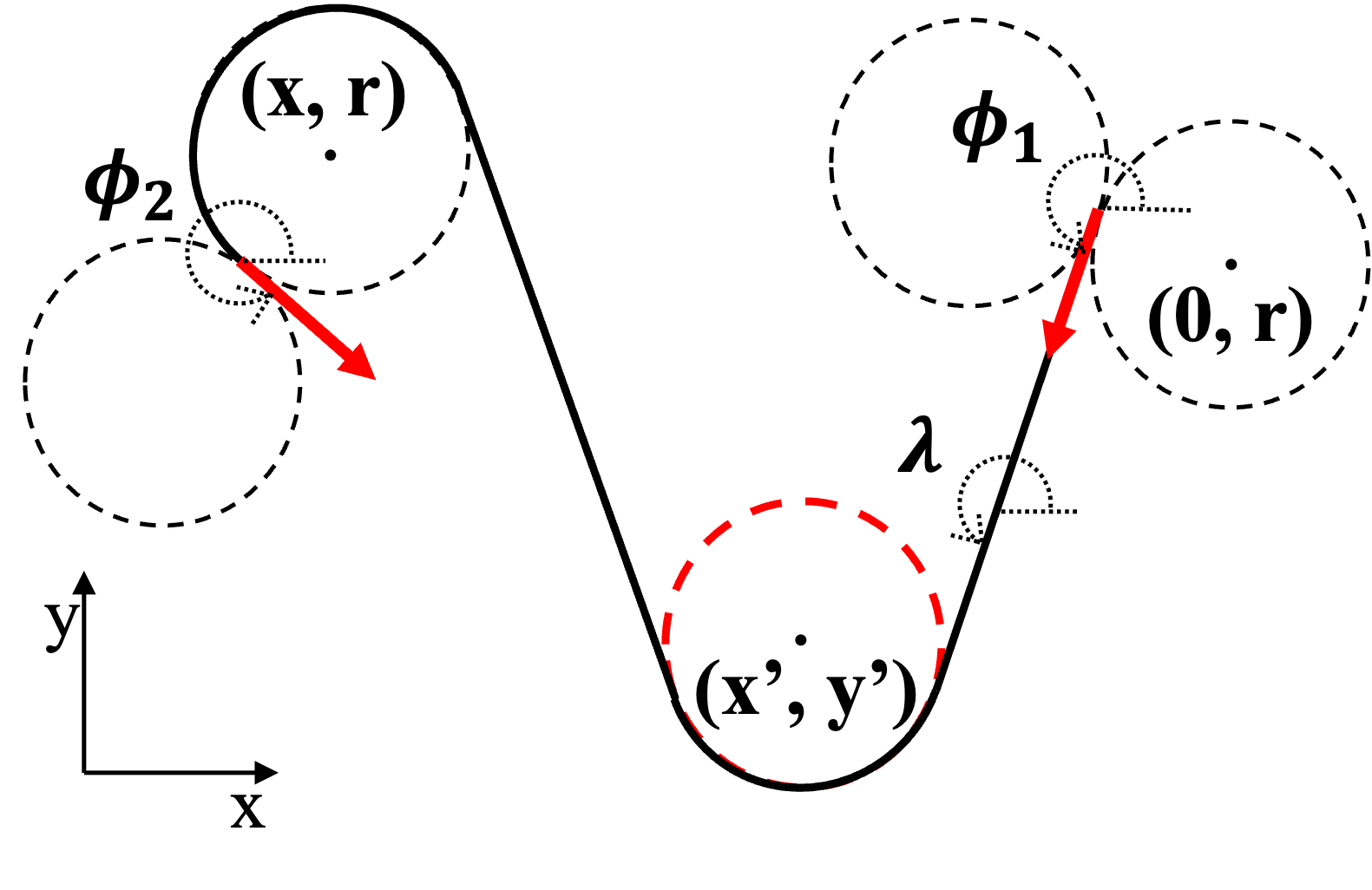}}
\caption{Illustration of the CSCSC curves that attain minimum $y$--coordinate} 
\label{fig_sec4_7}
\end{figure}

In this subsection, we present a method for determining the minimum $y$--coordinate, or the bottommost bound. Upmost and horizontal bounds constructed by non--CSCSC curves are discussed as well. Illustrative examples of curves of class CSCSC that achieve minimum $y$--coordinate in the left--left scenario is depicted in Fig.~\ref{fig_sec4_7}. A similar case study as the upmost bound scenario is possible. Scenario where the winding directions of all three C's are the same is depicted in Fig.~\ref{fig_sec4_7} (a) and (b). The other scenario is depicted in Fig.~\ref{fig_sec4_7} (c) and (d). For the first scenario, the length equation is given as Eq.~\eqref{eq:22}, and Eq.~\eqref{eq:21} for the second scenario. 
Then the lower bound obtained by solving the Eqs.~\eqref{eq:21} and~\eqref{eq:22} must be compared to the value at the boundary cases depicted in Fig.~\ref{fig_sec4_7} (b) and (d). 
%

\begin{figure}[htbp]
\centering
\subfigure{\centering \includegraphics[scale=0.4]{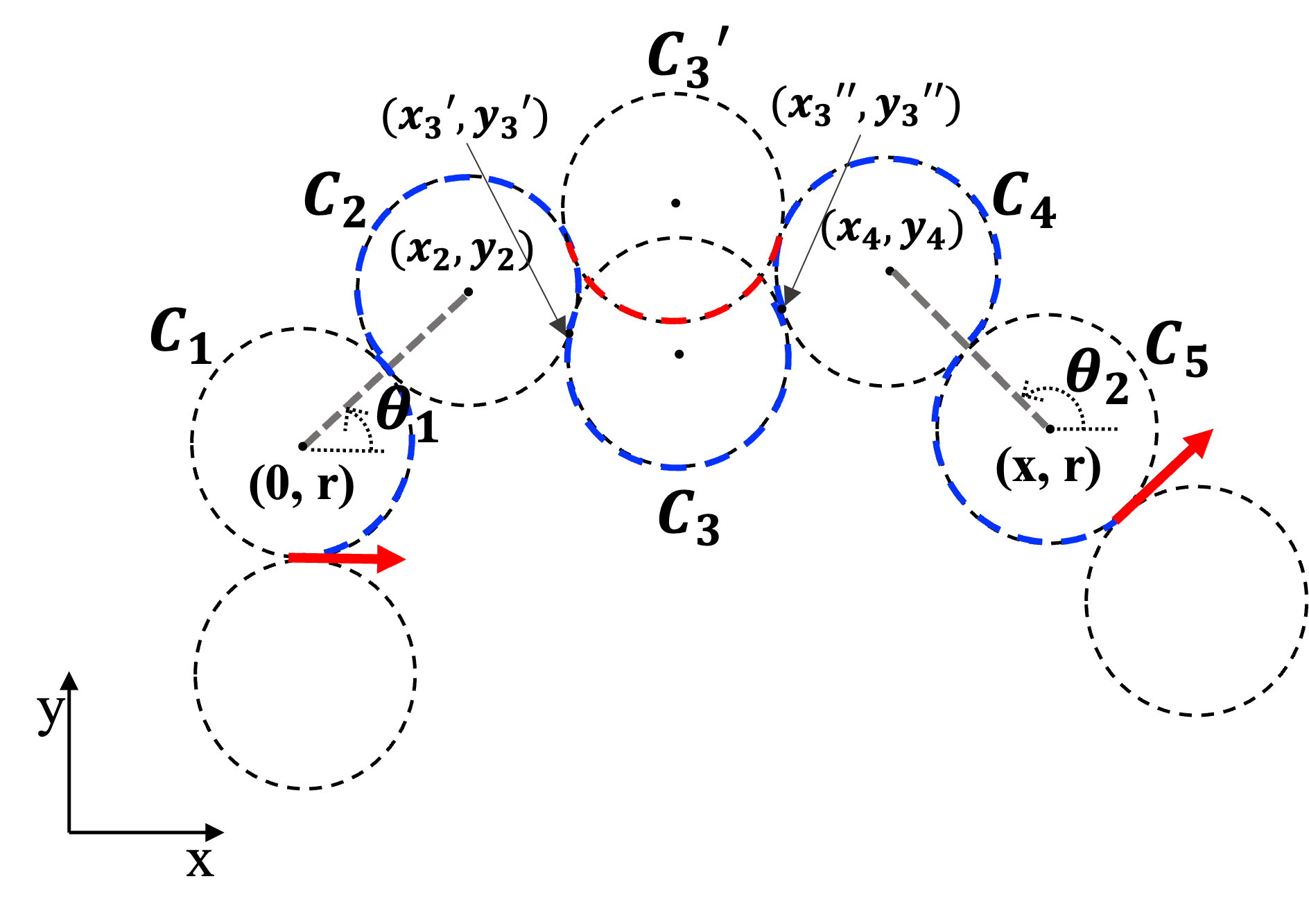}}
\caption{Illustration of CCCCC curve given $\theta_1$ and $\theta_2$} 
\label{fig_sec4_7e}
\end{figure}

For non--CSCSC curves, let us denote the $i$--th C component by $C_i$. The concatenation point of the two trajectories from problem [$\textbf{P3}$] must exist within the first, middle, or last C. Because the locations of the first and last C are fixed, it suffices to determine the location of the middle C. For the CCCCC curves, suppose the angles $\theta_1$ and $\theta_2$ in Fig.~\ref{fig_sec4_7e} are given. Then the centers of $C_2$ and $C_4$, denoted as $(x_2, y_2)$ and $(x_4, y_4)$ respectively, are determined and hence the two possible locations of $C_3$ as well. Two such C's are indicated as $C_3$ and $C_3'$ in Fig.~\ref{fig_sec4_7e}. Let us denote its center by $(x_3(\theta_1, \theta_2), y_3(\theta_1, \theta_2))$ and the boundary points by $(x_3'(\theta_1, \theta_2), y_3'(\theta_1, \theta_2))$ and $(x_3''(\theta_1, \theta_2), y_3''(\theta_1, \theta_2))$. Then $(x_3(\theta_1, \theta_2), y_3(\theta_1, \theta_2)) = \left(\frac{x_2 + x_4}{2}, \frac{y_2 + y_4}{2}\right) \pm \boldsymbol{h}$ where $\boldsymbol{d} = (x_4, y_4) - (x_2, y_2)$, $\boldsymbol{h} = \sqrt{ 4r^2 - \left( \frac{|\boldsymbol{d}|}{2} \right)^2 } R_{90}(\hat{\boldsymbol{d}})$, and $R_{90}$ represents the counterclockwise $90^{\circ}$ rotation in $\mathbb{R}^2$. The length of the curve can also be obtained explicitly by similar means in the previous subsections. 
Then numerical techniques(e.g., Newton--Raphson) can be applied to find the bounds of $x_3$, $x_3'$, $x_3''$, $y_3$, $y_3'$, and $y_3''$, and hence the bound of the envelope of the middle C. Analogous approach is available for the CSCCC and CCCSC curves as well. 

\begin{figure}[htbp]
\centering
\subfigure{\centering \includegraphics[scale=0.4]{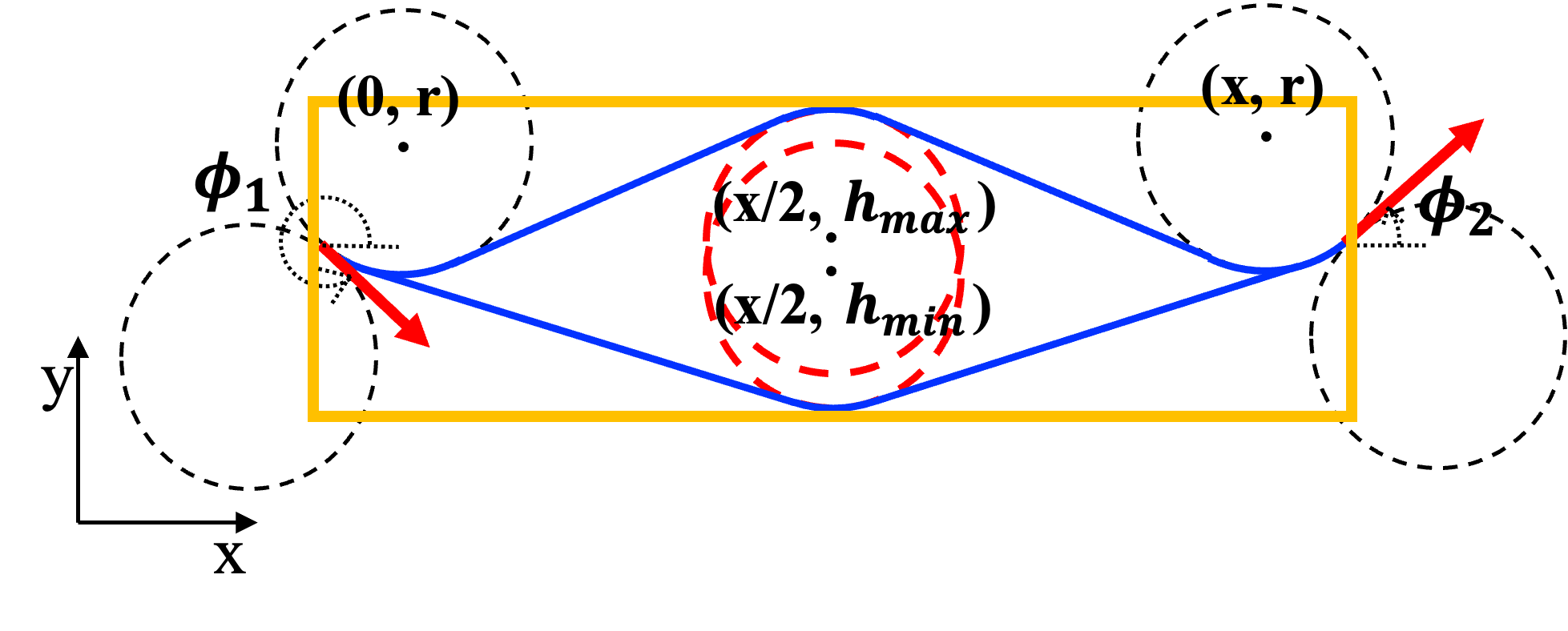}}
\caption{Schematic illustration of a rectangular patch in a left--left scenario} 
\label{fig_sec4_8}
\end{figure}

Building on the previous discussions, a schematic illustration of a rectangular patch in a left--left scenario is shown in Fig.~\ref{fig_sec4_8}. The boundary of the rectangular patch is depicted by a yellow solid line. 
%
%
%

%
\subsection{Mesh Refinement Algorithm and the Definiteness Theorem}
Based on the previous discussions, the comprehensive mesh refinement algorithm integrated with NLP solvers is outlined as below Algorithm~\ref{alg:01}. 
\begin{algorithm}[htbp]
\caption{Mesh Refinement Algorithm(NLP Solver)}\label{alg:01}
\begin{algorithmic}[1]
\Require Initial mesh points $\{t_j\}_{1:N^{(0)}}$, tolerance $\varepsilon > 0$
\For{k = 1:MaxIteration}

\State Solve the discretized problem, Eq.~\eqref{eq:2}, with mesh points $\{t_j\}_{1:N^{(k-1)}}$ 

\For{i = 1:N-1}
	
	\State Obtain rectangular patches at the $i$--th mesh interval. 
	
	\If{Any rectangular patch intrudes the forbidden region more than $\varepsilon$} 
		\State Divide the mesh interval $\left[ t_{i}, t_{i+1} \right]$ into two equi-sized subintervals. 
	\EndIf

\EndFor

\State Update $N^{(k)}$ 
\State Update mesh points, $\{t_j\}_{1:N^{(k)}}$ 

\If{No mesh interval is refined in current loop}
 \State \textbf{break}
\EndIf

\EndFor

\end{algorithmic}
\end{algorithm}
The process continues until a sufficient number and locations of mesh points are obtained to ensure that the rectangular patches do not intersect the forbidden regions. Importantly, this algorithm does not introduce a significant computational burden that could impact real-time feasibility. The following theorem proves that the process of mesh refinement does not continue indefinitely for any given tolerance $\varepsilon > 0$. It is important to highlight that the assumption that the mesh points remain outside the forbidden region is essentially a restatement of the path constraint applied to the discretized problem, as shown in Eq.~\eqref{eq:2}. In practical applications, the forbidden region that is imposed as a path constraint should be expanded by an amount of $\varepsilon$ compared to the original forbidden region that is intended. 
\begin{thm} \label{thm:02}
	Suppose that the dynamics function $f$ in Eq.~\eqref{eq:1} is bounded within the mesh intervals $[\tau_j, \tau_{j+1}]$. If the state variables at the mesh points remain outside of the forbidden region, then for any $\varepsilon > 0$, it is possible to achieve mesh points that ensure a violation of the forbidden region between adjacent mesh points by an amount less than $\varepsilon$. Moreover, such mesh points can be obtained through a finite number of refinements. 
\end{thm}
\begin{pf*}{Proof.}
	Boundedness of $f$ implies Lipschitz continuity of the state variables. Hence, there exists an upper bound, $V_M$, for the speed that the trajectory can travel within the mesh interval. Consequently, for an arbitrary mesh interval, if $\Updelta \tau < \frac{\varepsilon}{V_M}$, then the amount of violation into the forbidden region cannot exceed $V_M \cdot \Updelta \tau < \varepsilon$. Hence, the interval $[\tau_j, \tau_{j+1}]$ can be refined into multiple mesh intervals such that the maximum interval size is less than $\frac{\varepsilon}{V_M}$. Such condition can be achieved by at most $N = \left\lceil \log_{2} \frac{V_M\Updelta\tau_j}{\varepsilon} \right\rceil$ number of refinements. 
\qed
\end{pf*}
Furthermore, integration of the mesh refinement algorithm into iterative methods such as SCP framework in~\cite{malyuta2022} and \cite{zhou2021sequential} is presented in the below Algorithm~\ref{alg:02}. The refinement process can be incorporated during the iterations while the solution is progressively converging to optimal. 
\begin{algorithm}[htbp]
\caption{Mesh Refinement Algorithm(SCP Framework)}\label{alg:02}
\begin{algorithmic}[1]
\Require Initial mesh points $\{t_j\}_{1:N^{(0)}}$, initial guess $\boldsymbol{z}^{(0)}$, tolerances $\varepsilon, \varepsilon_{TRC} > 0$
\For{k = 1:MaxIteration} 

\State Solve the convexified discrete problem using mesh points $\{t_j\}_{1:N^{(k-1)}}$ and initial guess $\boldsymbol{z}^{(k-1)}$ 

\For{i = 1:$N^{(k-1)}-1$}
	
	\State Obtain rectangular patches at the $i$--th mesh interval. 
	
	\If{Any rectangular patch intrudes the forbidden region more than $\varepsilon$} 
		\State Divide the mesh interval $\left[ t_{i}, t_{i+1} \right]$ into two equi-sized subintervals. 
	\EndIf

\EndFor

\State Update $N^{(k)}$
\State Update mesh points, $\{t_j\}_{1:N^{(k)}}$ 
\State Update initial guess, $\boldsymbol{z}^{(k)}$

\If{ $\left|\boldsymbol{z}^{(k)} - \boldsymbol{z}^{(k-1)}\right| < \varepsilon_{TRC}$ \textbf{and} No mesh interval is refined in current loop} 
 \State \textbf{break}
\EndIf

\EndFor

\end{algorithmic}
\end{algorithm}
The \textit{convexified discrete problem} in Algorithm~\ref{alg:02} refers to the problem Eq.~\eqref{eq:2} with its component functions convexified, typically through linearization around the initial guess. Detailed convexification process in a practical example is outlined in Appendix.~\ref{appendix:01}. 

Upon updating the initial guess with the expanded mesh points, suitable interpolation is necessary due to the newly appended meshes. While more sophisticated methods may exist for general hp-adaptive mesh refinement, we utilize straightforward linear interpolation for our purposes in trapezoidal method. In fact, any interpolation methods are acceptable for obtaining the updated initial guess. This is because the interpolation step is merely reconstruction step of the initial guess, not the actual solution process. 
%
%
\section{Numerical Demonstration}\label{sec:05}
Throughout this section, the numerical demonstration of the proposed mesh refinement algorithm is presented. For an illustrative example, the dynamics of the fixed wing unmanned aerial vehicles(UAVs) is considered, which is a typical example of curvature bounded dynamics applied under presence of holonomic constraints. Other practical examples include missile guidance problems~\cite{chlee2013} and autonomous underwater vehicles~\cite{yanji2017}. 
\subsection{Path Planning for Fixed--Wing UAVs with No-Fly Zones}
The problem is formulated with path constraints composed of multiple bounded regions, known as \textit{No--Fly Zones (NFZs)}. These regions represent areas that the vehicle is prohibited from traversing. Similar problems have been extensively explored in various studies, such as \cite{xiaoliang2017}. In the context of this problem, the objective function is to minimize energy consumption, while circular NFZs are considered as forbidden regions. The problem can be formulated in the structure of Eq.~\eqref{eq:1} as follows: 

minimize
\begin{equation} \label{eq:23}
	\enspace J  =  \int_{0}^{t_f} \|u(t)\|^2 \enspace dt
\end{equation}
subject to 
\begin{equation} \label{eq:24}
	\begin{cases}
		\dot{x} 		= 	V\cos\gamma \\
		\dot{y} 		= 	V\sin\gamma \\
		\dot{\gamma} 	= 	\frac{u}{V}
	\end{cases}
\end{equation}
\begin{equation} \label{eq:25}
\begin{split}
	& \left[ {x(0) ,\;y(0) ,\;\gamma(0) ,\;x(t_f) ,\;y(t_f) ,\; \gamma(t_f) \; } \right] \\
	& = \left[ {x_0 ,\;y_0 ,\;\gamma_0 ,\;x_f ,\;y_f ,\; \gamma_f \; } \right]
\end{split}
\end{equation}
\begin{equation} \label{eq:26}
	-u_{max} \leq u \leq u_{max}
\end{equation}
\begin{equation} \label{eq:27}
	(x - x_{c, i})^2 + (y - y_{c, i})^2 \geq r_{i}^2, \quad i = 1,\dots,N_{NFZ}
\end{equation}
where $t_f$ is considered as a free variable. 
The total number of NFZs is represented as $N_{NFZ}$, and the center and the radius of the $i$--th NFZ are denoted as $x_{c, i}$, $y_{c, i}$, and $r_{i}$, respectively. To further validate the real--time effectiveness of the proposed algorithm, we utilize the SCP framework for the direct formulation of this problem. As the convex formulation of this problem falls outside the scope of this paper, the direct formulation of this problem within the SCP framework is provided in Appendix~\ref*{appendix:01}. 

The numerical demonstration is conducted using Matlab 2019b on an Apple Macbook Air with an Apple M2 processor. A state--of--the--art dual--primal interior point algorithm is employed by calling the MOSEK solver~\cite{mosek} within Matlab. The maximum number of iterations for the SCP framework is set to 100. The initial guess for the state variables were obtained through linear interpolation between the initial and terminal values. The initial guess for the flight time was set at 300 seconds. Detailed information about the model parameters is summarized in Table~\ref{tbl:01}. 
\begin{table}[htbp]
\centering
\caption{Model Parameters} \label{tbl:01}
\begin{tabular}{|c|c|} 
	\hline 
	Parameter			 		& Value 						\\ 
	\hline
	$u_{max}$ 					& 100 	$[m/s^2]$ 			\\
	$V$ 						& 300 	$[m/s]$ 				\\
	$x_0$ 						& 0 	$[km]$ 				\\
	$x_f$  						& 50 	$[km]$ 				\\
	$y_0$  						& 0 	$[km]$ 				\\
	$y_f$  						& 50 	$[km]$ 				\\
	$\gamma_0$  				& $\frac{\pi}{4}$ 			\\
	$\gamma_f$  				& $\frac{\pi}{2}$ 			\\
	$t_f$  						& Free 		 				\\
	NFZ Tolerance 				& 50 	$[m]$ 	 			\\
	$[$NFZ 1 Location, Radius$]$ 	& $[$(8, 30), 	10$]$ $[km]$ 	\\
	$[$NFZ 2 Location, Radius$]$ 	& $[$(33, 30), 	12$]$ $[km]$ 	\\
	$[$NFZ 3 Location, Radius$]$ 	& $[$(18, 7), 	10$]$ $[km]$ 	\\
	$[$NFZ 4 Location, Radius$]$ 	& $[$(41, 41), 	6$]$ $[km]$ 		\\
	$[$NFZ 5 Location, Radius$]$ 	& $[$(3, 8), 	2.5$]$ $[km]$ 	\\
	$[$NFZ 6 Location, Radius$]$ 	& $[$(35, 15), 	18$]$ $[km]$ 	\\
	\hline
\end{tabular}
\end{table}
\subsection{Results}
\begin{figure}[ht]
	\begin{center}
	\resizebox{84mm}{!}{\includegraphics{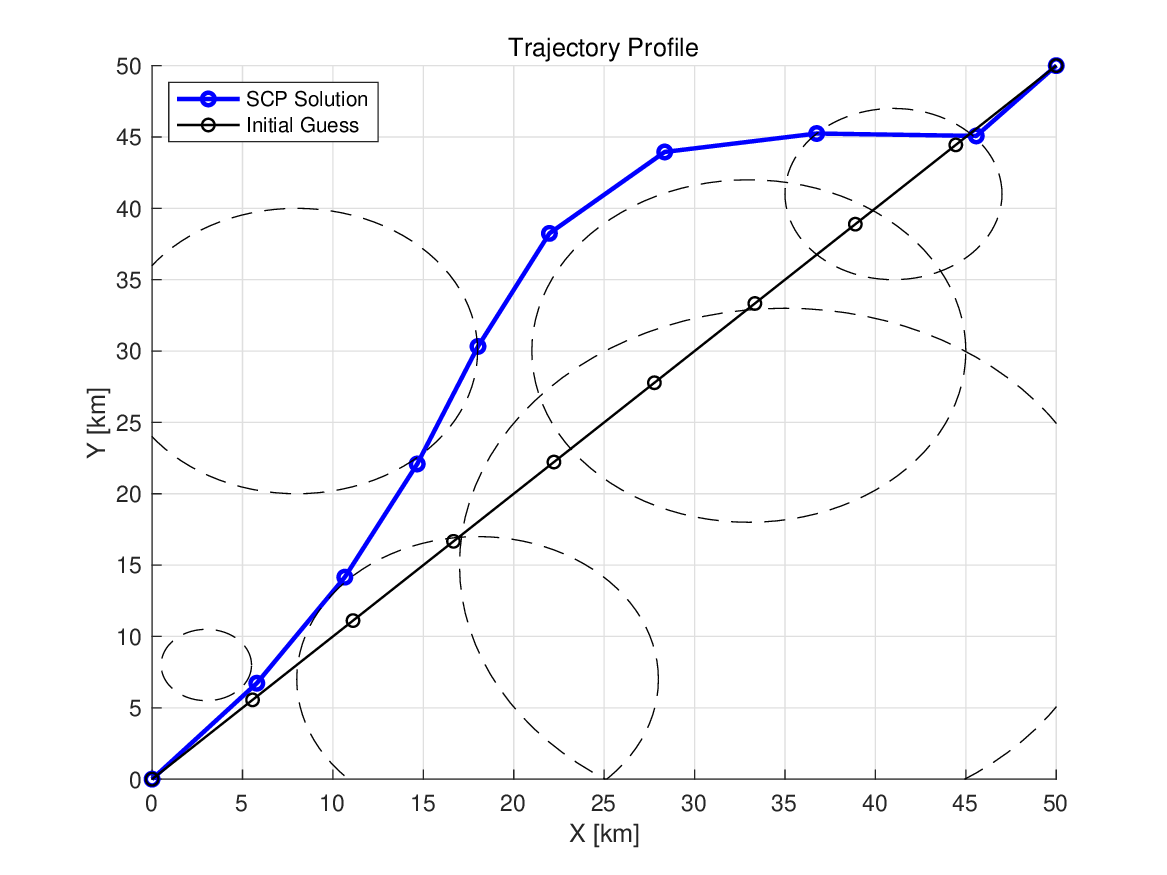}}
	\resizebox{84mm}{!}{\includegraphics{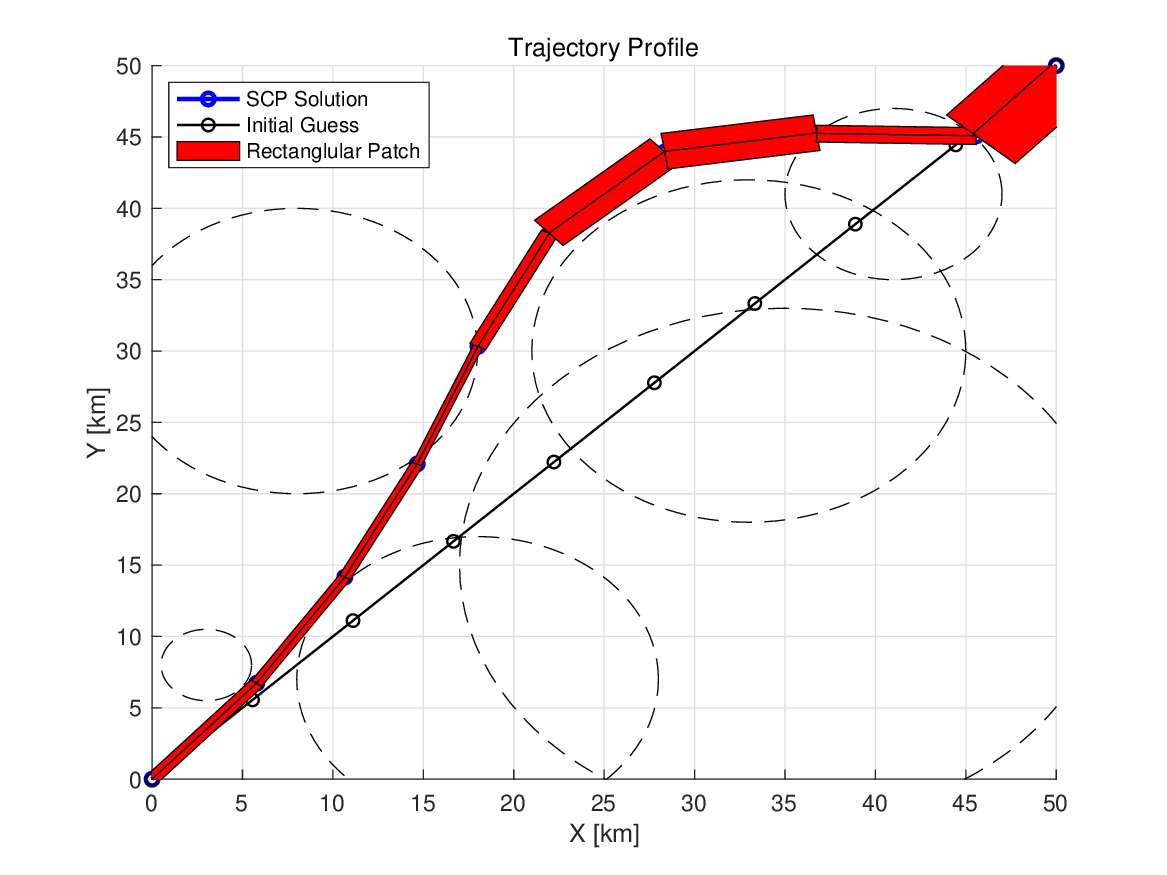}}
	\caption{Results Without the Proposed Mesh Refinement Algorithm} \label{demo:nonadaptive_all} 
	\end{center}
\end{figure}

\begin{figure}[ht]
	\begin{center}
	\resizebox{84mm}{!}{\includegraphics{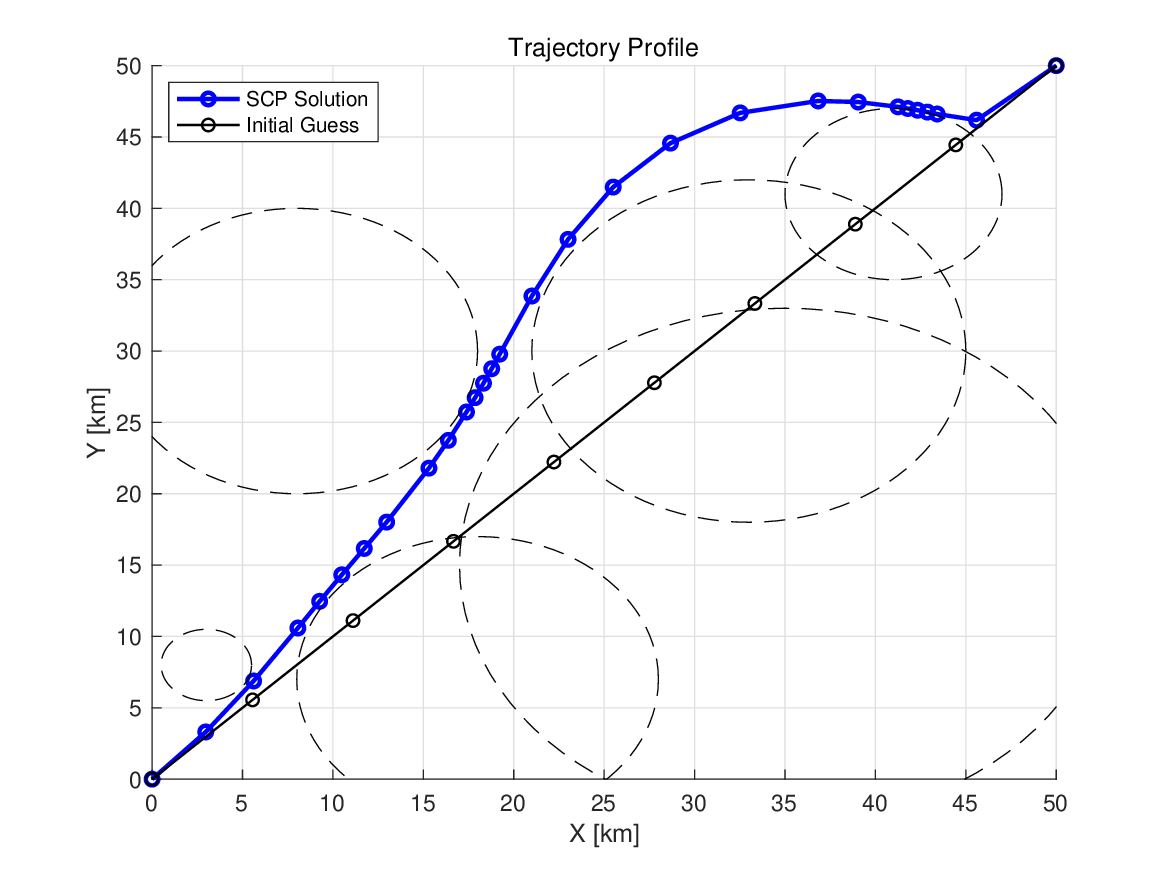}}
	\resizebox{84mm}{!}{\includegraphics{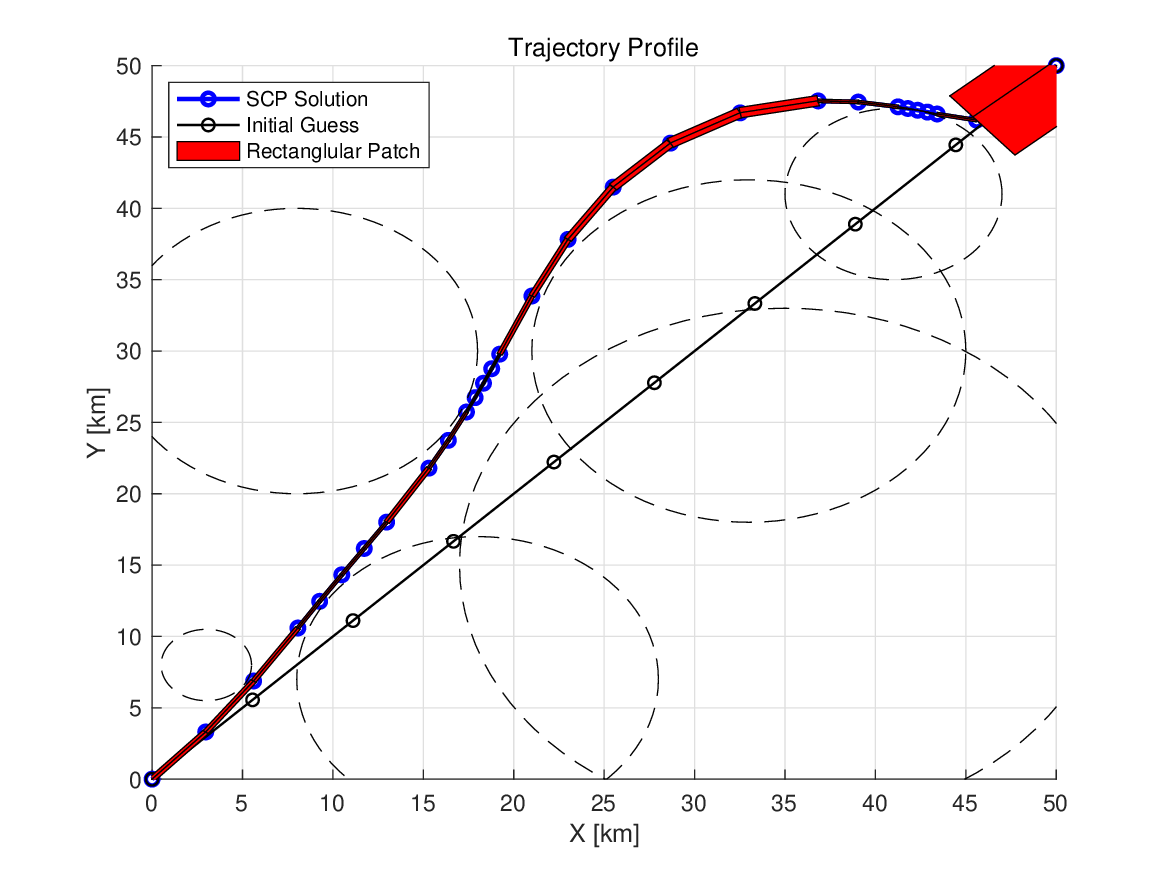}}
	\caption{Results With the Proposed Mesh Refinement Algorithm} \label{demo:adaptive_all} 
	\end{center}
\end{figure}

\begin{figure}[ht]
	\begin{center}
	\resizebox{84mm}{!}{\includegraphics{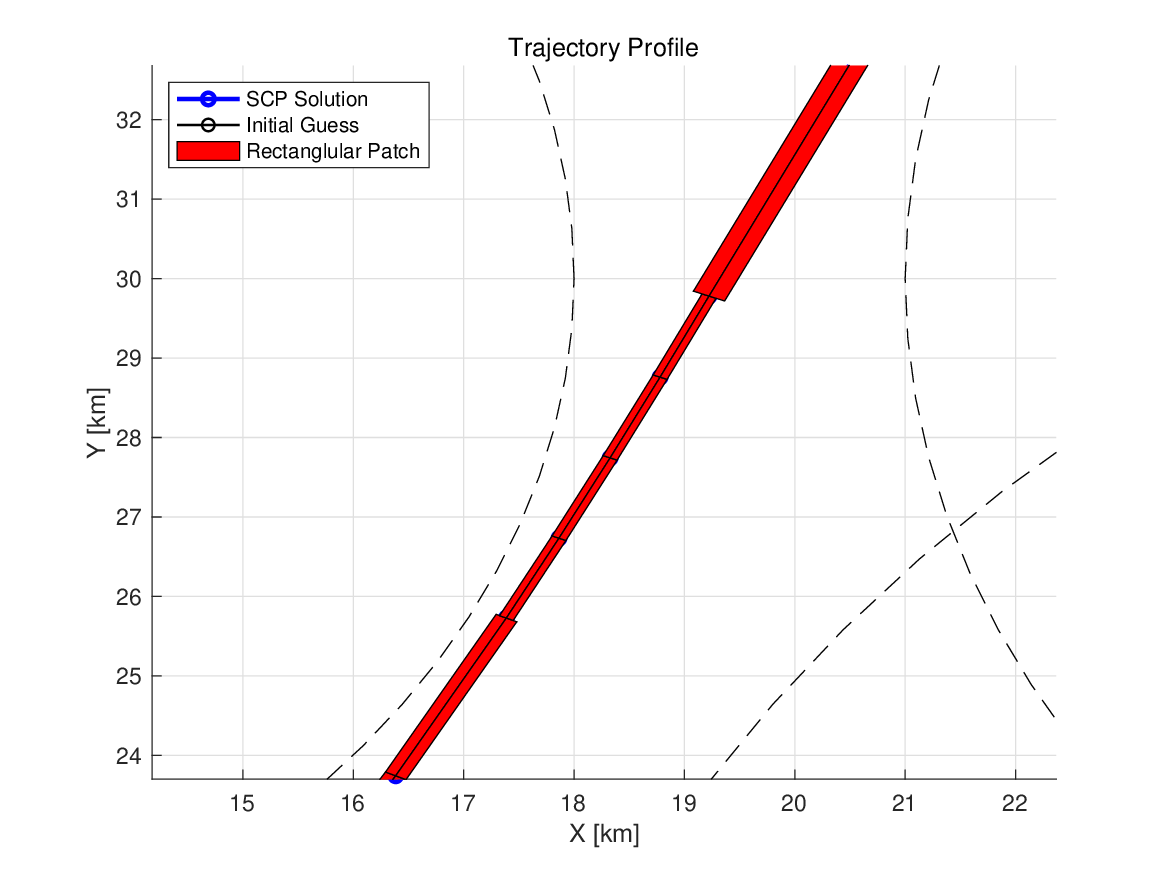}}
	\resizebox{84mm}{!}{\includegraphics{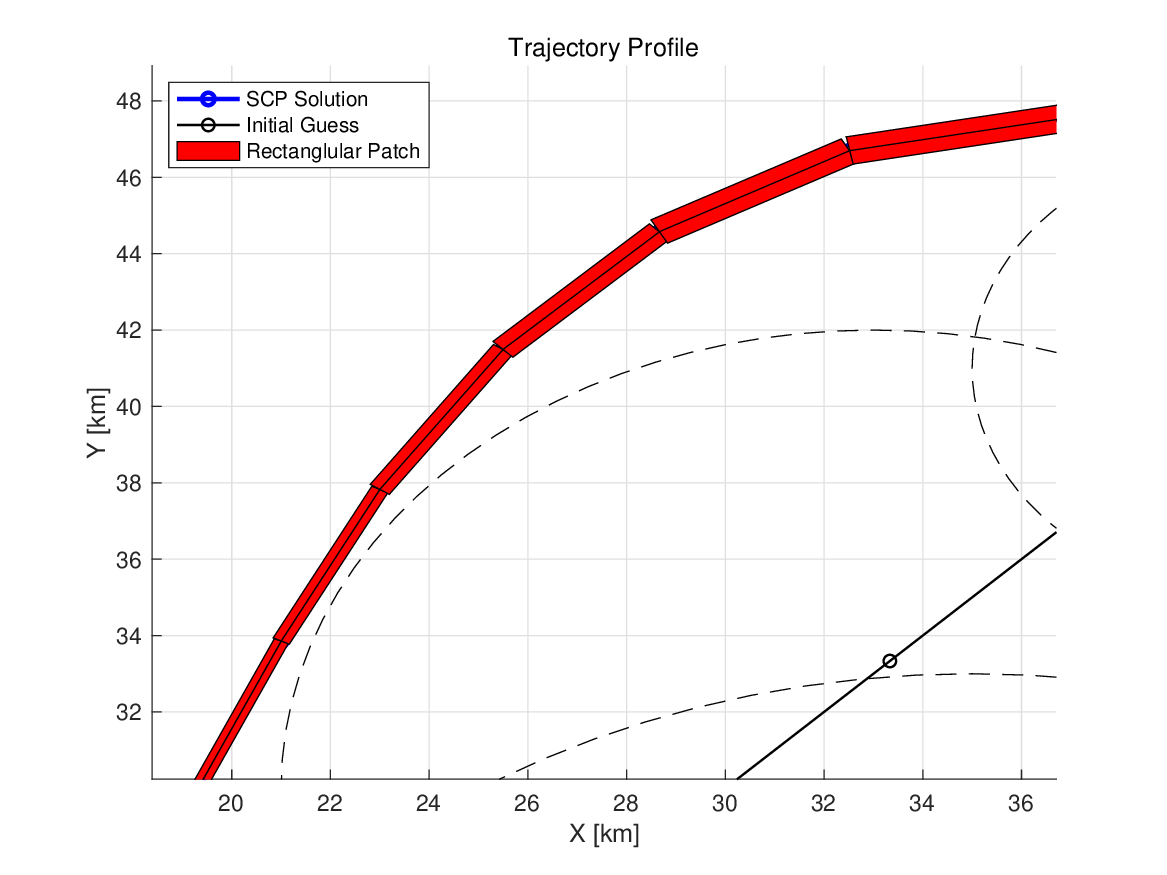}}
	\caption{Rectangular Patches} \label{demo:adaptive_patches}
	\end{center}
\end{figure}

A comparison result between the solutions obtained through the SCP framework with and without the proposed mesh refinement algorithm is presented throughout this subsection. The initial mesh is given by 10 equi--spaced mesh points. It is essential to recognize that the prevailing literature on mesh refinement algorithms, exemplified by studies such as~\cite{liu2015adaptive} and \cite{ma2016trajectory}, intimates that a mere 10 mesh points may not suffice to attain precise solutions. Nonetheless, for illustrative purposes, this analysis employs an initial mesh of 10 equi--spaced mesh points to accentuate the significance of the proposed algorithm. The converged trajectories obtained from the SCP framework are depicted in Figs.~\ref{demo:nonadaptive_all} and \ref{demo:adaptive_all}. The rectangular patches are depicted in Fig.~\ref{demo:adaptive_patches}. 
In the absence of the proposed mesh refinement algorithm, it required 6 iterations for the SCP framework to converge. It is evident that the mesh points depicted in Fig.~\ref{demo:nonadaptive_all} exhibits potential collision with the NFZs. Conversely, using the proposed method, the entire SCP framework converged within 8 iterations. Notably, a convergent mesh consisting of 29 mesh points is attained after the 5--th iteration. 
The following points necessitate more detailed explanations:
\begin{enumerate}
	\item At the vicinity of the NFZ boundaries, high mesh point density is often observed, but not consistently. 
	\item There are typically two rectangular patches at each mesh intervals that contribute to forming the overall bound. 
\end{enumerate}
The first aspect can be attributed as follows. The proposed algorithm does not eliminate mesh points, even when the mesh density appears sufficient to prevent violations into forbidden regions. This cautious approach is adopted because the removal of a mesh point requires careful consideration; it could exacerbate interpolation errors and lead to inaccuracies in the solution. Inadvertent elimination might even result in unintended violations into forbidden regions. On the other hand, these concerns do not apply when adding a mesh point, as this typically enhances the solution's accuracy without introducing such risks. Therefore, as the refinement process is incorporated during the iterations when the solution is converging to the optimum, those points that were previously in close proximity to the boundary of the forbidden region could exhibit high mesh densities, even when it is not near the forbidden region in the convergent solution. In this regards, the development of more sophisticated algorithms for eliminating unnecessary mesh points warrants further investigation in future works. 

The second aspect concerns the ratio between the prescribed lengths and the distance between the mesh intervals. The lengths of the curves were relatively short compared to the spacing between the mesh points, typically falling within a ratio scale of $1.00\mathrm{xx}$. Consequently, the initial and terminal points establish the horizontal boundaries as depicted in Fig.~\ref{fig_sec4_6} (c), whereas the CSCSC curves define the vertical boundaries. This leads to a situation where two CSCSC curves which determine the vertical boundaries originate from two of the four possible scenarios of left and right combination. As a result, only the two scenarios out of the four typically contribute to the formation of the overall boundary. 
\subsection{Analysis on Computational Time}
To evaluate the viability of onboard implementation of the entire algorithm for real--time applications, examination of computational load imposed by the proposed algorithm is imperative. The average computational time over iterations is detailed in Table.~\ref{tbl:02}. In the table, the mesh refinement component, corresponding to lines $3 \sim 11$ in Algorithm~\ref{alg:02}, is referred to as \textit{MR}, an abbreviation for mesh refinement. The outcomes demonstrate that the proposed algorithm does not introduce significant computational overhead that might hinder real--time feasibility. Moreover, the inclusion of additional mesh points did not significantly elevate the computational burden, resulting in similar CPU times for the SCP subproblems, both with and without the mesh refinement. Given that traditional NLP algorithms~\cite{gpops} require considerably more computational cost than SCP algorithms, the proposed algorithm can be seamlessly integrated with a range of direct method algorithms without incurring substantial computational expenses. 
\begin{table}[htbp]
\centering
\caption{Average Computational Time Over Iterations}
\begin{tabular}{|c|c|c|} 
\hline 
Average Time [sec] 			& MR 						& SCP Subproblem 		\\
\hline
Without MR  				& --   						& 0.0011					\\ 
\hline
With MR  					& 0.0040  					& 0.0019					\\ 
\hline
\end{tabular}
\label{tbl:02}
\end{table}
%
%
\section{Conclusion}\label{sec:06}
This paper introduces a novel strategy to resolve inter--sample collision problem of 2--dimensional state constraints by dint of mesh refinement. Compliance of the path constraint in between the mesh points is proved, as well as the convergence of the mesh refinement process. The proposed method can be applied to general trajectory optimization problems formulated in direct method, independent of the solver types: NLP or convex. Through numerical demonstration on fixed wing dynamics problem, the potential of real--time application is confirmed as well. Throughout this development, we propose and prove a method for constructing the reachable region by curves with a prescribed length, curvature bound, and specified initial and terminal locations and directions. This result expands the upon concept of the reachability set as covered in the literature, including \cite{patsko2003}, \cite{dubins1957}, \cite{cockayne1975}, and \cite{chen2023}, extending these concepts into their envelope. 
%
%
\begin{ack}                               
We gratefully acknowledge Nearthlab for providing wholehearted support and assistance to this paper. 
\end{ack}
%
%
%
\appendix
\section{Convex Formulation of the Fixed--Wing UAV Problem}\label{appendix:01}
The non--convex constraints in the fixed--wing UAV problem are Eqs.~\eqref{eq:24} and~\eqref{eq:27}. These constraints are casted into suitable constraints of a convex optimization problem through linearization. Following the standard linearization process in~\cite{mceowen2023}, we first linearize the nonlinear dynamics function. The dynamics function in Eq.~\eqref{eq:24} is in control affine form of $f(\boldsymbol{z}, u) = \tilde{f}(\boldsymbol{z}) + B u$ where $\tilde{f}(\boldsymbol{z}) = [ V\cos\gamma ,\; V\sin\gamma ,\; 0]^T$ and $B = \left[ 0 ,\; 0 ,\; \frac{1}{V} \right]^T$ for state variable $\boldsymbol{z} = [ x ,\; y ,\; \gamma]^T$. Then for the solution pair, $\left\{ \boldsymbol{z}^{(k)}, u^{(k)} \right\}$, obtained from the previous $k$--th iteration, define 
\begin{equation} \label{eq:28}
	A\left( \boldsymbol{z}^{(k)} \right) \equiv \left. \frac{\partial \tilde{f}}{\partial \boldsymbol{z}} \right\vert_{\boldsymbol{z} = \boldsymbol{z}^{(k)}} = \begin{pmatrix} 0 & 0 & -V \sin\gamma^{(k)} \\ 0 & 0 & V \cos\gamma^{(k)} \\ 0 & 0 & 0 \end{pmatrix}
\end{equation}
\begin{equation} \label{eq:29}
	b\left( \boldsymbol{z}^{(k)} \right) = \tilde{f}\left( \boldsymbol{z}^{(k)} \right) - A\left( \boldsymbol{z}^{(k)} \right) \boldsymbol{z}^{(k)}
\end{equation}
Then the nonlinear dynamics is linearized as below: 
\begin{equation} \label{eq:30}
	f(\boldsymbol{z}, u) \approx A\left( \boldsymbol{z}^{(k)} \right) \boldsymbol{z} + Bu + b\left( \boldsymbol{z}^{(k)} \right)
\end{equation}
The nonconvex path inequality constriant, Eq.~\eqref{eq:26}, is linearized as below:
\begin{equation} \label{eq:31}
\begin{split}
	& 2\left( x^{(k)} - x_{c, i} \right) \left( x - x^{(k)} \right) + 2\left( y^{(k)} - y_{c, i} \right) \left( y - y^{(k)} \right) \\
	& - r_{i}^2 \geq 0, \quad i = 1,\dots,N_{NFZ}
\end{split}
\end{equation}
After such linearization, the trust region constraint(TRC) is added to prevent unboundedness of the convexified problem as below~\cite{mceowen2023} : 
\begin{equation} \label{eq:32}
	| \boldsymbol{z} - \boldsymbol{z}^{(k)} | \leq \varepsilon_{TRC}
\end{equation}
where $\varepsilon_{TRC} \in \mathbb{R}^3$ is a constant representing termination criterion. The inequality is applied componentwise. 

To summarize, the optimal control problem described by Eqs.~\eqref{eq:23} $\sim$~\eqref{eq:27} is linearized through two steps. Firstly, the nonlinear functions given by Eqs.~\eqref{eq:24} and~\eqref{eq:27} are replaced with their linearized counterparts, as shown in Eqs.~\eqref{eq:30} and~\eqref{eq:31}. Secondly, the trust region constraint, as represented by Eq.~\eqref{eq:32}, is introduced. 
Then the subsequent discretization process is as described in Sec.~\ref{sec:02}. 
%
%
%
\bibliographystyle{plain}        
\bibliography{Ref_Papers.bib}           

\begin{thebibliography}{10}

\bibitem{mosek}
Erling Andersen and Knud Andersen.
\newblock The mosek interior point optimizer for linear programming: An
  implementation of the homogeneous algorithm.
\newblock 33, 01 1999.

\bibitem{accikmecse2011lossless}
Behçet Açıkmeşe and Lars Blackmore.
\newblock Lossless convexification of a class of optimal control problems with
  non-convex control constraints.
\newblock {\em Automatica}, 47(2):341--347, 2011.

\bibitem{bae2022}
Juho Bae, Sang-Don Lee, Young-Won Kim, Chang-Hun Lee, and Sung-Yug Lim.
\newblock Convex optimization-based entry guidance for spaceplane.
\newblock {\em International Journal of Control, Automation and Systems}, 2022.

\bibitem{chen2023}
Zheng Chen, Kun Wang, and Heng Shi.
\newblock Elongation of curvature-bounded path.
\newblock {\em Automatica}, 151:110936, 2023.

\bibitem{cockayne1975}
E.~J. Cockayne and G.~W.~C. Hall.
\newblock Plane motion of a particle subject to curvature constraints.
\newblock {\em SIAM Journal on Control}, 13(1):197--220, 1975.

\bibitem{dubins1957}
L.~E. Dubins.
\newblock On curves of minimal length with a constraint on average curvature,
  and with prescribed initial and terminal positions and tangents.
\newblock {\em American Journal of Mathematics}, 79(3):497--516, 1957.

\bibitem{dueri2017}
Daniel Dueri, Yuanqi Mao, Zohaib Mian, Jerry Ding, and Behçet Açıkmeşe.
\newblock Trajectory optimization with inter-sample obstacle avoidance via
  successive convexification.
\newblock pages 1150--1156, 12 2017.

\bibitem{garg2010unified}
Divya Garg, Michael Patterson, William~W. Hager, Anil~V. Rao, David~A. Benson,
  and Geoffrey~T. Huntington.
\newblock A unified framework for the numerical solution of optimal control
  problems using pseudospectral methods.
\newblock {\em Automatica}, 46(11):1843--1851, 2010.

\bibitem{gusev2017}
Mikhail Gusev and Igor Zykov.
\newblock On extremal properties of boundary points of reachable sets for a
  system with integrally constrained control.
\newblock {\em IFAC-PapersOnLine}, 50:4082--4087, 07 2017.

\bibitem{hartl1995survey}
Richard~F. Hartl, Suresh~P. Sethi, and Raymond~G. Vickson.
\newblock A survey of the maximum principles for optimal control problems with
  state constraints.
\newblock {\em SIAM Review}, 37(2):181--218, 1995.

\bibitem{chlee2013}
Chang-Hun Lee, Tae-Hun Kim, Min-Jea Tahk, and Ick-Ho Whang.
\newblock Polynomial guidance laws considering terminal impact angle and
  acceleration constraints.
\newblock {\em IEEE Transactions on Aerospace and Electronic Systems},
  49(1):74--92, 2013.

\bibitem{Lee1967}
Ernest~Bruce Lee and Lawrence Markus.
\newblock Foundations of optimal control theory.
\newblock 1967.

\bibitem{liu2015adaptive}
Fengjin Liu, William~W. Hager, and Anil~V. Rao.
\newblock Adaptive mesh refinement method for optimal control using
  nonsmoothness detection and mesh size reduction.
\newblock {\em Journal of the Franklin Institute}, 352(10):4081--4106, 2015.

\bibitem{liu2017adaptive}
Fengjin Liu, William~W. Hager, and Anil~V. Rao.
\newblock Adaptive mesh refinement method for optimal control using decay rates
  of legendre polynomial coefficients.
\newblock {\em IEEE Transactions on Control Systems Technology},
  26(4):1475--1483, 2018.

\bibitem{liu2016entry}
Xinfu Liu, Zuojun Shen, and Ping Lu.
\newblock Entry trajectory optimization by second-order cone programming.
\newblock {\em Journal of Guidance, Control, and Dynamics}, 39(2):227--241,
  2016.

\bibitem{yanji2017}
Yanji Liu, Jie Ma, Ning Ma, and Guichen Zhang.
\newblock Path planning for underwater glider under control constraint.
\newblock {\em Advances in Mechanical Engineering}, 9(8):1687814017717187,
  2017.

\bibitem{ma2016trajectory}
Lin Ma, Zhijiang Shao, Weifeng Chen, and Zhengyu Song.
\newblock Trajectory optimization for lunar soft landing with a
  hamiltonian-based adaptive mesh refinement strategy.
\newblock {\em Advances in Engineering Software}, 100:266--276, 2016.

\bibitem{malyuta2022}
Danylo Malyuta, Taylor Reynolds, Michael Szmuk, Thomas Lew, Riccardo Bonalli,
  Marco Pavone, and Behçet Açıkmeşe.
\newblock Convex optimization for trajectory generation : A tutorial on
  generating dynamically feasible trajectories reliably and efficiently.
\newblock {\em IEEE Control Systems}, 42(5):40--113, 10 2022.

\bibitem{mao2018}
Yuanqi Mao, Michael Szmuk, and Behçet Açıkmeşe.
\newblock Successive convexification: A superlinearly convergent algorithm for
  non-convex optimal control problems.
\newblock 04 2018.

\bibitem{mceowen2023}
Skye Mceowen, Abhinav~G Kamath, Purnanand Elango, Taewan Kim, Samuel~C Buckner,
  and Behcet Acikmese.
\newblock High-accuracy 3-dof hypersonic reentry guidance via sequential convex
  programming.
\newblock In {\em AIAA SCITECH 2023 Forum}, page 0300, 2023.

\bibitem{oh2019}
Young-Jae Oh, Heekun Roh, and Min-Jea Tahk.
\newblock Fast trajectory optimization using sequential convex programming with
  no-fly zone constraints.
\newblock {\em IFAC-PapersOnLine}, 52(12):298--303, 2019.
\newblock 21st IFAC Symposium on Automatic Control in Aerospace ACA 2019.

\bibitem{patsko2003}
Valerii Patsko, S.G. Pyatko, and Andrey Fedotov.
\newblock Three-dimensional reachability set for a nonlinear control system.
\newblock {\em Journal of Computer and Systems Sciences International}, 42, 05
  2003.

\bibitem{patterson2015ph}
Michael~A Patterson, William~W Hager, and Anil~V Rao.
\newblock A ph mesh refinement method for optimal control.
\newblock {\em Optimal Control Applications and Methods}, 36(4):398--421, 2015.

\bibitem{gpops}
Michael~A. Patterson and Anil~V. Rao.
\newblock Gpops-ii: A matlab software for solving multiple-phase optimal
  control problems using hp-adaptive gaussian quadrature collocation methods
  and sparse nonlinear programming.
\newblock 41(1), oct 2014.

\bibitem{sagliano2018pseudospectral}
Marco Sagliano.
\newblock Pseudospectral convex optimization for powered descent and landing.
\newblock {\em Journal of Guidance, Control, and Dynamics}, 41(2):320--334,
  2018.

\bibitem{sagliano2019generalized}
Marco Sagliano.
\newblock Generalized hp pseudospectral-convex programming for powered descent
  and landing.
\newblock {\em Journal of Guidance, Control, and Dynamics}, 42(7):1562--1570,
  2019.

\bibitem{sagliano2016onboard}
Marco Sagliano, Erwin Mooij, and Stephan Theil.
\newblock Onboard trajectory generation for entry vehicles via adaptive
  multivariate pseudospectral interpolation.
\newblock {\em Journal of Guidance, Control, and Dynamics}, 40(2):466--476,
  2016.

\bibitem{szmuk2017successive}
Michael Szmuk, Utku Eren, and Behcet Acikmese.
\newblock Successive convexification for mars 6-dof powered descent landing
  guidance.
\newblock In {\em AIAA Guidance, Navigation, and Control Conference}, page
  1500, 01 2017.

\bibitem{szmuk2020}
Michael Szmuk, Taylor Reynolds, and Behçet Açıkmeşe.
\newblock Successive convexification for real-time six-degree-of-freedom
  powered descent guidance with state-triggered constraints.
\newblock {\em Journal of Guidance, Control, and Dynamics}, 43(8):1399--1413,
  06 2020.

\bibitem{tordesillas2019}
Jesus Tordesillas, Brett~T. Lopez, and Jonathan~P. How.
\newblock Faster: Fast and safe trajectory planner for flights in unknown
  environments.
\newblock In {\em 2019 IEEE/RSJ International Conference on Intelligent Robots
  and Systems (IROS)}, pages 1934--1940, 2019.

\bibitem{xiaoliang2017}
Xiaoliang Wang, Peng Jiang, Deshi Li, and Tao Sun.
\newblock Curvature continuous and bounded path planning for fixed-wing uavs.
\newblock {\em Sensors}, 17(9), 2017.

\bibitem{wang2017constrained}
Zhenbo Wang and Michael~J. Grant.
\newblock Constrained trajectory optimization for planetary entry via
  sequential convex programming.
\newblock {\em Journal of Guidance, Control, and Dynamics}, 40(10):2603--2615,
  2017.

\bibitem{wang2020improved}
Zhenbo Wang and Ye~Lu.
\newblock Improved sequential convex programming algorithms for entry
  trajectory optimization.
\newblock {\em Journal of Spacecraft and Rockets}, 57(6):1373--1386, 2020.

\bibitem{zhou2021sequential}
Xiang Zhou, Rui-Zhi He, Hong-Bo Zhang, Guo-Jian Tang, and Wei-Min Bao.
\newblock Sequential convex programming method using adaptive mesh refinement
  for entry trajectory planning problem.
\newblock {\em Aerospace Science and Technology}, 109:106374, 2021.

\end{thebibliography}



\end{document}